\documentclass[12pt]{article}
\usepackage{a4wide,epsfig,psfrag,amsmath,amssymb,cite,scalefnt}
\usepackage{color}

\graphicspath{ {figs/} }

\allowdisplaybreaks

\begin{document}

\title{\vskip-3cm{\baselineskip14pt
    \begin{flushleft}
      \normalsize DESY 16-106, TTP16-023
    \end{flushleft}} \vskip1.5cm 
  $\overline{\rm MS}$-on-shell quark mass
  relation up to four loops in QCD and a general SU$(N)$ gauge group}

\author{
  Peter Marquard$^{a}$,
  Alexander V. Smirnov$^{b}$,
  \\
  Vladimir A. Smirnov$^{c}$,
  Matthias Steinhauser$^{d}$,
  David Wellmann$^{d}$
  \\[1em]
  {\small\it (a) Deutsches Elektronen-Synchrotron, DESY}
  {\small\it  15738 Zeuthen, Germany}
  \\
  {\small\it (b) Research Computing Center, Moscow State University}\\
  {\small\it 119991, Moscow, Russia}
  \\  
  {\small\it (c) Skobeltsyn Institute of Nuclear Physics of Moscow State University}\\
  {\small\it 119991, Moscow, Russia}
  \\
  {\small\it (d) Institut f{\"u}r Theoretische Teilchenphysik,
    Karlsruhe Institute of Technology (KIT)}\\
  {\small\it 76128 Karlsruhe, Germany}  
}
  
\date{}

\maketitle

\thispagestyle{empty}

\begin{abstract}

  We compute the relation between heavy quark masses defined in
  the modified minimal subtraction and the on-shell schemes. Detailed results are
  presented for all coefficients of the SU$(N_c)$ colour factors.  The
  reduction of the four-loop on-shell integrals is performed for a general QCD
  gauge parameter. Altogether there are about 380 master integrals. Some
  of them are computed
  analytically, others with high numerical precision using Mellin-Barnes
  representations, and the rest numerically with the help of {\tt FIESTA}.  We
  discuss in detail the precise numerical evaluation of the four-loop master
  integrals.  Updated relations between various short-distance masses and the
  $\overline{\rm MS}$ quark mass to next-to-next-to-next-to-leading order accuracy
  are provided for the charm, bottom and top quarks. We discuss the dependence
  on the renormalization and factorization scale.

  \medskip


\end{abstract}

\thispagestyle{empty}


\newpage


\section{Introduction}

Quark masses are fundamental parameters of Quantum Chromodynamics
(QCD) and thus it is mandatory to determine their numerical values as
precisely as possible. Furthermore, it is important to have precise
relations at hand which relate the masses in 
different renormalization schemes.

The renormalization scheme for the quark masses has to be fixed once
quantum corrections are considered.  In QCD there are two distinct
renormalization schemes for the quark masses: the on-shell (OS) scheme,
which is motivated by the physical interpretation of the mass
parameter, and the modified minimal subtraction ($\overline{\rm MS}$)
scheme which is very convenient for many practical calculations, in
particular in high-energy processes.

In the case of the lighter quarks (up, down and strange) the meson
masses are in general much larger than the masses of the quarks.
Thus, the concept of the on-shell scheme is not applicable to light
quark flavours; their numerical values are usually given in the
$\overline{\rm MS}$ scheme.  On the other hand, the masses of the
mesons involving
charm and bottom quarks are essentially dominated by the quark masses.
For this reason, the quantum corrections considered in this paper are
mainly relevant for the three heavy quarks, charm, bottom and top.

The top quark plays a special role in this context. Due to its large
width it decays before hadronization and thus can be considered as
an almost free quark. As a consequence it can be expected that
the on-shell value for the top quark can be determined with 
a relatively small uncertainty. This aspect has been studied in detail
in recent papers~\cite{Nason:2016tiy,Beneke:2016cbu}.
It has been shown that the on-shell top quark can be computed from the
$\overline{\rm MS}$ mass with an irreducible uncertainty of about
70~MeV~\cite{Beneke:2016cbu}.

There are various methods which can be used to obtain numerical values 
for the quark masses. Some of them determine directly the $\overline{\rm MS}$
quark mass (see, e.g., Ref.~\cite{Chetyrkin:2009fv}) and thus do not
suffer from the inherent renormalon ambiguity.  However, the highest
sensitivity to the quark masses is in general obtained from physical
quantities evaluated at energies close to the quark mass.
In such situations it is convenient to
introduce so-called threshold masses to parametrize the physical quantities.
Among the most
prominent ones are the potential subtracted (PS)~\cite{Beneke:1998rk},
1S~\cite{Hoang:1998hm,Hoang:1998ng,Hoang:1999zc}, renormalon
subtracted (RS)~\cite{Pineda:2001zq} and the kinetic
mass~\cite{Czarnecki:1997sz}.\footnote{Note that the relation of
  the kinetic mass to the on-shell mass is currently only known to
  NNLO. For this reason it will not be considered in the following.} They
allow for a precise determination of the heavy $\overline{\rm MS}$
mass without explicit reference to the pole quark mass. However, at intermediate
stages the pole mass and, in particular, the relation between the pole
and the $\overline{\rm MS}$ mass is still needed.

In the following we describe three typical examples where the 
four-loop terms in the mass relations turn out to be important.
\begin{itemize}

\item At the TEVATRON and the LHC the top quark mass is measured with
  an uncertainty below 1~GeV. For example, the combination of results from
  ATLAS, CDF, CMS and D0 from March 2014~\cite{ATLAS:2014wva}
  leads to
  \begin{eqnarray}
    M_t &=& 173.34 \pm 0.27({\rm stat}) \pm 0.71({\rm syst})~\mbox{GeV}
    \label{eq::mtmass}
    \,,
  \end{eqnarray}
  with a total uncertainty of 760~MeV.
  The value in Eq.~(\ref{eq::mtmass}) is often called ``Monte-Carlo mass''
  and there are several attempts which suggest methods to relate it to
  the on-shell mass (see, e.g.,
  Refs.~\cite{Skands:2007zg,Kawabataa:2014osa,Kieseler:2015jzh}).
  In case Eq.~(\ref{eq::mtmass}) is interpreted as the
  on-shell quark mass it has to be converted to the 
  $\overline{\rm MS}$ top quark mass. Note that the three-loop term
  in the conversion formulae contributes approximately $500$~MeV
  which is of the same order as the experimental uncertainties
  in Eq.~(\ref{eq::mtmass}).

\item From measurements of the top quark pair production cross section
  close to threshold at a future linear collider it will be possible
  to determine the top quark threshold mass with an accuracy below
  100~MeV (see, e.g., Refs.~\cite{Beneke:2015kwa,Simon:2016htt}).  
  In the conversion to the $\overline{\rm MS}$
  definition there is a contribution of about 150-200~MeV from the
  three-loop term in the mass relations which contributes significantly to
  the final uncertainty of the $\overline{\rm MS}$ mass (see
  Section~\ref{sub::thr} for precise numbers). With the help
  of the four-loop $\overline{\rm MS}$-on-shell relation this
  uncertainty can be drastically reduced.

  For the sake of completeness let us mention that there is an approach
    to determine directly the $\overline{\rm MS}$ top quark mass from
    the threshold cross section (see, e.g., Ref.~\cite{Kiyo:2015ooa}).
    In future it will be interesting to compare the top quark mass
    values obtained with different methods.

\item The bottom quark mass can be extracted from $\Upsilon$ sum rules
  (see Refs.~\cite{Penin:2014zaa,Beneke:2014pta} for recent N$^3$LO
  analyses) and from 
  $M(\Upsilon(1S))$~\cite{Penin:2002zv,Ayala:2014yxa,Kiyo:2015ufa,Ayala:2016sdn}. Usually,
  in a first step a threshold mass is obtained.  To be able to compare
  with the $\overline{\rm MS}$ quark mass (as, e.g., extracted from
  low-moment sum rules~\cite{Chetyrkin:2009fv}) one has to apply the
  corresponding conversion formula. At three loops the contribution is
  of the order of 30~MeV, which is of the same order of magnitude
  (in some cases even larger) than the combination of all
  other uncertainties involved.

\end{itemize}

These examples show that the three-loop contribution is sizeable and
a reliable estimate of the uncertainty is only obtained once the
four-loop corrections are available. Furthermore, note that
for the PS, 1S and RS masses one knows the relation to the pole mass to
N$^3$LO. However, due to strong cancellations (see below) the N$^3$LO term
cannot be used unless four-loop corrections to the $\overline{\rm MS}$ and
on-shell quark mass are available.

The remainder of the paper is organized as follows: In the next
Section we introduce the conversion factor between the on-shell and
the $\overline{\rm MS}$ mass and discuss the colour decomposition of
the four-loop term. Furthermore, we provide several technical details
and discuss, in particular, the numerical accuracy of the master
integrals. Section~\ref{sec::MSOS} is devoted to the results of the
$\overline{\rm MS}$-on-shell relation which we discuss for the
physical limit, i.e. $N_c=3$ and fixed number of massless quarks, $n_l$,
but also for generic $N_c$ and even for general SU$(N_c)$ colour factors.
Several applications of the $\overline{\rm MS}$-on-shell relation are
discussed in Section~\ref{sec::appl} and our conclusions are contained
in Section~\ref{sec::concl}.


\section{Technicalities}


\subsection{Mass relations}

\begin{figure}[t]
  \centering
    \begin{tabular}{cccc}
    \includegraphics[width=.22\textwidth]{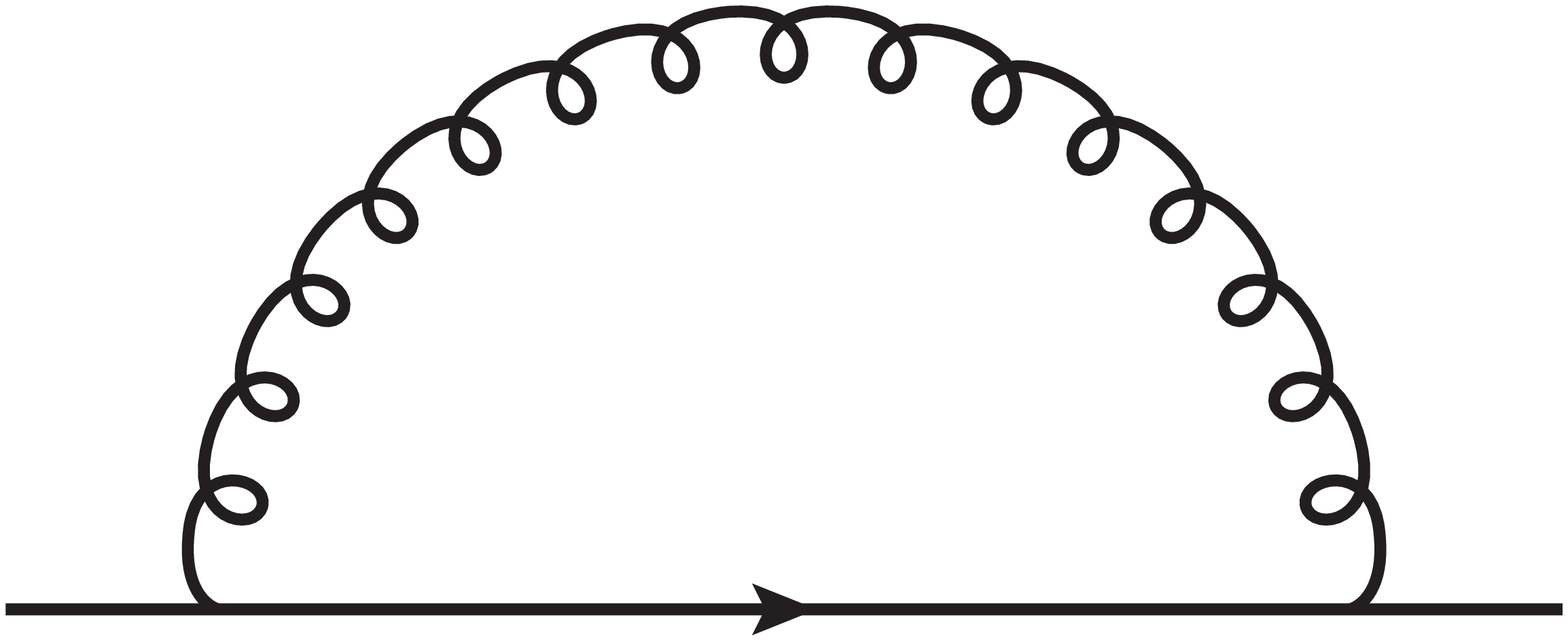} &
    \includegraphics[width=.22\textwidth]{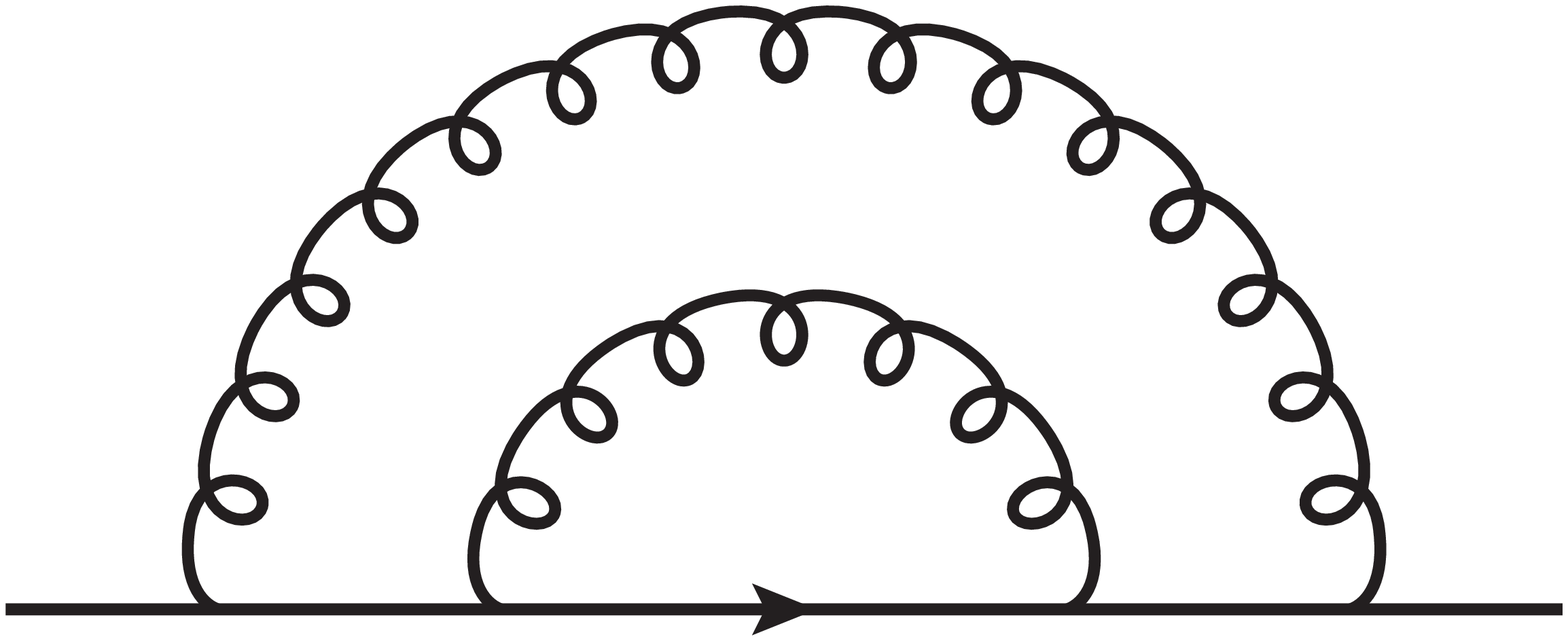} &
    \includegraphics[width=.22\textwidth]{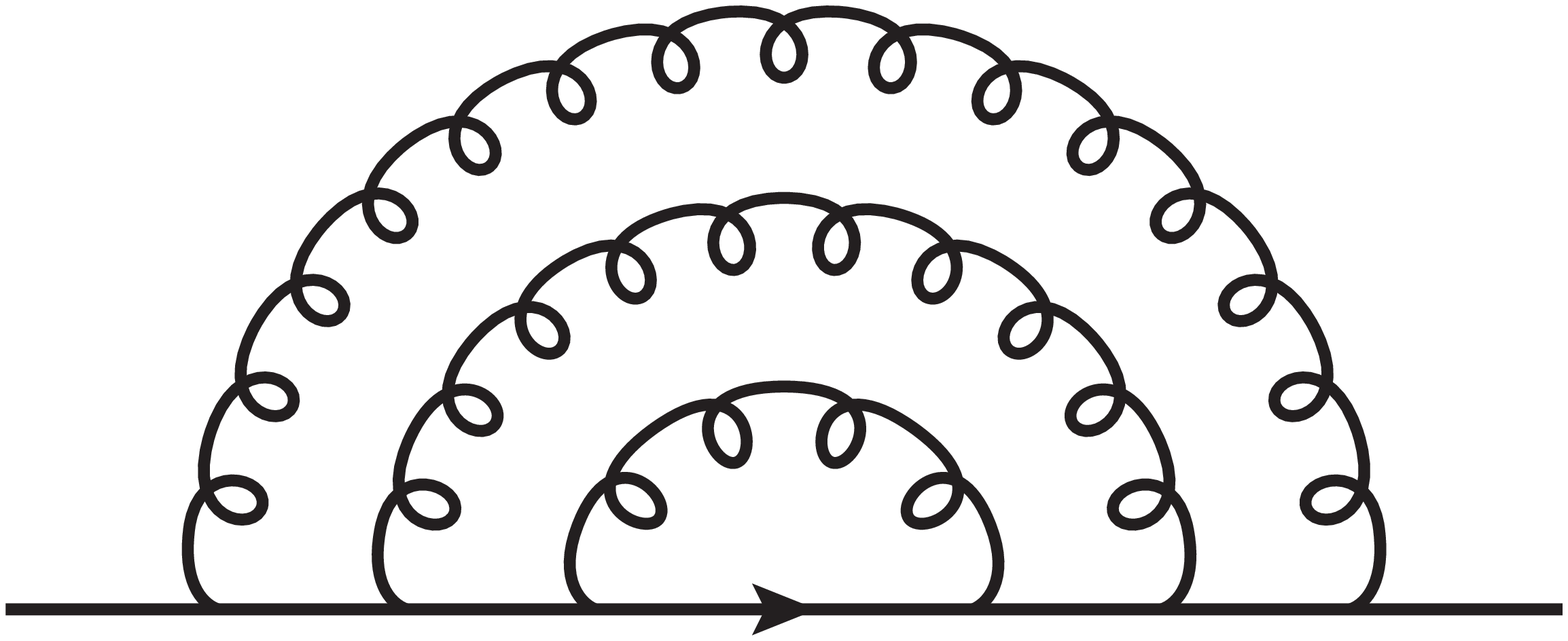} &
    \includegraphics[width=.22\textwidth]{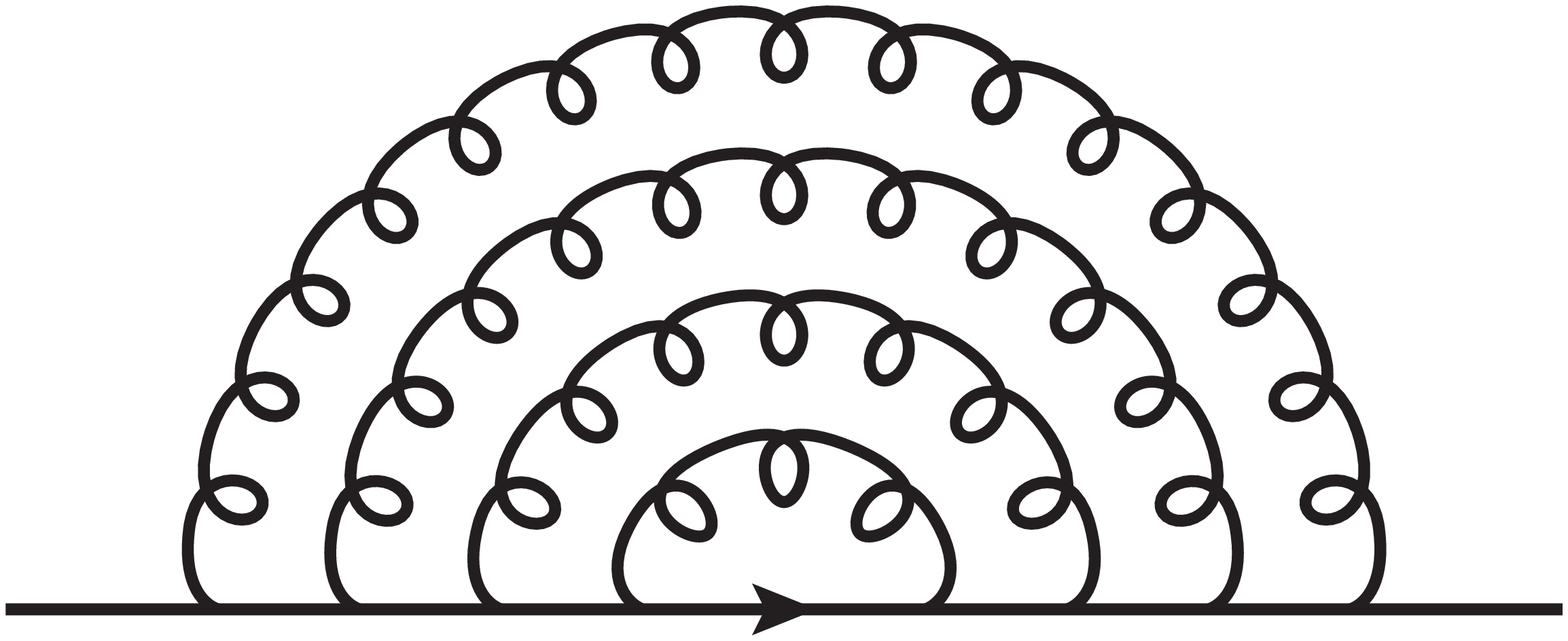} \\
    \\
    \includegraphics[width=.22\textwidth]{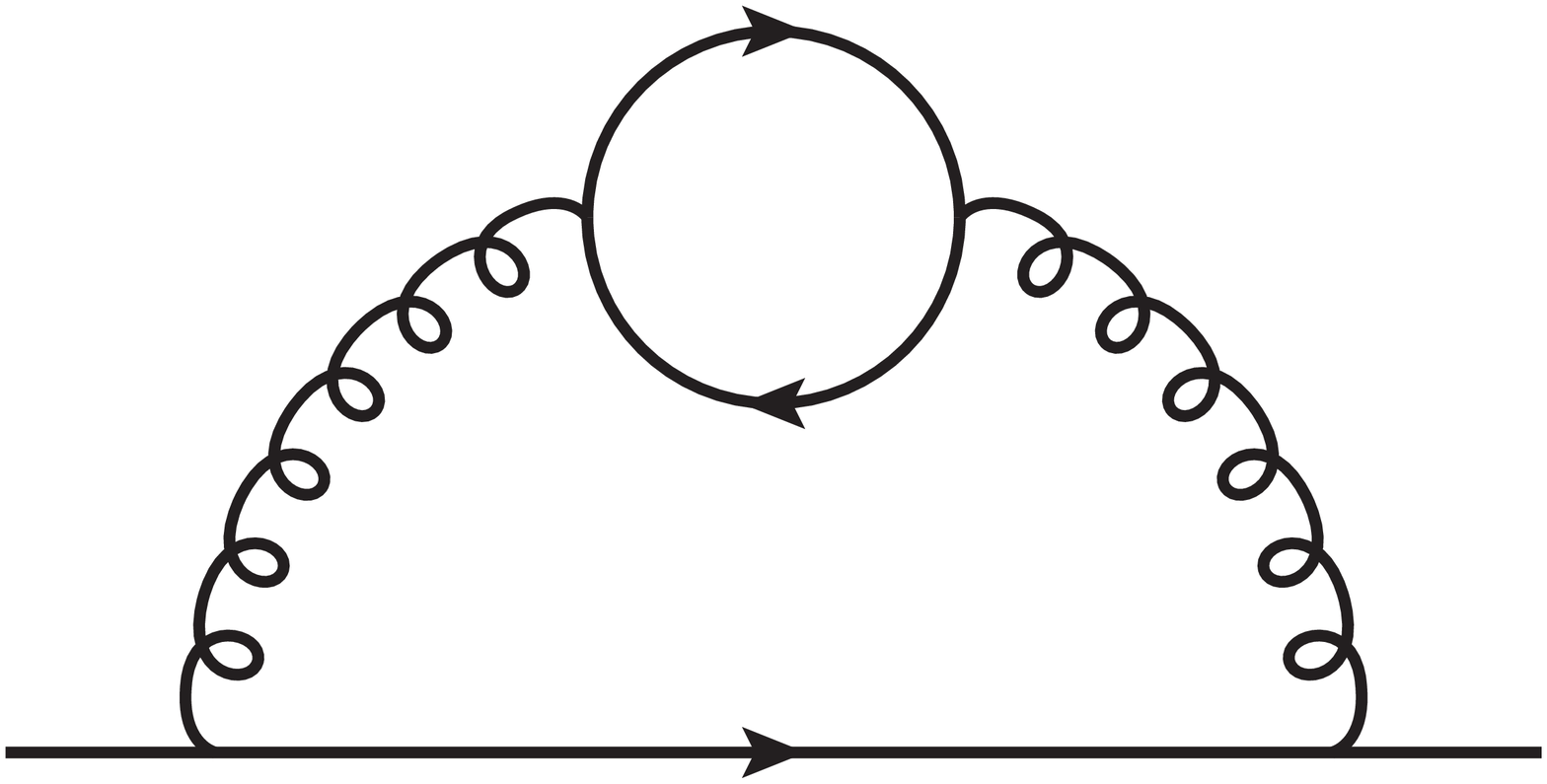} &
    \includegraphics[width=.22\textwidth]{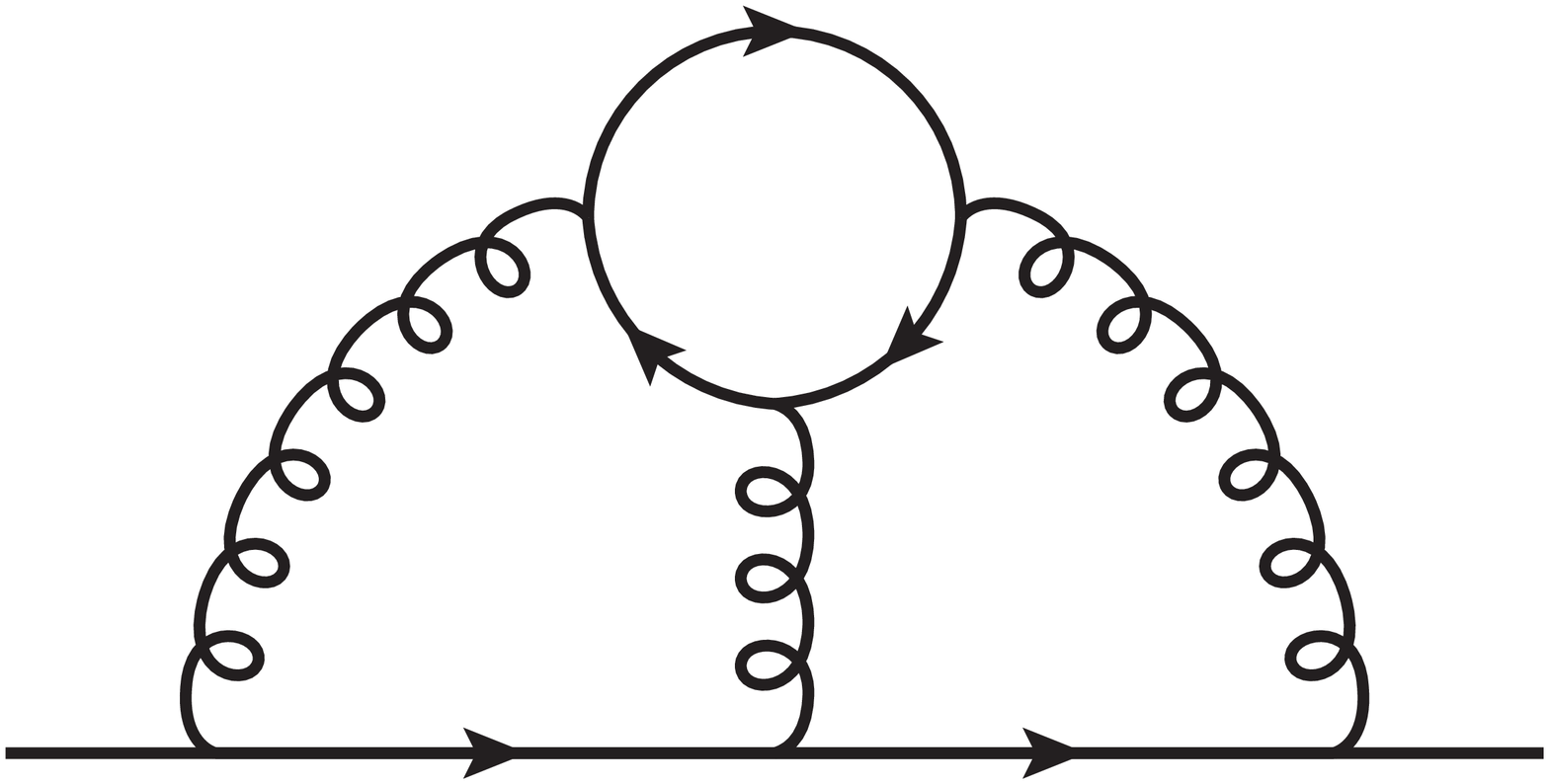} &
    \includegraphics[width=.22\textwidth]{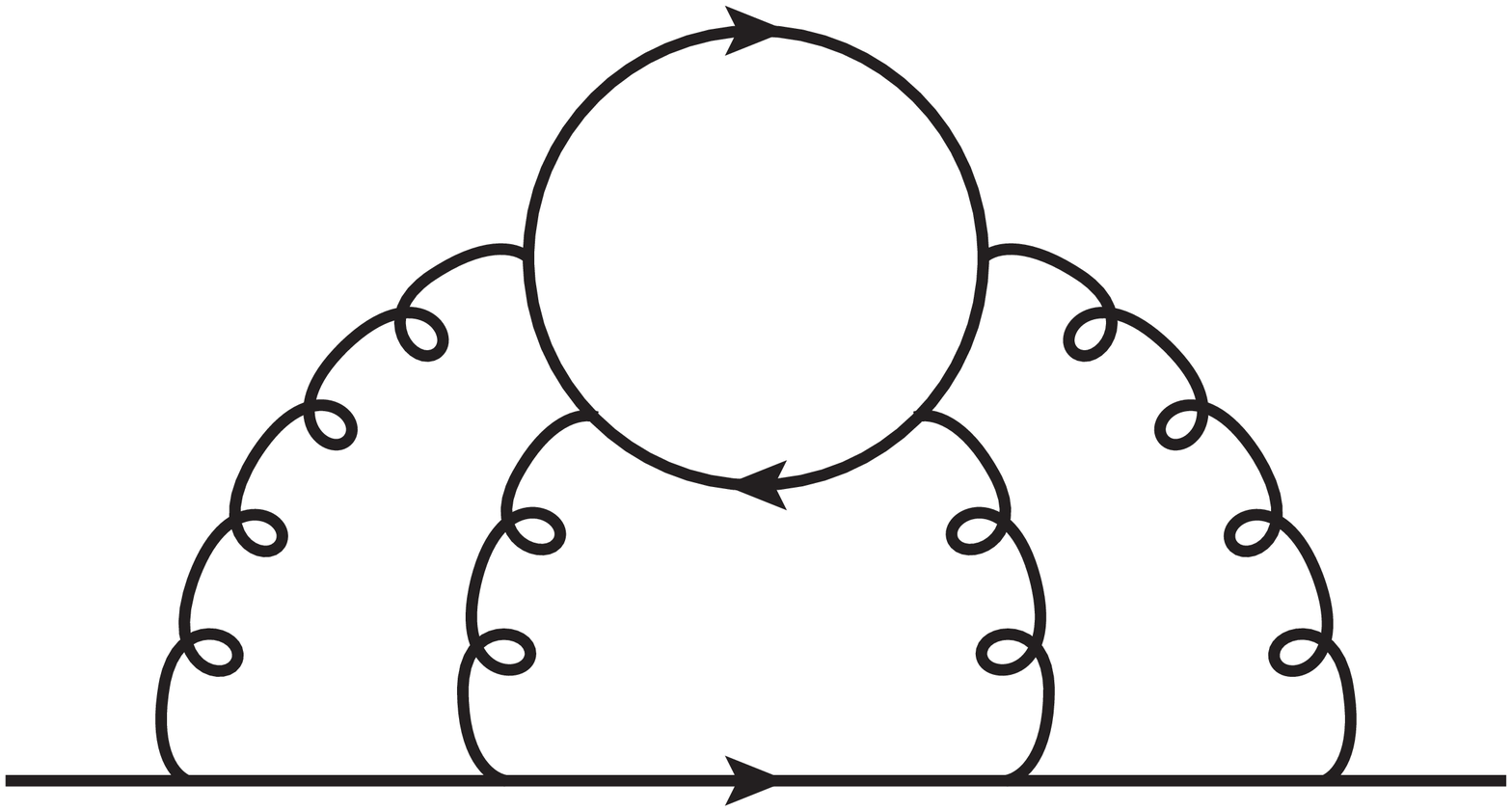} &
    \includegraphics[width=.22\textwidth]{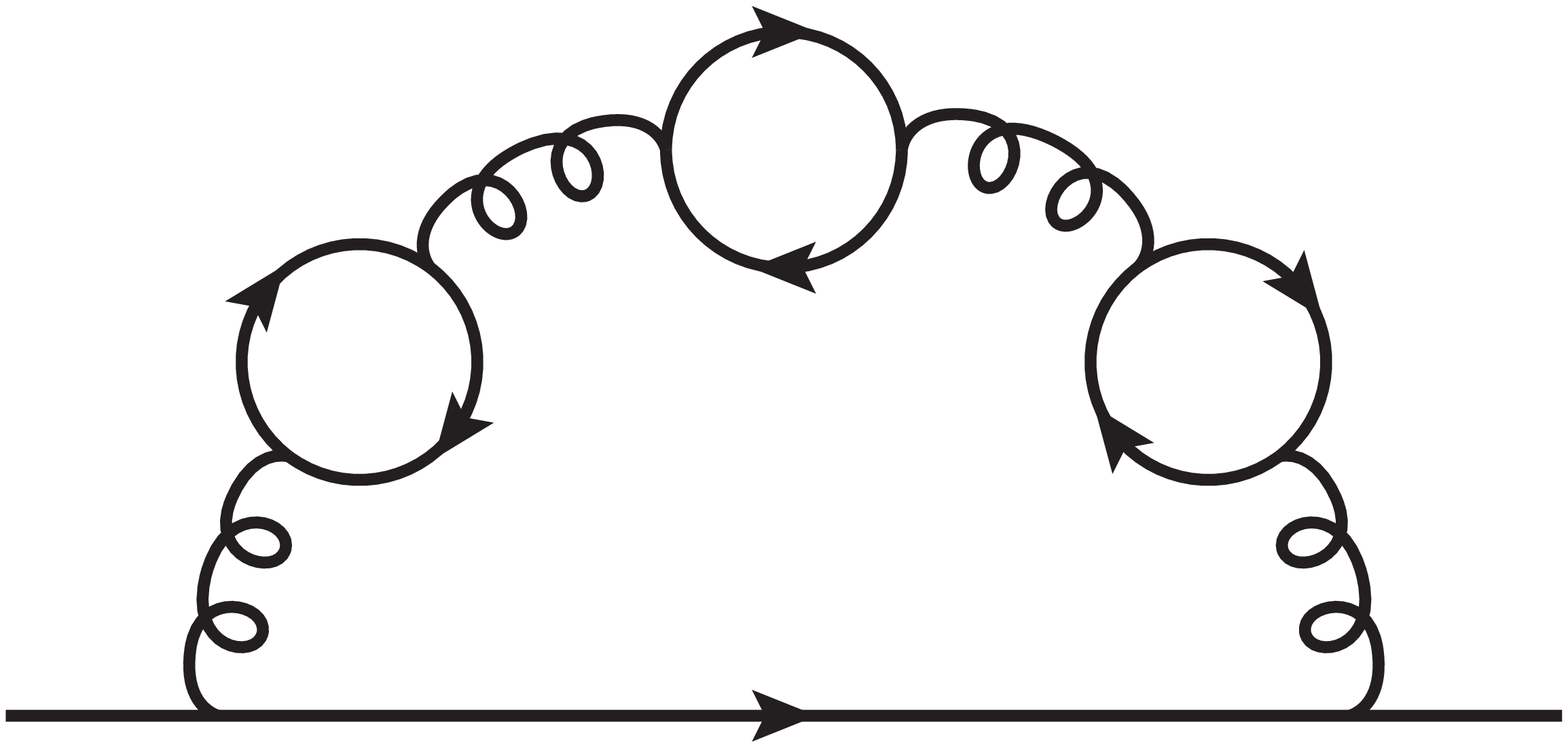} \\
    \\
    \includegraphics[width=.22\textwidth]{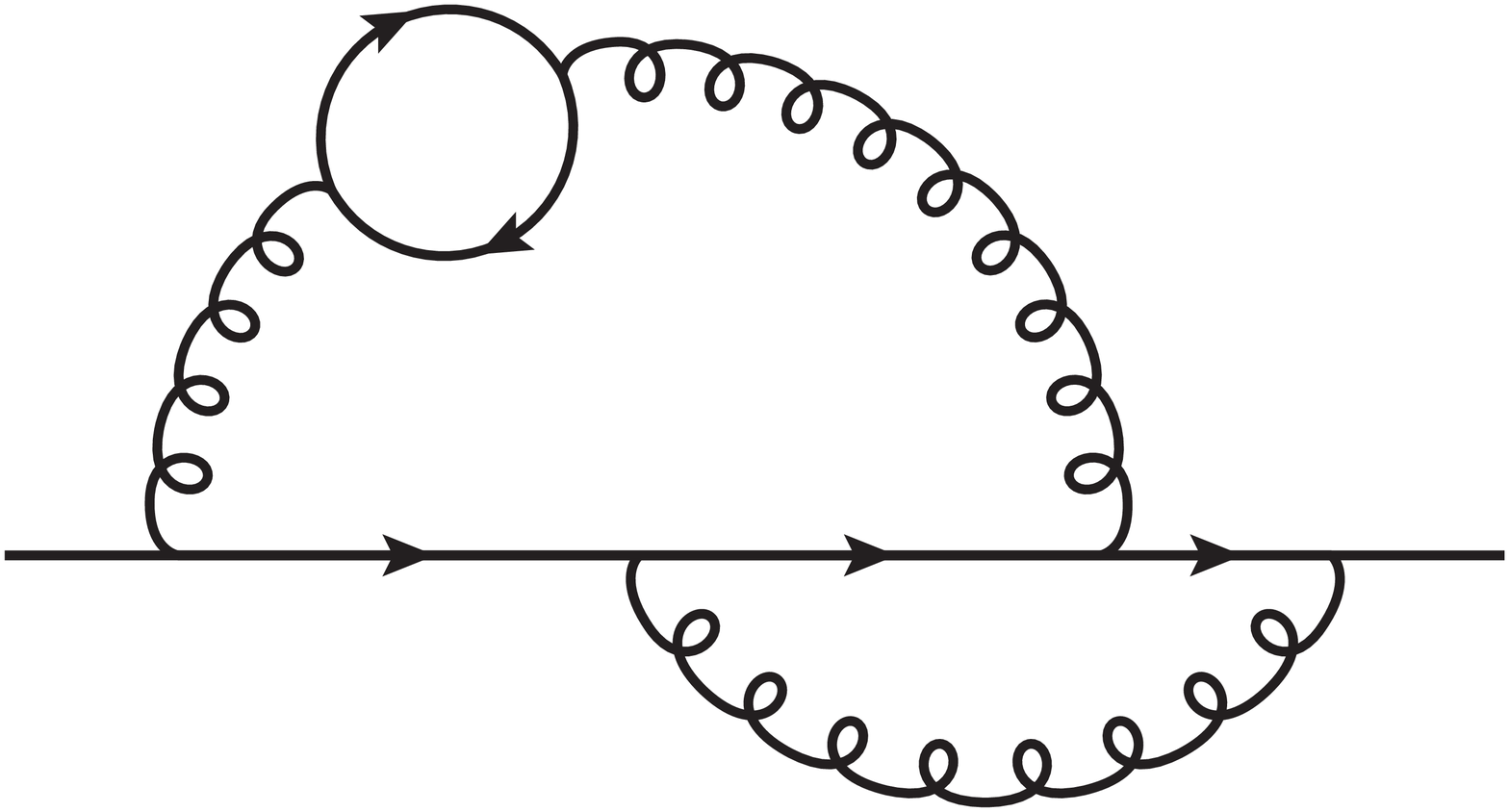} &
    \includegraphics[width=.22\textwidth]{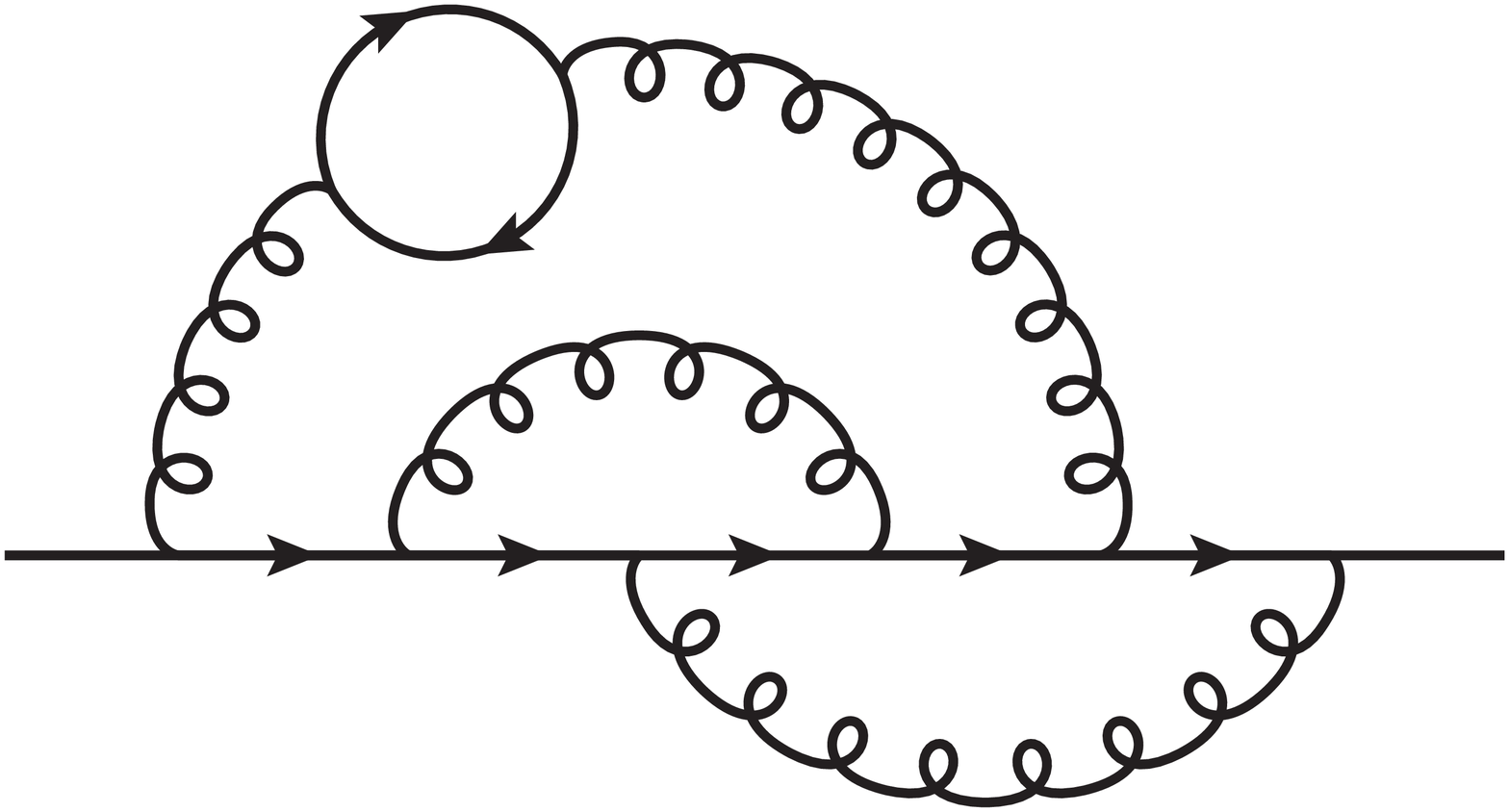} &
    \includegraphics[width=.22\textwidth]{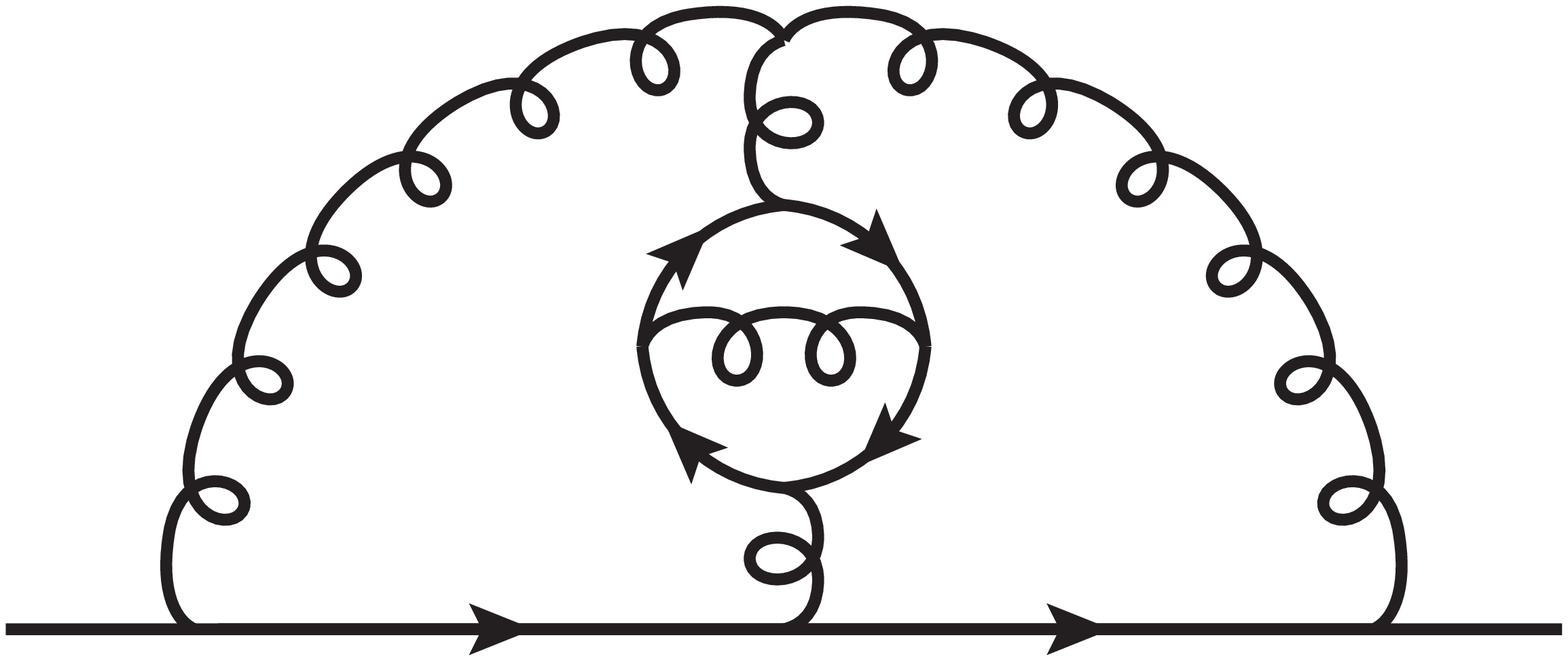} &
    \includegraphics[width=.22\textwidth]{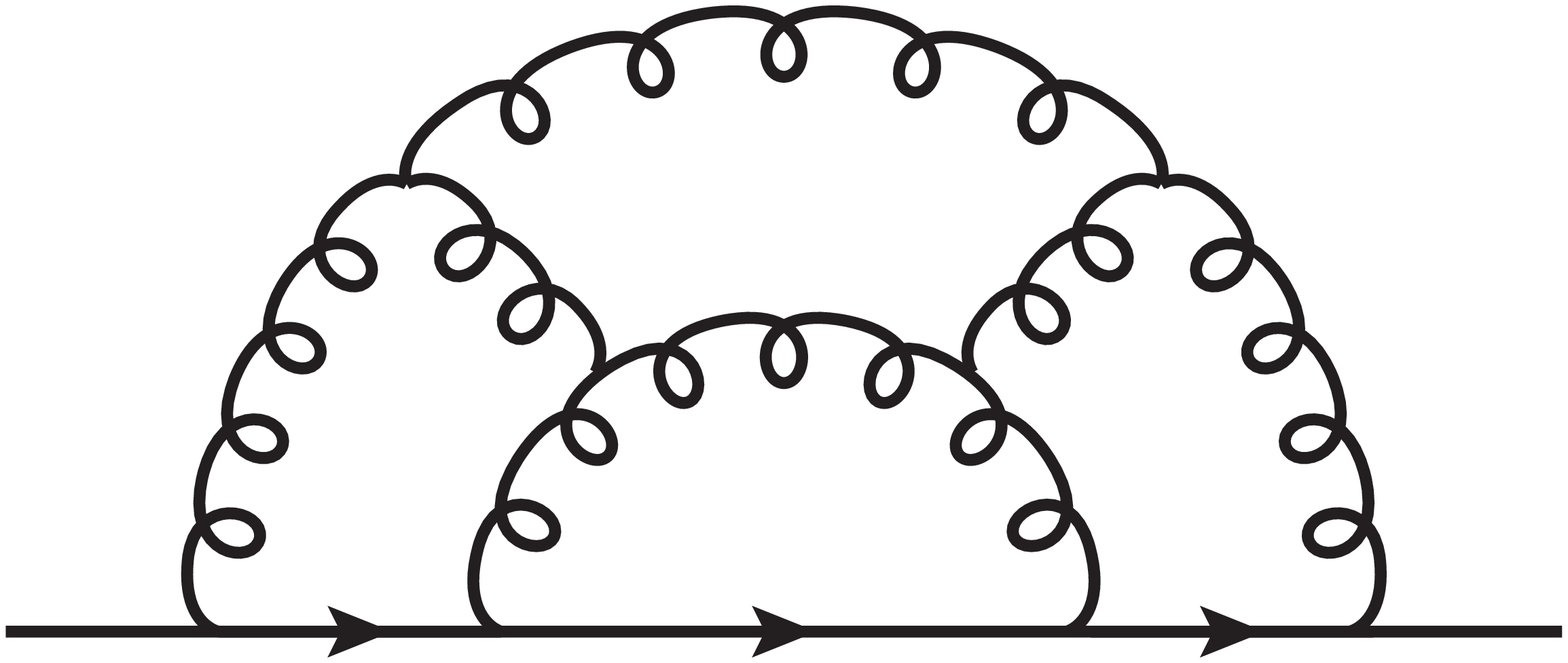} \\
    \end{tabular}
    \caption[]{\label{fig::diags}Sample Feynman diagrams contributing
      to  $\Sigma_S$ and $\Sigma_V$ at one-, two-, three- and
      four-loop order. The solid lines represent quarks and the curly lines
      gluons.}
\end{figure}

The relation between the bare ($m_0$) and renormalized mass in the
$\overline{\rm MS}$ scheme ($m$) is given by
\begin{eqnarray}
  m_0 = Z_m^{\overline{\rm MS}} m
  \,,
  \label{eq::ZmMS}
\end{eqnarray}
where $Z_m^{\overline{\rm MS}}$ only contains poles in $\epsilon$.  It is
obtained by requiring that the renormalized propagator is finite. Note that in
QCD the fermion propagator contains two Lorenz structures (scalar and vector).
Thus next to $Z_m^{\overline{\rm MS}}$ also the $\overline{\rm MS}$ wave
function renormalization constant is determined.  $Z_m^{\overline{\rm MS}}$
has been computed to five-loop order in Ref.~\cite{Baikov:2014qja}; for our
calculation only four-loop
corrections~\cite{Chetyrkin:1997dh,Vermaseren:1997fq,Chetyrkin:2004mf} are
needed. We use $Z_m^{\overline{\rm MS}}$ expressed 
for generic SU$(N_c)$ colour factors which
can be extracted from the anomalous dimension given
in~\cite{Vermaseren:1997fq}.  For convenience we present the result for
${Z_m^{\overline{\rm MS}}}$ in Appendix~\ref{app::ZmMS}. Note that the
$\overline{\rm MS}$-renormalized mass $m$ depends on the renormalization scale
$\mu$ which is suppressed in Eq.~(\ref{eq::ZmMS}).  $Z_m^{\overline{\rm MS}}$
depends on $\mu$ via the strong coupling constant $\alpha_s(\mu)$.

In the on-shell renormalization scheme one requires that the quark two-point
function vanishes at the position of the on-shell mass $M$ which fixes the
renormalization constant $Z_m^{\rm OS}$ introduced via
\begin{eqnarray}
  m_0 = Z_m^{\rm OS} M
  \,.
\end{eqnarray}
Note that $m_0$ and $M$ are $\mu$-independent and $Z_m^{\rm OS}$ contains
$\alpha_s(\mu)$ and $\log(\mu/M)$ terms.  The on-shell wave function
renormalization constant is determined from the requirement that the quark
propagator has a residue $-i$ at $q^2= M^2$. This leads to a formula for
$Z_2^{\rm OS}$ which is independent of $Z_m^{\rm OS}$. This is different in
the $\overline{\rm MS}$ scheme where $Z_m^{\overline{\rm MS}}$ and
$Z_2^{\overline{\rm MS}}$ have to be determined simultaneously.

A formula for $Z_m^{\rm OS}$ is conveniently derived by considering the 
renormalized quark propagator
\newcommand{\qsla}{q\!\!\!/\,\,}
\begin{eqnarray}
  S_F(q) &=& \frac{-i Z_2^{\rm OS}}{\qsla - m_{q,0} + \Sigma(q,M)}
  \label{eq::defs::fullprop1}
  \,,
\end{eqnarray}  
$\Sigma(q,M)$ is the (amputated) quark self energy which can be split into a
scalar and vector contribution 
\begin{eqnarray}
  \Sigma(q,M) &=&
  M\, \Sigma_S(q^2,M) + \qsla \, \Sigma_V(q^2,M)\,,
  \label{eq::defs::sigmadecomp}
\end{eqnarray}
where $\Sigma_S$ and $\Sigma_V$ only depend of $q^2$, the (renormalized) quark
mass and $\mu$ (which is again suppressed).  They are obtained from the self
energy $\Sigma$ with the help of the projectors
\begin{eqnarray}
  \Sigma_S((M^2,M) &=&
  \frac{1}{4M}\,\mbox{Tr}\,\left(\Sigma(q,M)\right)\Bigg|_{q^2=M^2}\,,\\ 
  \Sigma_V((M^2,M) &=&
  \frac{1}{4q^2}\,\mbox{Tr}\,\left(\qsla\,\Sigma(q,M)\right)\Bigg|_{q^2=M^2}\,. 
\end{eqnarray}
Sample Feynman diagrams contributing to $\Sigma(q,M)$
are shown in Fig.~\ref{fig::diags}.

Requiring that the inverse quark propagator, $\left[S_F(q)\right]^{-1}$
vanishes at the position of the on-shell mass, i.e. 
\begin{eqnarray}
  S_F(q)  &\stackrel{q^2\to M^2}{\longrightarrow}& \frac{-i}{\qsla - M}\,,
  \label{eq::defs::fullprop2}
\end{eqnarray}
leads to
\begin{eqnarray}
  Z_m^{\rm OS} &=& 1 + \Sigma_S(M^2,M) + \Sigma_V(M^2,M)
  \,.
  \label{eq::ZmOS}
\end{eqnarray}
Thus, for the evaluation of the $n$-loop contribution to $Z_m^{\rm OS}$
$n$-loop on-shell integrals have to be computed.

In this paper we present results for the finite quantity
\begin{eqnarray}
   z_m(\mu) &=& \frac{m(\mu)}{M} \,\,=\,\, 
   \frac{ Z_m^{\rm OS} }{ Z_m^{\overline{\rm MS}} }
   \,,
  \label{eq::OS2MS}
\end{eqnarray}
which is obtained from Eqs.~(\ref{eq::ZmMS}) and~(\ref{eq::ZmOS}).
Note that $z_m(\mu)$ depends on $\alpha_s(\mu)$ and $\log(\mu/M)$
and has the following perturbative expansion
\begin{eqnarray}
  z_m(\mu) &=& \sum_{n\ge0} \left(\frac{\alpha_s(\mu)}{\pi}\right)^n z_m^{(n)}(\mu)
  \,,
  \label{eq::zm}
\end{eqnarray}
with $z_m^{(0)}=1$.
For later convenience we decompose $z_m^{(n)}(\mu)$ into 
\begin{eqnarray}
  z_m^{(n)}(\mu) &=& z_m^{(n)}(M) + z_m^{(n),\rm log}
  \,,
  \label{eq::zmmu}
\end{eqnarray}
where the second term on the right-hand side comprises the 
$\mu$-dependent terms which vanish for $\mu=M$. Analytic results
for $z_m^{(n),\rm log}$ are given in Appendix~\ref{app::ren}.

For later use we also introduce the inverted relation to Eq.~(\ref{eq::OS2MS})
\begin{eqnarray}
  c_m(\mu) &=& \frac{M}{m(\mu)}\,,
  \label{eq::MS2OS}
\end{eqnarray}
with
\begin{eqnarray}
  c_m(\mu) &=& \sum_{n\ge0} \left(\frac{\alpha_s(\mu)}{\pi}\right)^n c_m^{(n)}(\mu)
  \,,
  \label{eq::cm}
\end{eqnarray}
and $c_m^{(0)}=1$. $c_m^{(n)}(\mu)$ is a function of $\log(\mu/m(\mu))$.
Furthermore, we assume the similar decomposition of Eq.~(\ref{eq::zmmu})
with $c_m^{(n),\rm log}=0$ for $\mu=m(m)$.

In this paper we consider generic SU$(N_c)$ colour factors and present results
for the coefficients. The four-loop term of Eq.~(\ref{eq::zm}) can be
decomposed into 23 colour structures which are given by
\begin{eqnarray}
  z_m^{(4)} &=&
  C_F^4 z_m^{FFFF}
          +  C_F^3 C_A z_m^{FFFA}
          +  C_F^2 C_A^2 z_m^{FFAA}
          +  C_F C_A^3 z_m^{FAAA}
\nonumber\\&&\mbox{}
          +  \frac{d_F^{abcd}d_A^{abcd}}{N_c} z_m^{d_{FA}}
          +  n_l \frac{d_F^{abcd}d_F^{abcd}}{N_c} z_m^{d_{FF}L}
          +  n_h \frac{d_F^{abcd}d_F^{abcd}}{N_c} z_m^{d_{FF}H}
\nonumber\\&&\mbox{}
          +  C_F^3 T n_l z_m^{FFFL}
          +  C_F^2 C_A T n_l z_m^{FFAL}
          +  C_F C_A^2 T n_l z_m^{FAAL}
\nonumber\\&&\mbox{}
          +  C_F^2 T^2 n_l^2 z_m^{FFLL}
          +  C_F C_A T^2 n_l^2 z_m^{FALL}
          +  C_F T^3 n_l^3 z_m^{FLLL}
\nonumber\\&&\mbox{}
          +  C_F^3 T n_h z_m^{FFFH}
          +  C_F^2 C_A T n_h z_m^{FFAH}
          +  C_F C_A^2 T n_h z_m^{FAAH}
\nonumber\\&&\mbox{}
          +  C_F^2 T^2 n_h^2 z_m^{FFHH}
          +  C_F C_A T^2 n_h^2 z_m^{FAHH}
          +  C_F T^3 n_h^3 z_m^{FHHH}
\nonumber\\&&\mbox{}
          +  C_F^2 T^2 n_l n_h z_m^{FFLH}
          +  C_F C_A T^2 n_l n_h z_m^{FALH}
          +  C_F T^3 n_l^2 n_h z_m^{FLLH}
\nonumber\\&&\mbox{}
          +  C_F T^3 n_l n_h^2 z_m^{FLHH}
          \,,
          \label{eq::zm4_col}
\end{eqnarray}
where $C_F$ and $C_A$ are the eigenvalues of the quadratic Casimir operators
of the fundamental and adjoint representation for the SU$(N_c)$ colour group,
respectively, $T=1/2$ is the index of the fundamental representation, and
$n_l$ and $n_h$ count the number of massless and massive (with mass $M$)
quarks. In the applications discussed in Section~\ref{sec::appl} we will use
$n_h=1$. It is nevertheless convenient to keep the variable $n_h$ as a parameter.
$d_F^{abcd}$ and $d_A^{abcd}$ are the symmetrized traces of four generators in
the fundamental and adjoint representation, respectively.  The colour
structures in Eq.~(\ref{eq::zm4_col}) are related to $N_c$ via\footnote{
  Note that our results are also valid for other groups. The corresponding
  expressions can easily be obtained by the proper choice of the group theory
  factors, see, e.g., Ref.~\cite{vanRitbergen:1998pn}. We restrict ourselves to SU$(N_c)$
  since it is closely connected to QCD.}  (see, e.g.,
Ref.~\cite{vanRitbergen:1998pn})
\begin{eqnarray}
  &&C_F = \frac{N_c^2-1}{2N_c}\,,\qquad C_A = N_c\,,\qquad T=\frac{1}{2}\,,
  \nonumber\\
  &&d_F^{abcd}d_F^{abcd} = \frac{(N_c^2-1)(N_c^4 - 6 N_c^2 + 18)}{96 N_c^2}\,,
  \nonumber\\
  &&d_F^{abcd}d_A^{abcd} = \frac{N_c (N_c^2-1) (N_c^2+6)}{48}\,.
\end{eqnarray}
In the case of QCD we have $N_c=3$.

One-, two- and three-loop QCD results to $Z_m^{\rm OS}$ have been
computed in Refs.~\cite{Tarrach:1980up},~\cite{Gray:1990yh}
and~\cite{Chetyrkin:1999ys,Chetyrkin:1999qi,Melnikov:2000qh,Marquard:2007uj},
respectively, and electroweak effects have been considered in
Refs.~\cite{Hempfling:1994ar,Jegerlehner:2002em,Jegerlehner:2003py,Jegerlehner:2003sp,Faisst:2004gn,Martin:2005ch,Eiras:2005yt,Jegerlehner:2012kn,Kniehl:2015nwa,Martin:2016xsp}.
In Ref.~\cite{Marquard:2015qpa} the four-loop results for $z_m$ have
been presented for $N_c=3$ and $n_l=3,4$ and~$5$ with a numerical
precision of 3\% in the four-loop coefficient evaluated at
$\mu=M$. It is the aim of the present paper to generalize the findings
of~\cite{Marquard:2015qpa} to general $N_c$ and arbitrary
$n_l$. Furthermore, the precision is significantly improved.
In this paper we will not study light-quark mass effects which are
known at two~\cite{Gray:1990yh} and three loops~\cite{Bekavac:2007tk}.

The relations between the threshold and the $\overline{\rm MS}$  masses
are too long so that we refrain from printing them in explicit
form. For practical purposes it is convenient to use their
implementation in {\tt RunDec}~\cite{Chetyrkin:2000yt} and {\tt
  CRunDec}~\cite{Schmidt:2012az}. The construction of the relations
can be found in the original
literature~\cite{Beneke:1998rk,Hoang:1998hm,Hoang:1998ng,Hoang:1999zc,Pineda:2001zq},
a summary can be found in, e.g., Ref.~\cite{Marquard:2015qpa}.


\subsection{Reduction to master integrals}

For the calculation of $\Sigma_S$ and $\Sigma_V$ we use a highly automated and
well-established set-up based on {\tt qgraf}~\cite{Nogueira:1991ex}, {\tt q2e}
and {\tt exp}~\cite{Harlander:1997zb,Seidensticker:1999bb} and in-house {\tt
  Mathematica} and {\tt FORM}~\cite{Vermaseren:2000nd,Kuipers:2012rf} programs
which work hand-in-hand to minimize the manual interaction.  
Colour factors are computed with the help of {\tt color}~\cite{vanRitbergen:1998pn}.

\begin{figure}[t]
  \centering
  \includegraphics[bb=90 600 600 780,width=0.9\linewidth]{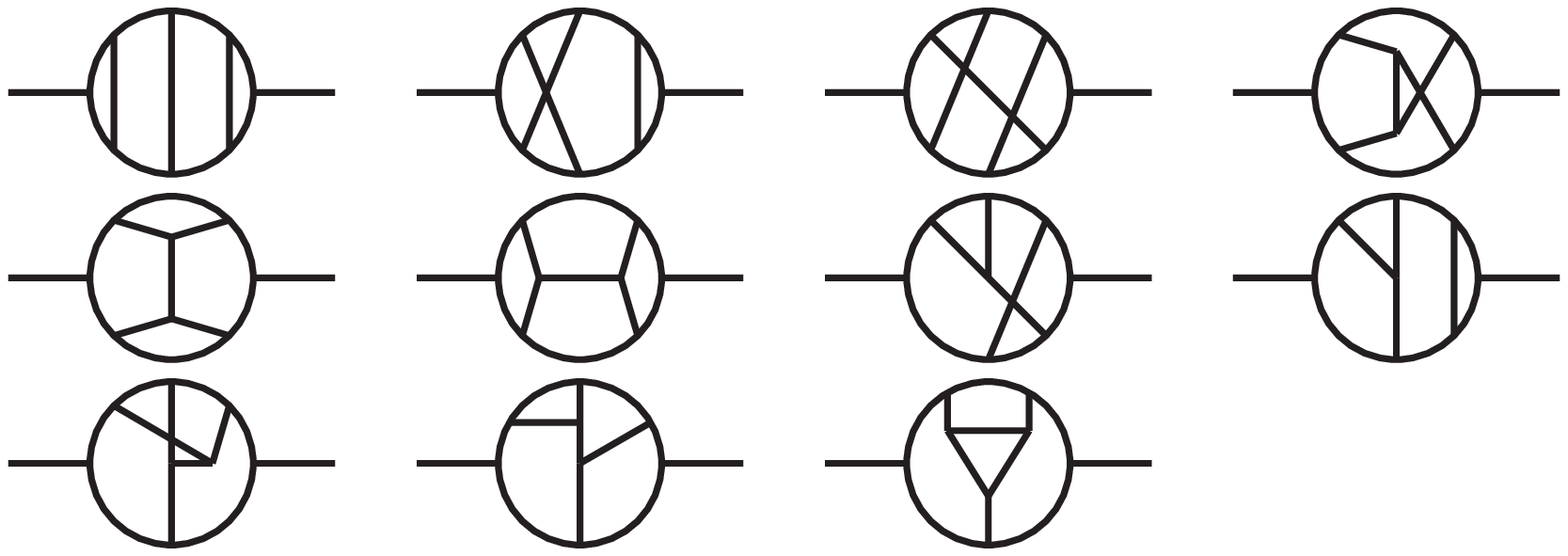}
  \caption{\label{fig::4l_prototype}
    Four-loop prototype families needed to generate the four-loop
    on-shell integral families shown in Appendix~\ref{app::intfam}.}
\end{figure}

We use {\tt qgraf} for the generation of the 3100 fermion self energy
amplitudes. They are converted to {\tt FORM} code
using {\tt q2e} and {\tt exp}.  A further task of the program {\tt exp} is to
map each diagram to one out of a set of 102 predefined integral families which
are shown in graphical form in Appendix~\ref{app::intfam}. To obtain these
families we start with the 11 prototypes shown in
Fig.~\ref{fig::4l_prototype}. They serve as the basis to generate the allowed
families by considering all possible routings of a massive line through the
diagrams. Diagrams with self-energy insertions can be obtained from the
ones in Fig.~\ref{fig::4l_prototype} by removing
some lines and raising the propagator powers of other lines.  For convenience
we show a pictorial representation for each family in
Appendix~\ref{app::intfam}. 
At four loops, they are labeled by 14 indices, that correspond to powers of
propagators and irreducible numerators. The maximal number of positive indices
is eleven.

We use in-house {\tt FORM} programs to apply the projectors
to the vector and scalar parts of the fermion propagator needed for the
calculation of the 
on-shell quark mass, to perform traces and to decompose the scalar products in
the numerator in terms of denominator factors.  As
an outcome our result is written as a linear combination of scalar
Feynman integrals which are related by integration-by-parts
identities~\cite{Chetyrkin:1981qh}. We apply to each family the Laporta
algorithm~\cite{Laporta:2001dd} as implemented in {\tt
  FIRE}~\cite{Smirnov:2008iw,Smirnov:2013dia,Smirnov:2014hma} and {\tt
  Crusher}~\cite{crusher} to perform a reduction to master integrals. 

We first work with each of the individual families and determine the
corresponding master integrals. It turns out that the primary sets of the
master integrals revealed with {\tt FIRE} are not minimal, i.e. there exist
additional relations among them.  Then, following
Ref.~\cite{Smirnov:2013dia}, we find additional relations using symmetries of
various integrals with indices $0,1$, and $2$.  For each sector\footnote{A
  sector is a subset
  of indices where some indices are positive and the other indices are
  non-positive.} one can estimate the number of the master integrals using the
code {\tt Mint}~\cite{Lee:2013hzt}.  There are, however, additional relations which
connect master integrals of partially overlapping sectors and they can be
revealed by the same procedure based on symmetries. The number of the master
integrals in a given family can be as large as 176.

One more criterion when looking for additional relations is the absence of
a spurious dependence of denominators in reduction relations on $d$. The general
analysis of singularities of Feynman integrals as functions of $d$ shows that
poles in $d$ can be only real rational numbers. So, if we observe a
non-factorizable polynomial
of second or higher degree in $d$ in a denominator this means that either
we miss a relation between the current master integrals or some master integrals
are chosen in an inappropriate way.  At least in all the cases in our
calculation, we managed to get rid of such spurious denominators by revealing
additional relations or making better choices of the master integrals.
However, with the sets of master integrals we have arrived at it is not
guaranteed that we have really minimal sets of master integrals, i.e.
bases of the corresponding linear spaces.

The next step was to find relations between master integrals of various
families. To do this, we used the Mathematica code {\tt tsort} which is part of
the latest {\tt FIRE} version~\cite{Smirnov:2014hma} and end up with 386
four-loop massive on-shell propagator integrals, i.e. with $q^2 = M^2$.

We have performed the calculation allowing for a general gauge
parameter $\xi$ keeping terms up to order $\xi^2$ in the expression we
give to the reduction routines. We have checked that $\xi$ drops out after
adding counterterm contributions from mass renormalization which is 
a welcome cross check on the consistency of our result.

As was mentioned above the algorithms we use to minimize the number of basis
integrals does not guarantee that we obtain all relations among the integrals
which appear as master integrals 
of the individual families. The fact that $\xi$ drops
out before using explicit results for the master integrals is a hint that we
are at least close to the minimal set.

We refrain from listing all master integrals but provide some examples
in the next subsection where the numerical accuracy of those integrals
is discussed which cannot be computed analytically.

Let us stress that up to this point our calculation is completely analytical.

  
\subsection{Computation of master integrals}

In this subsection we describe the methods that have been used
to obtain results for the master integrals.

All master integrals are computed numerically with the help of {\tt
  FIESTA}~\cite{Smirnov:2008py,Smirnov:2009pb,Smirnov:2013eza}.  {\tt FIESTA}
applies the sector decomposition algorithm which leads to a, in general,
multi-dimensional integral representation of the coefficients of the
$\epsilon$ expansion. The integration is performed using Monte-Carlo methods as
implemented in the {\tt CUBA}~\cite{Hahn:2004fe} library.  {\tt FIESTA} allows for a
highly parallel numerical integration and provides an almost linear scaling
behaviour. In fact, most of our calculations are performed at the High
Performance Computing Center Stuttgart (HLRS) and the Supercomputing Center of
Lomonosov Moscow State University which provide up to 1024 CPU cores or 64
Tesla GPUs for a single run.
The integral data base obtained with {\tt FIESTA} provides the
reference for the improvements for some of the integrals discussed in the
following.

\begin{figure}[t]
  \centering
  \begin{tabular}{cc}
    \raisebox{6em}{\includegraphics[width=.23\textwidth]{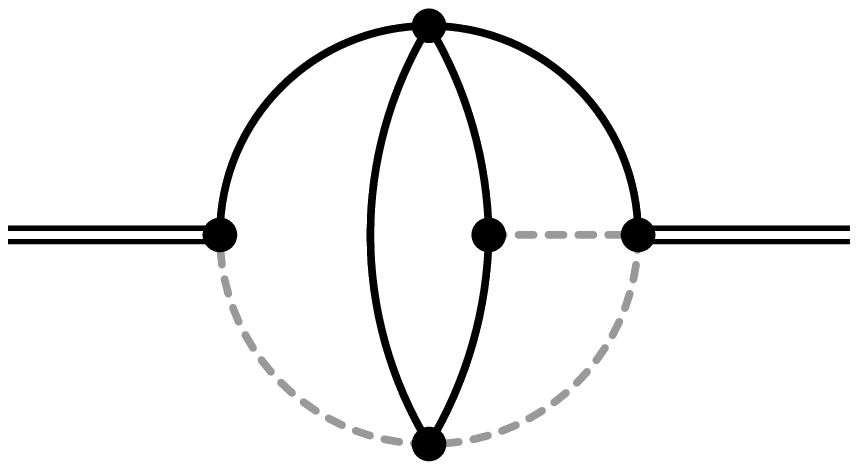}} &
    \includegraphics[width=0.45\linewidth]{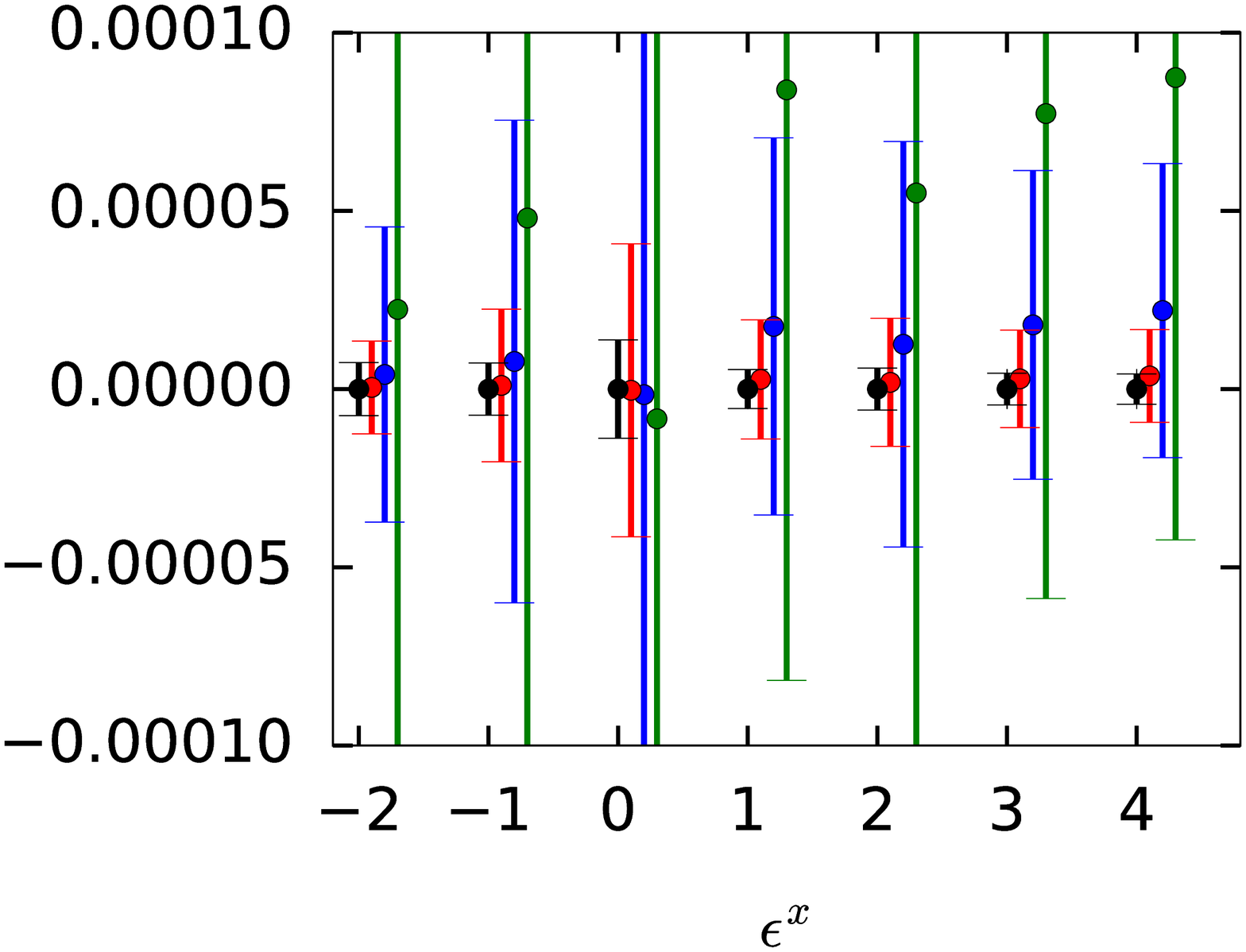}
    \\
    \raisebox{6em}{\includegraphics[width=.23\textwidth]{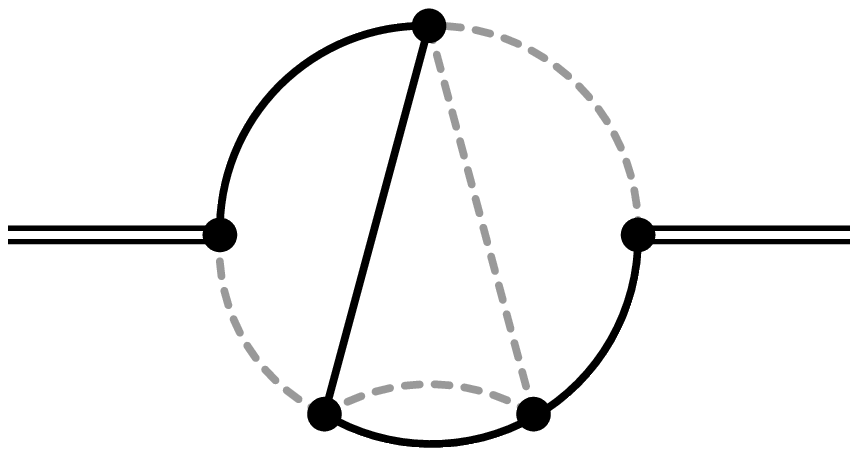}} &
    \includegraphics[width=0.45\linewidth]{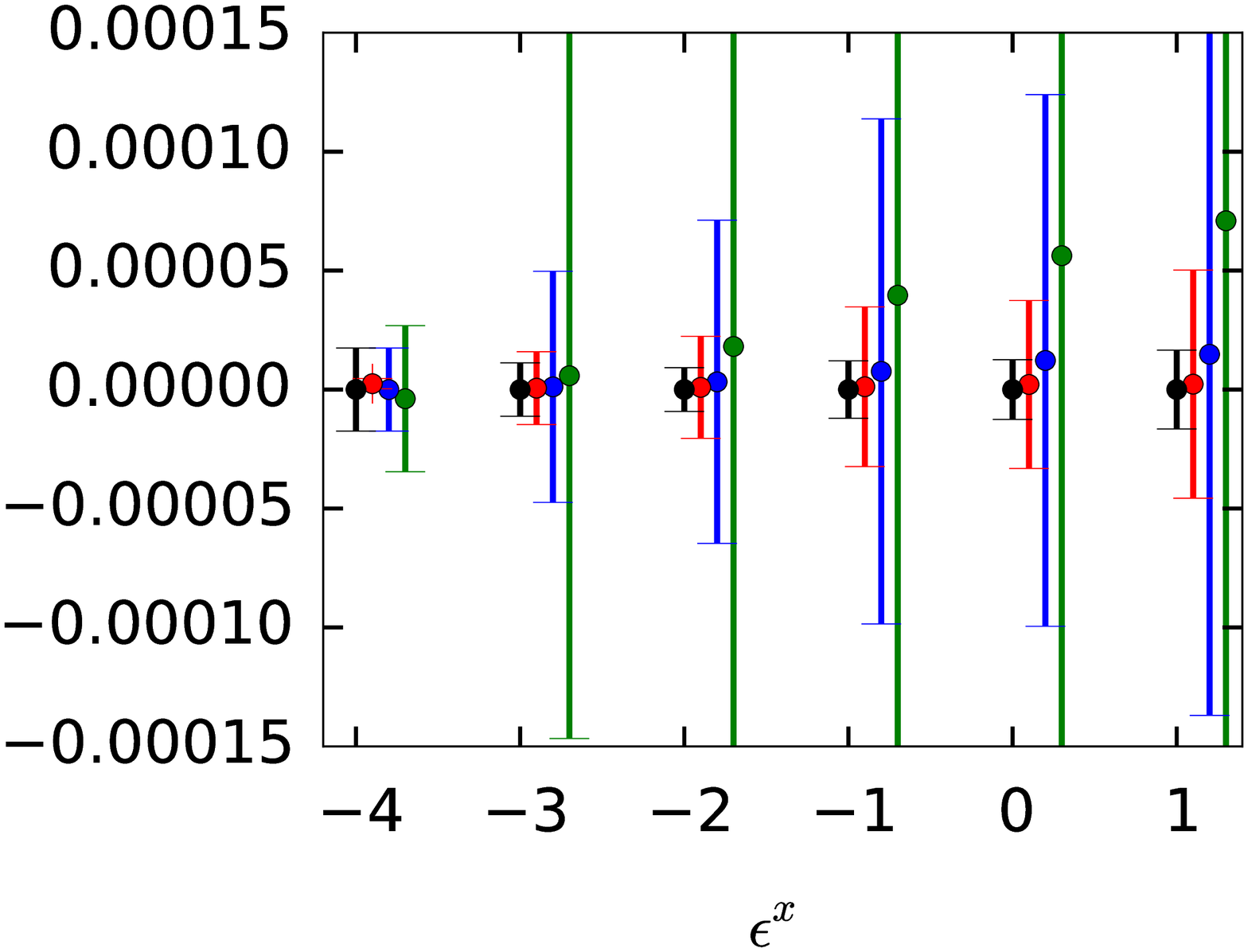}
    \\
    \raisebox{6em}{\includegraphics[width=.23\textwidth]{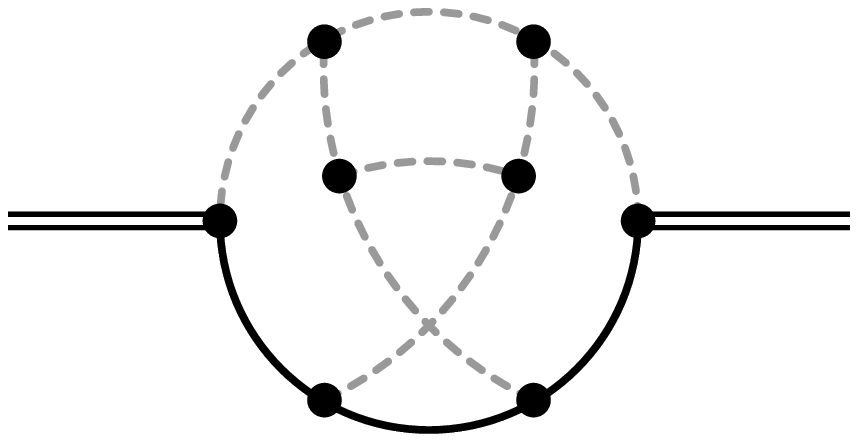}} &
    \includegraphics[width=0.45\linewidth]{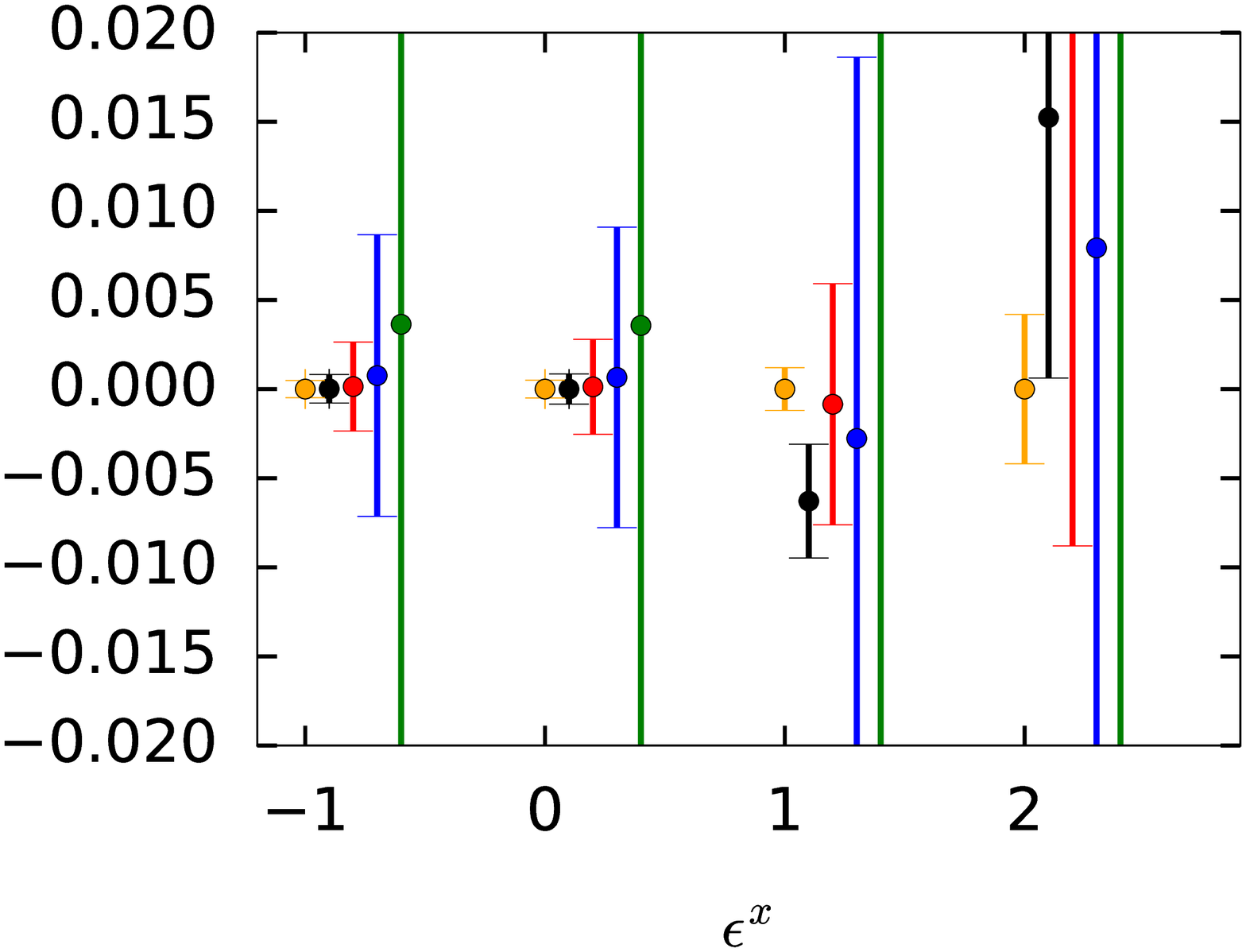}
  \end{tabular}
  \caption{\label{fig::MI_FIESTA_example}{\tt FIESTA} results for
    three typical integrals for various choices of $N$. The
    corresponding master integrals are shown to the left of the plots
    (see caption of Fig.~\ref{fig::MB_BB} for the meaning of the lines).
    In this plot the {\tt FIESTA} uncertainties have been multiplied by a
    factor ten. For each $\epsilon$ coefficient on the $x$ axis 
    results for different numbers of sampling points, $N$, are shown. For all
    plots we show results for $N=5\times 10^k$ with $k=5,6,7,8$.
    The bottom plot also contains results for $N=2\times 10^9$. In each case we
    normalize the results to the most precise one and then subtract 1.}
\end{figure}

We have computed all integrals using different statistics ranging from
$N=0.5\times10^6$ to $N=2\times10^{9}$ sampling points. We have
observed that the uncertainty decreases proportional to $1/\sqrt{N}$
according to the expectations for Monte-Carlo integrations.  In
Fig.~\ref{fig::MI_FIESTA_example} we show three typical master
integrals which are shown in graphical form to the left of the plot.
For each term of the $\epsilon$ expansion, which is indicated on the
$x$ axis, several data points are shown which correspond to different
numbers of sampling points.\footnote{For better readability the results
  for different sampling points are slightly displaced.}  The central
values are normalized to the most precise result and then we subtract
1 which explains why the central value of the leftmost data point is
equal to 0. The uncertainty bars correspond to the results where the
Monte-Carlo uncertainty based on {\tt Vegas}~\cite{Lepage:1980dq} is
multiplied by a factor ten (see also discussion below).

For the first two examples we observe that the central values of the
more precise calculation lie within the uncertainties of the less
precise ones. At the same time the uncertainty is significantly
reduced.\footnote{In those cases where the uncertainty does not become
  smaller after increasing the sampling points the requested precision
  is already reached for a smaller number of sampling points.}  The
third example behaves differently: For the $\epsilon^1$ and $\epsilon^2$ terms 
we observe a relatively big jumps
after increasing the sampling points from $N=5\times 10^7$ to
$N=5\times 10^8$ and then to
$N=2\times 10^9$. Furthermore, the more precise central value lies
partly outside the ten-sigma uncertainty bands. 

We have produced convergence plots as those in
Fig.~\ref{fig::MI_FIESTA_example} for all master integrals computed
with {\tt FIESTA}. Note that the one in the bottom panel of
Fig.~\ref{fig::MI_FIESTA_example} is among the integrals with the worst
behaviour. Altogether for about five master
integrals the five-sigma uncertainty
band is not sufficient to find agreement between the central values of the
high-precision results with the uncertainty band of the low-precision
results. For this reason we adopt a conservative attitude and multiply
the Monte-Carlo uncertainty of {\tt FIESTA} by a factor ten.
The reason for such a multiplication can also be justified by
the fact that each master integral leads
to thousands of individual sector integrals, and each of them produces
some error estimate. {\tt FIESTA} uses the mean-square norm when adding up
error estimates, but in unlucky situations this might be not enough for
a real error estimate.

It turns out that some of the master integrals determined with {\tt FIRE}, which have
usually indices equal to 1 and 0, are not optimal for the subsequent numerical
evaluation with {\tt FIESTA} and only a poor numerical precision is obtained.
In such situations, we tried to make a better
choice of the master integrals replacing master integrals of some sector by
other integrals which can have indices equal to 2. In some cases, we
successfully followed the strategy advocated in
Ref.~\cite{vonManteuffel:2014qoa} where the goal was to choose a finite or a
quasi-finite (in the sense that the only divergence comes from the overall
gamma function in Feynman parametrization) basis.

In particular, for our final result we replaced the integral shown in the
bottom panel of Fig.~\ref{fig::MI_FIESTA_example} by
an integral with numerators which shows a much better convergence
behaviour. Let us, however, stress that the final results (discussed in the next
Section) for the two different bases are consistent within the uncertainties.

For all factorizable integrals, we obtained analytic results from the known one-,
two- and three-loop results. In particular, we use the results of
Ref.~\cite{Lee:2010ik} where all three-loop master integrals have been obtained 
in an $\epsilon$ expansion up to the order typical to four-loop calculations.
For four of them, G43, G53, G62, and G65 (see Fig.~3
of~\cite{Lee:2010ik}) we had to add an additional
order in $\epsilon$ which is straightforward. In most cases one can
derive a one-dimensional Mellin-Barnes representation which converges
exponentially and thus ${\cal O}(1000)$ digits can easily be obtained.
In our calculation we encounter in total seven factorizable integrals.

For some master integrals, analytic results could be derived using a
straightforward loop-by-loop integration at general $d$, see, e.g.,
Fig.~\ref{fig::MIs_MB} (top, leftmost).  We also used analytical
results obtained for the 13 non-trivial four-loop on-shell master integrals
computed in our earlier paper~\cite{Lee:2013sx} (see Figs.~3 and~4
of~\cite{Lee:2013sx}).

At this point we adopt a practical attitude and generate an ordered list
which contains the $\epsilon$ coefficients of master integrals with large
contributions to the final uncertainty. This list is used as a starting
point to improve the accuracy of our result by increasing the
numerical precision of the corresponding master integral. Up to a certain
point this could be reached by simply increasing the statistics in the
approach based on {\tt FIESTA}. Of course, this approach is quite limited
since an increase of the number of sampling points by ten leads to an
uncertainty which is reduced by about a factor three.

A closer look at the generated list shows that the major contribution to the
uncertainty comes from master integrals containing two- or three-point sub-diagrams. 
For these integrals we proceed as follows:
\begin{itemize}
\item {\it Derive Mellin-Barnes representations for the subdiagrams.}\\
  This is achieved with the help of the formula
  \begin{eqnarray}
    \frac{1}{(X+Y)^\lambda} &=& \frac{1}{\Gamma(\lambda)}
    \,\frac{1}{2\pi i}\int_{-i\infty}^{+i\infty} {\rm d}z \, \frac{Y^z}{X^{\lambda+z}}
      \Gamma(\lambda+z)\Gamma(-z)
      \,,
  \end{eqnarray}
  which is used to split sums in the denominator raised to arbitrary
  power into products. In this way massive propagators can be
  transformed into massless ones at the cost of a Mellin-Barnes
  integration.  It is worth mentioning, that it does not need any
  specific hierarchy among the summands.  Depending on the other lines
  of the original diagram we use the Mellin-Barnes method such that
  the external momenta of the subdiagram are either massive or
  massless.  If possible, we apply on-shell conditions for external
  momenta.
 
  As an example we present our results for two typical ``building blocks''.
  \begin{itemize}
    \item The bubble integral with two massive lines (see
      Fig.~\ref{fig::MB_BB}, second diagram of the first row) and with
      massless external legs can be written in the following form
	  \begin{align}
		& \int \frac{{\rm d}^dk}{i\pi^\frac{d}{2}} 
		\frac{1}{[m^2-k^2]^{a_1}[m^2-(k+p)^2]^{a_2}} =
		\nonumber\\
		& \qquad \frac{1}{2\pi i} \int_{-i\infty}^{+i\infty} {\rm d}z \,
		\frac{(m^2)^{\frac{d}{2}-a_1-a_2-z}}{(-p^2)^{-z}}
		\nonumber \\
		& \qquad \times \frac{\Gamma(-z)\Gamma(a_1 + z)\Gamma(a_2 + z)
		\Gamma(a_1 + a_2 - \frac{d}{2} + z)} 
              {\Gamma(a_1)\Gamma(a_2)\Gamma(a_1 + a_2 + 2z) }\,.
		\label{eq::bb1}
	  \end{align}

    \item The triangle integral with two massive internal 
      lines and one massive, one 
      massless and one on-shell leg (see first diagram of second row
      in Fig.~\ref{fig::MB_BB}) is given by
	  \begin{align}
		& \int \frac{{\rm d}^dk}{i\pi^\frac{d}{2}} 
		\frac{1}{[m^2-(k+p_1)^2]^{a_1}[m^2-(k+p_1+p_2)^2]^{a_2}[-k^2]^{a_3}} =
		\nonumber\\
		& \qquad \frac{1}{(2\pi i)^2} \int_{-i\infty}^{+i\infty} {\rm d}z_1{\rm d}z_2 \,
		\frac{(m^2)^{\frac{d}{2}-a_1-a_2-a_3-z_1-z_2}}{(m^2-p_1^2)^{-z_1}(-p_2^2)^{-z_2}}
		\nonumber \\
		& \qquad \times \frac{\Gamma(-z_1)\Gamma(-z_2)\Gamma(a_2 + z_2)
		\Gamma(a_3 + z_1)\Gamma(a_1 + z_1+z_2)} 
                {\Gamma(a_1)\Gamma(a_2)\Gamma(a_3)\Gamma(d-a_1-a_2-a_3)}
                \nonumber \\
                & \qquad \times \frac{\Gamma(d-a_1-a_2-2a_3-z_1) \Gamma(a_1+a_2+a_3-\frac{d}{2}+z_1+z_2)} 
                {\Gamma(a_1+a_2+z_1+2z_2)}\,.
		\label{eq::bb2}
	  \end{align}      
  
  \end{itemize}
  Note that the exponents in Eqs.~(\ref{eq::bb1}) and~(\ref{eq::bb2})
  need not be integer but may also depend on $\epsilon$.
  
  \begin{figure}[t]
    \centering
    \includegraphics[width=.9\textwidth]{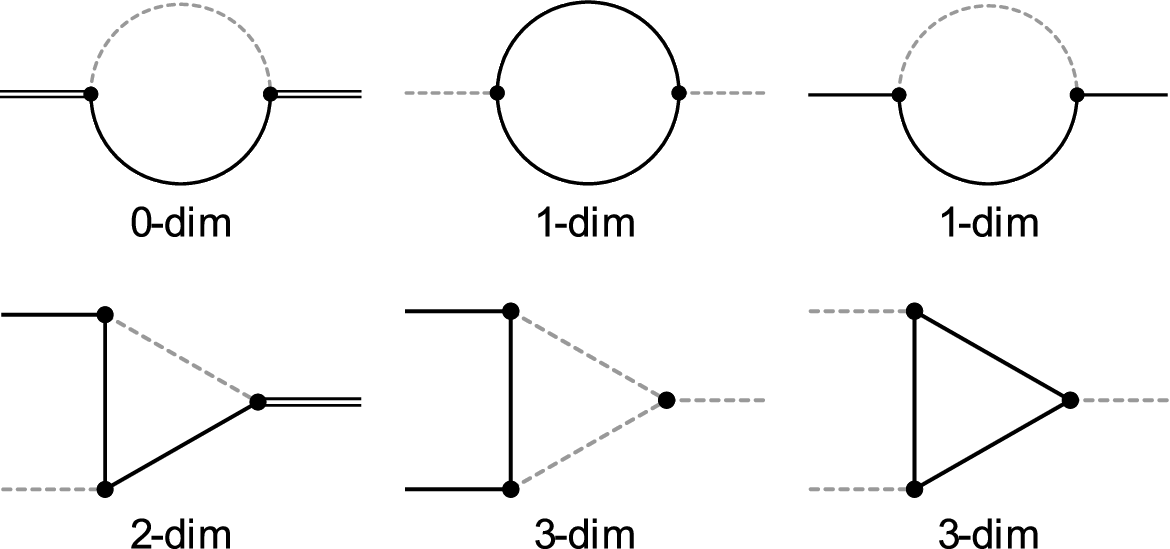}
    \caption[]{\label{fig::MB_BB}
      Sample building blocks for the loop-by-loop approach with
      three different types of external legs: dashed or solid lines
      denote massless- or massive propagators of general
      momentum $p^2$ respectively. Their general complex
      powers can depend on the dimensional regularization parameter
      $\epsilon$ and Mellin-Barnes integration variable $z_i$. Double lines
      are on-shell with the condition $p^2=m^2$. The dimension
      of the Mellin-Barnes integration is specified below the
      diagrams.

}
  \end{figure}

\item {\it Decompose integral into products of building blocks.}\\ 
  The derived building blocks are applied step-by-step until all
  momentum integrations are replaced by Mellin-Barnes integrals.  For
  simple integrals one ends up with a two- or three-dimensional
  integration (cf. Fig.~\ref{fig::MIs_MB}). In theses cases a
  precision of about nine digits is achieved for the $\epsilon^6$
  terms. The coefficients of the lower $\epsilon$-orders are more precise.
  We also encountered higher-dimensional integrals which lead to a
  lower precision. Some examples with five, six or even seven
  dimensional integrations can be found in Fig.~\ref{fig::MIs_MB}. For
  these cases one obtains about five digits for the $\epsilon^0$ and
  two to three digits for the $\epsilon^3$ term.

  It is interesting to note that the decomposition into building blocks
  is not unique. In fact, different representations may have significantly
  different convergence properties which we exploited for some of the integrals.  

\end{itemize}

\begin{figure}[t]
  \centering
  \includegraphics[width=.9\textwidth]{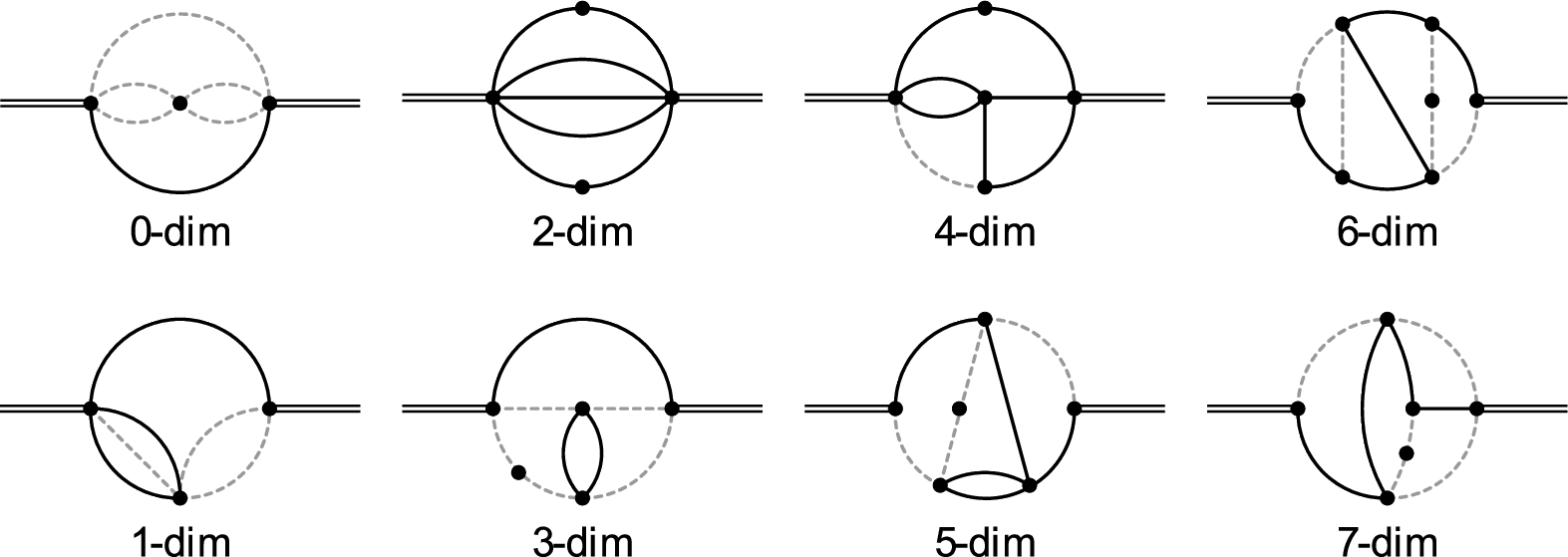}
  \caption[]{\label{fig::MIs_MB}Sample master integrals which are treated
    with the Mellin-Barnes method. The dimension of the
    Mellin-Barnes integration is specified below the diagrams.
  }
\end{figure}

Altogether we have treated 80 master integrals with the help of the
described method.  The results of the Mellin-Barnes integrals are
usually quite precise for lower orders of the $\epsilon$ expansion and
give several digits more than {\tt FIESTA} provides.  For 16 out of 80
integrals FIESTA produced more precise results for the higher orders in
$\epsilon$ and we chose to compose ``hybrid'' results where the lower
orders were taken from the Mellin Barnes (MB) integrals and the $\epsilon^3$ or higher
terms came from {\tt FIESTA}.

For the preparation of the Mellin-Barnes integrals we use the package
{\tt MB}~\cite{Czakon:2005rk} together with its extensions discussed
in Ref.~\cite{Smirnov:2009up}. For the numerical integration we use
the integrator {\tt cuhre} as implemented in the {\tt CUBA}
library~\cite{Hahn:2004fe}. As far as our experience goes the estimated
uncertainty of {\tt cuhre} is too small which can be seen by comparing
the results of the numerical integration to (analytically) known
results. Thus, we multiply the uncertainty by a factor 100 to be on
the conservative side.  For the higher-dimensional integrals we have
also tried to use {\tt vegas}, however, could not increase the
precision~\cite{msc::wellmann}.

We have compared all 80 master integrals computed with the
Mellin-Barnes method with the {\tt FIESTA} results and found good
agreement for almost all $\epsilon$ coefficients within three standard
deviations. However, in a few cases deviations up to seven sigma are
observed which once again justifies the use of a conservative limit of ten
sigma for the Monte-Carlo uncertainty of {\tt
  FIESTA}~\cite{msc::wellmann}.
The systematic application of the Mellin-Barnes method is
the main source for the improvements as compared to the results
presented in Ref.~\cite{Marquard:2015qpa}.

The described procedure can, of course, only be applied to a subset of all
master integrals. However, as mentioned above, in our basis these
integrals provide the substantial part of the uncertainty to $z_m$
in case we use the results based on {\tt FIESTA}.

For the remaining 259 integrals (i.e. the ones which are
neither known analytically nor treated with the Mellin-Barnes method)
we use the {\tt FIESTA} result.  When inserting the master integrals
we keep track of all uncertainties and combine them in quadrature in
the final expression.  We interpret the resulting uncertainty as a
standard deviation and multiply it by ten (as justified above) in the
final result for the relation between the $\overline{\rm MS}$ and
on-shell quark mass.  Note that, if we add the uncertainties from the
individual contributions linearly we obtain an uncertainty which is
about five times larger than the uncertainty resulting from the
quadratic combination.
For example, $z_m^{(4)}$ for $N_c=3$ and $n_l=5$
reads $-871.732 \pm 0.180$ for quadratic and 
$  -871.732 \pm   0.872$ for the linear combination
(without security factor 10).



\section{\label{sec::MSOS}Results for the $\overline{\rm MS}$-on-shell relation}

As an outcome of the procedure discussed in the previous Section we obtain
bare four-loop results for $\Sigma_V(q^2=M_q^2)+\Sigma_S(q^2=M_q^2)$ which
still contain fourth-order poles in the regularization parameter
$\epsilon$. Furthermore, uncertainties from each $\epsilon$ order of the
numerically evaluated master integrals are present in the expression.  The
individual uncertainties shall eventually be combined quadratically to obtain
the overall uncertainty. It is obvious that the latter is sensitive to 
the following choices:
\begin{itemize}
\item Set $N_c=3$ (and optionally also a value for $n_l$)
  before combining the uncertainties from the 
  master integrals.
\item Parametrize $\Sigma_V+\Sigma_S$ in terms of generic $N_c$ and $n_l$.
\item Parametrize $\Sigma_V+\Sigma_S$ in terms of SU$(N_c)$ Casimir invariants.
\end{itemize}
In this Section we will discuss the three options.
Note that we interpret the final uncertainty as a standard deviation which we
multiply by a factor ten to be on the conservative side.

It is convenient to present results for the finite relation between the
$\overline{\rm MS}$ and on-shell mass. It is obtained after renormalization of
the quark mass in the on-shell and the strong coupling constant in the
$\overline{\rm MS}$ scheme using three-loop renormalization constants. Whereas
$\alpha_s$ is renormalized by a simple multiplicative factor it is convenient
to generate the mass counterterm contribution at the same time as the
lower-order contributions. A finite quantity is obtained after dividing the
(parameter renormalized) $Z_m^{\rm OS}$ by $Z_m^{\overline{\rm MS}}$, as
discussed around Eq.~(\ref{eq::OS2MS}).

To get a sense of the quality of the cancellations of the poles
we present in the following table results for three typical 
contributions to $z_m^{(4)}$ ($\mu^2=M^2$)
\\ \\
\begin{tabular}{c|ccc|ccc|ccc}
  & \multicolumn{3}{c|}{$z_m^{(4)}$ for $N_c=3$, $n_l=5$} & 
  \multicolumn{3}{c|}{coef. of $N_c^4$ term} &  
  \multicolumn{3}{c}{coef. of $C_F^4$ term} \\
  \hline
$\epsilon^{-4}$&$  -0.00001 $&$\pm$&$    0.00002$ &$-0.0000002
  $&$\pm$&$  0.0000002$ &$ -0.000006 $&$\pm$&$   0.000013$ \\
$\epsilon^{-3}$&$    0.0003 $&$\pm$&$     0.0002$ &$  0.000002
  $&$\pm$&$   0.000002$ &$    0.0001 $&$\pm$&$     0.0001$ \\
$\epsilon^{-2}$&$   -0.0002 $&$\pm$&$     0.0018$ &$  0.000001
  $&$\pm$&$   0.000016$ &$   -0.0007 $&$\pm$&$     0.0009$ \\
$\epsilon^{-1}$&$    0.0044 $&$\pm$&$     0.0191$ &$   0.00002
  $&$\pm$&$    0.00018$ &$    0.0005 $&$\pm$&$     0.0081$ \\
$\epsilon^{0}$&$  -871.732 $&$\pm$&$      0.180$ &$   -51.181
  $&$\pm$&$      0.002$ &$    -6.983 $&$\pm$&$      0.081$ \\
\end{tabular}
\\ \\ \\
Note that the uncertainties are the ones returned from the numerical
integration without introducing any security factor.
Still, all pole coefficients are zero within one standard deviation
which shows that the factor ten applied to the final results presented below
is conservative.

From now on we only consider $\epsilon^0$ terms and
choose $\mu^2=M^2$ (for $z_m$) or 
$\mu^2=m^2(\mu^2)$ (for $c_m$). 
The renormalization scale dependent terms can be computed analytically using
renormalization group techniques; they are given in Appendix~\ref{app::ren}.


\subsection{\label{sub::nc3}Results for $N_c=3$}

We start with specifying both $N_c$ and $n_l$ before combining the
uncertainties of the master integrals. The results for $z_m^{(4)}(M)$ and
$c_m^{(4)}(m)$ for $N_c=3$ are shown in Table~\ref{tab::zm_nc3}.
Note that for the physically interesting cases $n_l=4,5$ and $6$
we find a relative uncertainty between 0.1\% and 0.2\%.

\begin{table}[t]
  \begin{center}
    \begin{tabular}{c|rcl|rcl}
$n_l$ &         &$z_m^{(4)}(M)$&          &              &$c_m^{(4)}(m)$ & \\
\hline
0 & $  -3654.15 $&$ \pm $&$       1.64$ & $   3567.60 $&$ \pm $&$1.64$ \\
1 & $  -2940.01 $&$ \pm $&$       1.67$ & $   2864.60 $&$ \pm $&$1.67$ \\
2 & $  -2308.77 $&$ \pm $&$       1.70$ & $   2244.32 $&$ \pm $&$1.70$ \\
3 & $  -1756.36 $&$ \pm $&$       1.74$ & $   1702.70 $&$ \pm $&$1.74$ \\
4 & $  -1278.70 $&$ \pm $&$       1.77$ & $   1235.66 $&$ \pm $&$1.77$ \\
5 & $   -871.73 $&$ \pm $&$       1.80$ & $    839.14 $&$ \pm $&$1.80$ \\
6 & $   -531.39 $&$ \pm $&$       1.84$ & $    509.07 $&$ \pm $&$1.84$ \\
7 & $   -253.59 $&$ \pm $&$       1.87$ & $    241.37 $&$ \pm $&$1.87$ \\
8 & $    -34.28 $&$ \pm $&$       1.91$ & $     31.99 $&$ \pm $&$1.91$ \\
9 & $    130.62 $&$ \pm $&$       1.94$ & $   -123.15 $&$ \pm $&$1.94$ \\
10 & $    245.17 $&$ \pm $&$       1.98$ & $   -228.12 $&$ \pm $&$1.98$ \\
11 & $    313.45 $&$ \pm $&$       2.01$ & $   -286.98 $&$ \pm $&$2.01$ \\
12 & $    339.51 $&$ \pm $&$       2.05$ & $   -303.81 $&$ \pm $&$2.05$ \\
13 & $    327.44 $&$ \pm $&$       2.08$ & $   -282.68 $&$ \pm $&$2.08$ \\
14 & $    281.30 $&$ \pm $&$       2.12$ & $   -227.64 $&$ \pm $&$2.12$ \\
15 & $    205.16 $&$ \pm $&$       2.16$ & $   -142.78 $&$ \pm $&$2.16$ \\
16 & $    103.09 $&$ \pm $&$       2.19$ & $    -32.15 $&$ \pm $&$2.19$ \\
17 & $    -20.85 $&$ \pm $&$       2.23$ & $    100.16 $&$ \pm $&$2.23$ \\
18 & $   -162.58 $&$ \pm $&$       2.26$ & $    250.10 $&$ \pm $&$2.26$ \\
19 & $   -318.03 $&$ \pm $&$       2.30$ & $    413.59 $&$ \pm $&$2.30$ \\
20 & $   -483.15 $&$ \pm $&$       2.34$ & $    586.56 $&$ \pm $&$2.34$ \\
    \end{tabular}
    \caption{\label{tab::zm_nc3}Results for $z_m^{(4)}(M)$ and $c_m^{(4)}(m)$
      for $N_c=3$ and $0\le n_l \le 20$.}
  \end{center}
\end{table}

\begin{figure}[t]
  \centering
  \includegraphics[width=0.8\linewidth]{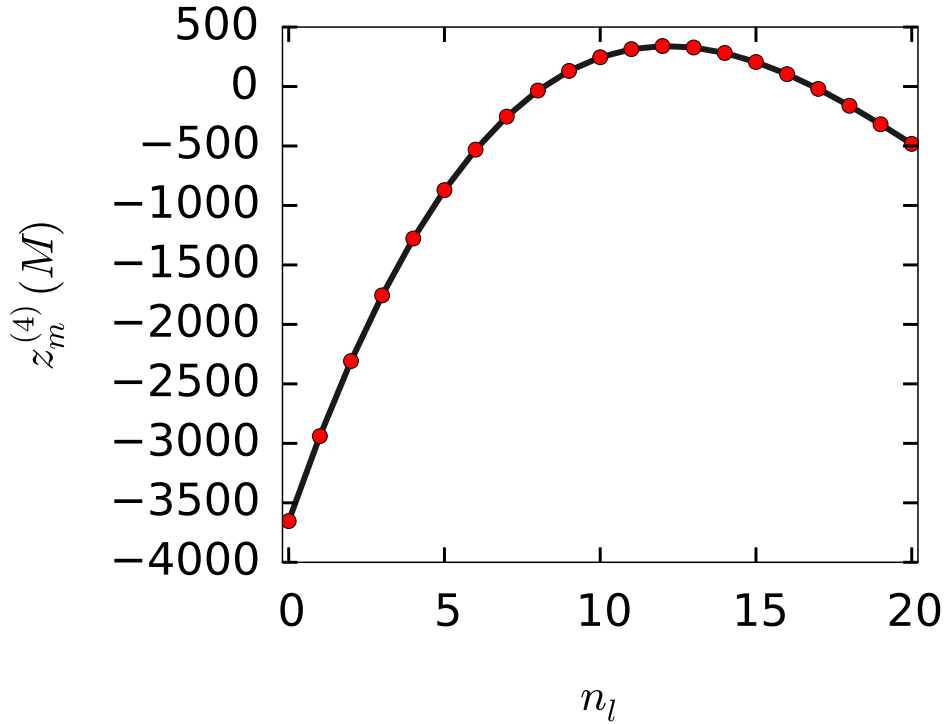}
  \caption{\label{fig::zm_nl_dep}$n_l$-dependence of $z_m^{(4)}(M)$.}
\end{figure}

From Table~\ref{tab::zm_nc3} one observes that the uncertainty has only a very
mild dependence on $n_l$. Thus, to a good approximation we can write
$z_m^{(4)}$ in the form\footnote{Note, that this is not a fit to
  Table~\ref{tab::zm_nc3}.}
\begin{eqnarray}
  z_m^{(4)} &=&
  -3654.15 \pm 1.64
  + (756.942 \pm 0.040 ) n_l - 43.4824 n_l^2 + 0.678141 n_l^3
  \,.
  \label{eq::zm_nc3_nl}
\end{eqnarray}
In Fig.~\ref{fig::zm_nl_dep} we plot Eq.~(\ref{eq::zm_nc3_nl}) for $n_l$ between 0
and 20 and combine the data points for integer $n_l$ to guide the
eye. It is interesting to note that the four-loop coefficient 
$z_m^{(4)}$ becomes positive between $n_l=9$ and $n_l=16$. Close to
these values of $n_l$ (i.e. for $n_l=8$ and $n_l=17$) the absolute 
value of $z_m^{(4)}$ is quite small and thus the relative uncertainty 
exceeds 5\%.
The range for $n_l$ where the four-loop coefficient changes
  sign coincides with the one for the so-called Banks-Zaks fixed
  point for the QCD beta function~\cite{Banks:1981nn}. However,
  we are not aware of a deeper connection which might be a subject for further
  studies.

The four-loop coefficient of the inverted relation, $c_m^{(4)}$, which
is basically obtained from negative $z_m^{(4)}$ plus some products of
lower order contributions, shows a similar behaviour except for the
overall sign. It has, in particular, the same uncertainty, as can be
seen in the last column of Table~\ref{tab::zm_nc3}.
The explicit $n_l$ dependence reads
\begin{eqnarray}
  c_m^{(4)} &=&
  3567.60 \pm 1.64
  - (745.721 \pm 0.040 ) n_l + 43.3963 n_l^2 - 0.678141 n_l^3
  \,.
  \label{eq::zm_nc3_nl_cm}
\end{eqnarray}

For some applications it is useful to have control over all
fermionic contributions, including the ones from closed fermion loops
of mass $M$ which we label by $n_h$.  The corresponding result
is shown in Table~\ref{tab::zm_n3_nl_nh} where we present the
coefficients of $n_l^i n_h^j$ for $i,j=0,1,2,3$ with $i+j\le3$.

\begin{table}[t]
  \centering
  \begin{tabular}{c|ccc}
    \hline
$n_l^0 n_h^0$ & $  -3678.28 $&$\pm$&$       1.63$\\
$n_l^0 n_h^1$ & $     23.63 $&$\pm$&$       0.12$\\
$n_l^0 n_h^2$ & $    0.5273 $&$\pm$&$     0.0027$\\
$n_l^0 n_h^3$ & $  -0.02484 $&$\pm$&$    0.00000$\\
$n_l^1 n_h^0$ & $    757.64 $&$\pm$&$       0.04$\\
$n_l^1 n_h^1$ & $   -0.6646 $&$\pm$&$     0.0004$\\
$n_l^1 n_h^2$ & $  -0.03617 $&$\pm$&$    0.00000$\\
$n_l^2 n_h^0$ & $    -43.47 $&$\pm$&$       0.00$\\
$n_l^2 n_h^1$ & $  -0.01720 $&$\pm$&$    0.00000$\\
$n_l^3 n_h^0$ & $    0.6781 $&$\pm$&$     0.0000$\\
    \hline
  \end{tabular}
\caption{\label{tab::zm_n3_nl_nh}{$z_m^{(4)}$} decomposed into coefficients of 
  $n_l^i n_h^j$.}
\end{table}


\subsection{Results for generic $N_c$}

In a next step we do not specify numerical values for $N_c$ and $n_l$
which leads to 23 non-zero colour structures. For the corresponding
coefficients we obtain
\begin{eqnarray}
z_m^{LLL 1/N_c^1} &=&   -0.25430 \,,\nonumber\\
          z_m^{LLL N_c^1} &=&    0.25430 \,,\nonumber\\
          z_m^{LL 1/N_c^2} &=&   -0.14090 \,,\nonumber\\
          z_m^{LL 1/N_c^1} &=&    0.00645 \,,\nonumber\\
          z_m^{LL N_c^0} &=&    5.58971 \,,\nonumber\\
          z_m^{LL N_c^1} &=&   -0.00645 \,,\nonumber\\
          z_m^{LL N_c^2} &=&   -5.44881 \,,\nonumber\\
          z_m^{L 1/N_c^3} &=&    0.1788 \pm    0.0333\,,\nonumber\\
          z_m^{L 1/N_c^2} &=&   -0.18076 \pm    0.00000\,,\nonumber\\
          z_m^{L 1/N_c^1} &=&    0.9282 \pm    0.0445\,,\nonumber\\
          z_m^{L N_c^0} &=&    0.28392 \pm    0.00005\,,\nonumber\\
          z_m^{L N_c^1} &=&  -32.7991 \pm    0.0109\,,\nonumber\\
          z_m^{L N_c^2} &=&   -0.10316 \pm    0.00005\,,\nonumber\\
          z_m^{L N_c^3} &=&   31.69215 \pm    0.00124\,,\nonumber\\
          z_m^{ 1/N_c^4} &=&   -0.4364 \pm    0.0503\,,\nonumber\\
          z_m^{ 1/N_c^3} &=&    0.821 \pm    0.121\,,\nonumber\\
          z_m^{ 1/N_c^2} &=&    0.1739 \pm    0.0738\,,\nonumber\\
          z_m^{ 1/N_c^1} &=&    0.645 \pm    0.161\,,\nonumber\\
          z_m^{ N_c^0   } &=&    -0.614  \pm    0.175\,,\nonumber\\
          z_m^{ N_c^1} &=&   -2.6228 \pm    0.0415\,,\nonumber\\
          z_m^{ N_c^2} &=&  52.0579 \pm    0.0808\,,\nonumber\\
          z_m^{ N_c^3} &=&    1.15654 \pm    0.00424\,,\nonumber\\
          z_m^{ N_c^4} &=&   -51.1812 \pm    0.0161\,,
\label{eq::zm_nc}
\end{eqnarray}
where the notation used for the superscripts is
self-explanatory.\footnote{Example: $z_m^{L 1/N_c^2}$ is the
  coefficient of $n_l/N_c^2$; ``$L$'' counts the factors $n_l$.}  The $n_l^3$
and $n_l^2$ terms are known analytically and can be found in
Ref.~\cite{Beneke:1994qe,Lee:2013sx} (see Appendix~\ref{app::ana_res}).  Both
for the linear-$n_l$ and the $n_l$-independent contribution one obtains small
(relative) uncertainties for the positive powers in $N_c$ which dominate in
the physical limit $N_c=3$.  This explains the small uncertainties of
coefficients in the previous subsection.

From Eq.~(\ref{eq::zm_nc}) one learns that for $N_c=3$ the dominant 
uncertainty originates from $z_m^{ N_c^2}$, followed by the
$N_c$-independent term $z_m^{ N_c^0}$.

As a cross check we choose $N_c=3$, fix $n_l$ and use the coefficients of
Eqs.~(\ref{eq::zm_nc}) to compute $z_m^{(4)}$ combining all uncertainties again
quadratically. We obtain the following results
\begin{center}
\begin{tabular}{c|c}
  $n_l$ & $z_m^{(4)}$\\
  \hline
  3 & $-1756.36 \pm 1.52$\\
  4 & $-1278.70 \pm 1.53$\\
  5 & $-871.73 \pm 1.53$
\end{tabular}
\end{center}
The central values are by construction identical to the corresponding
entries in Table~\ref{tab::zm_nc3}, the uncertainties are even
slightly smaller. {This might happen since the uncertainties are added
  linearly when setting $N_c$ and $n_l$ to
  numerical values before combining the uncertainties from the individual
  $\epsilon$ terms (cf. Subsection~\ref{sub::nc3}). As compared to adding the
  uncertainties in quadrature this 
  might lead to larger (as in the case at hand) or smaller (see next
  subsection) uncertainties.}


\subsection{\label{sub::SUN}Results in terms of Casimir colour factors}

This subsection is devoted to the most general
results, namely $z_m$ in the form of Eq.~(\ref{eq::zm4_col}).
For the coefficients of the 23 colour structures we obtain
\begin{eqnarray}
          z_m^{FFFF} &=&   -6.983 \pm    0.805\,,\nonumber\\
          z_m^{FFFA} &=&   13.40 \pm    2.07\,,\nonumber\\
          z_m^{FFAA} &=&  -11.17 \pm    1.74\,,\nonumber\\
          z_m^{FAAA} &=&  -99.272 \pm    0.493\,,\nonumber\\
          z_m^{d_{FA}} &=&    0.39 \pm    1.07\,,\nonumber\\
          z_m^{d_{FF}L} &=&   -0.937 \pm    0.178\,,\nonumber\\
          z_m^{d_{FF}H} &=&   -3.924 \pm    0.642\,,\nonumber\\
          z_m^{FFFL} &=&   -0.05094 \pm    0.00298\,,\nonumber\\
          z_m^{FFAL} &=&    9.26642 \pm    0.00454\,,\nonumber\\
          z_m^{FAAL} &=&  122.1872 \pm    0.0100\,,\nonumber\\
          z_m^{FFLL} &=&   -2.25441 \,,\nonumber\\
          z_m^{FALL} &=&  -42.46326 \,,\nonumber\\
          z_m^{FLLL} &=&    4.06885 \,,\nonumber\\
          z_m^{FFFH} &=&   -1.3625 \pm    0.0132\,,\nonumber\\
          z_m^{FFAH} &=&   14.9800 \pm    0.0334\,,\nonumber\\
          z_m^{FAAH} &=&   -2.3597 \pm    0.0342\,,\nonumber\\
          z_m^{FFHH} &=&    1.65752 \pm    0.00031\,,\nonumber\\
          z_m^{FAHH} &=&   -0.20934 \pm    0.00273\,,\nonumber\\
          z_m^{FHHH} &=&   -0.14902 \pm    0.00000\,,\nonumber\\
          z_m^{FFLH} &=&   -2.89209 \pm    0.00010\,,\nonumber\\
          z_m^{FALH} &=&    0.62076 \pm    0.00042\,,\nonumber\\
          z_m^{FLLH} &=&   -0.10321 \,,\nonumber\\
          z_m^{FLHH} &=&   -0.21703 \pm    0.00000\,.
\label{eq::zm_cacf}
\end{eqnarray}
The $n_l^3$ and $n_l^2$ terms are known analytically and can be found
in Appendix~\ref{app::ana_res}. The linear-$n_l$ term is dominated by
$z_m^{FAAL}$ which has an uncertainty below 0.01\%. On the other hand,
for $z_m^{FFFL}$ the precision is only about 4\%, however, the 
numerical impact is small, even for $N_c=2$.

The contributions involving closed heavy quark loops are generally
small and known to a precision of about 10\% or better, the numerically
dominant $z_m^{FFAH}$ contribution even to about 1.3\%.

There are five non-fermionic contributions,
$z_m^{FFFF}$, $z_m^{FFFA}$, $z_m^{FFAA}$, $z_m^{FAAA}$
and $z_m^{d_{FA}}$. The most precise one, $z_m^{FAAA}$,
has by far the largest coefficient and furthermore 
the largest colour factor. The three coefficients
$z_m^{FFFF}$, $z_m^{FFFA}$ and $z_m^{FFAA}$ have an uncertainty 
between 11\% and 15\%.
$z_m^{d_{FA}}$ is the worst known coefficient. Actually, within our
precision we cannot claim whether it is positive or negative.
Note, however, that not only the coefficient itself but also 
the colour factor is numerically small
as compared to others. For example, for $N_c=3$ we have
$d^{abcd}_F d^{abcd}_A/N_c = 15/6 = 2.5$ whereas $C_F C_A^3=36$.
The current uncertainty of $z_m^{d_{FA}}$ is dominated
by master integrals where we rely on the {\tt FIESTA} results.

As a cross check we insert the results from Eq.~(\ref{eq::zm_cacf})
into Eq.~(\ref{eq::zm4_col}) and specify the colour factors
to their numerical values with $N_c=3$. We add all uncertainties
in quadrature and obtain
\begin{center}
\begin{tabular}{c|c}
  $n_l$ & $z_m^{(4)}$\\
  \hline
  3 & $-1756.36 \pm 36.3$\\
  4 & $-1278.70 \pm 36.3$\\
  5 & $-871.73 \pm 36.3$
\end{tabular}
\end{center}
which has to be compared with the corresponding entries in
Table~\ref{tab::zm_nc3} where $N_c=3$ is chosen
before combining the uncertainties from the individual master integrals.
As expected, one observes the same central value, however, the
uncertainties are significantly larger.



\section{\label{sec::appl}Applications}


\subsection{$\overline{\rm MS}$-on-shell transformation formulae}

In the following we discuss the relation between the $\overline{\rm MS}$ and
on-shell quark mass and specify the number of massless quarks to the top,
bottom and charm case.

Let us start with the version where the on-shell mass is computed from the
$\overline{\rm MS}$ mass. We use as input the following $\overline{\rm MS}$
masses
\begin{eqnarray}
  m_t(m_t) &=& 163.508~\mbox{GeV}\,,\nonumber\\
  m_b(m_b) &=& 4.163~\mbox{GeV}\,,\nonumber\\
  m_c(3~\mbox{GeV}) &=& 0.986~\mbox{GeV}\,,
  \label{eq::input}
\end{eqnarray}
where $m_t(m_t)$ is computed from
$M_t=173.34$~GeV~\cite{ATLAS:2014wva} using four-loop accuracy.
The $\overline{\rm MS}$ masses for charm and bottom are taken from
Ref.~\cite{Chetyrkin:2009fv}.

The values for the strong coupling are given by
$\alpha_s^{(6)}(m_t)=0.1085$, $\alpha_s^{(5)}(m_b)=0.2253$, and
$\alpha_s^{(4)}(3~\mbox{GeV})=0.2540$.  They have been computed from
$\alpha_s^{(5)}(M_Z)=0.1181$ using {\tt
  RunDec}~\cite{Chetyrkin:2000yt,Schmidt:2012az}.  In the case of the
charm quark we also provide results for $\mu=m_c(m_c)$ using the input
values $m_c(m_c) = 1.279$~GeV and $\alpha_s^{(4)}(m_c)=0.3872$.  Note
that the choice $\mu=3$~GeV is preferable since it has the advantage
that low renormalization scales $\mu\approx m_c$ are avoided.

In the following equations we list the results for the relations
which convert the $\overline{\rm MS}$ to the on-shell mass.
For simplicity we set here and in the remainder of this section
the uncertainty of the four-loop coefficient to
0.2\% although it is for charm and bottom slightly smaller (see
Table~\ref{tab::zm_nc3}). We obtain
\begin{eqnarray}
  M_t &=& m_t(m_t)\left(
    1 + 0.4244 \,{\alpha_s} + 0.8345 \,{\alpha_s^2} + 2.375 \,{\alpha_s^3}
    + (8.615 \pm 0.017) \,{\alpha_s^4}
  \right)
  \nonumber\\&=&\mbox{}
  163.508 + 7.529 + 1.606 + 0.496
  + (0.195 \pm 0.0004)~\mbox{GeV}
  \,,
  \label{eq::mt}
\\
  M_b &=& m_b(m_b)\left(
    1 + 0.4244 \,{\alpha_s} + 0.9401 \,{\alpha_s^2} + 3.045 \,{\alpha_s^3}
    + (12.685 \pm 0.025) \,{\alpha_s^4}
  \right)
  \nonumber\\&=&\mbox{}
  4.163 + 0.398 + 0.199 + 0.145
  + (0.136 \pm 0.0003)~\mbox{GeV}
  \,,
  \label{eq::mb}
\\
  M_c &=& m_c(3~\mbox{GeV})\left(
    1 + 1.133 \,{\alpha_s} + 3.119 \,{\alpha_s^2} + 10.981 \,{\alpha_s^3}
    + (51.419 \pm 0.102) \,{\alpha_s^4}
  \right)
  \nonumber\\&=&\mbox{}
  0.986 + 0.284 + 0.198 + 0.177
  + (0.211 \pm 0.0004)~\mbox{GeV}
  \,,
  \label{eq::mc3}
\\
  M_c &=& m_c(m_c)\left(
    1 + 0.4244 \,{\alpha_s} + 1.0456 \,{\alpha_s^2} + 3.757 \,{\alpha_s^3}
    + (17.480 \pm 0.035) \,{\alpha_s^4}
  \right)
  \nonumber\\&=&\mbox{}
  1.279 + 0.210 + 0.200 + 0.279
  + (0.503 \pm 0.001)~\mbox{GeV}
  \,,
  \label{eq::mc}
\end{eqnarray}
where the renormalization scale of $\alpha_s$ in each equation is
identical to the one specified for the $\overline{\rm MS}$ quark mass
in the prefactor of the first lines in each equation.

One observes a good convergence for the top quark where the 
coefficients steadily decrease; the four-loop coefficient is more
than a factor two smaller than the three-loop one. This is different
for charm and bottom where the two-, three- and four-loop coefficients 
are of the same order of magnitude. In Eq.~(\ref{eq::mc}) (where
$\mu^2=m_c^2$ has been chosen) the four-loop coefficient is even almost
twice as large as the three-loop coefficient.

For convenience we also present the inverted relation of
Eq.~(\ref{eq::mt}) which is given by
\begin{eqnarray}
  m_t(m_t) &=& M_t\left(
    1 - 0.4244 \,{\alpha_s} - 0.9246 \,{\alpha_s^2} - 2.593\,{\alpha_s^3}
    - (8.949 \pm 0.018) \,{\alpha_s^4}
  \right)
  \nonumber\\&=&\mbox{}
  173.34 - 7.924 - 1.859 - 0.562
  - (0.209 \pm 0.0004)~\mbox{GeV}
  \,,
  \label{eq::mt2}
\end{eqnarray}
where $\alpha_s\equiv \alpha_s(M_t)=0.1077$.
We refrain from providing the analogue equations for charm and bottom
since this would require specifying the pole masses.


\subsection{\label{sub::thr}Relation between $\overline{\rm MS}$ and threshold mass}

The threshold masses are constructed such that the relation to
the $\overline{\rm MS}$ mass is well behaved in perturbation theory.
It is illustrating to examine the cancellations which take
place between the coefficients in the $\overline{\rm MS}$-OS relation
and the ones in the relation of the OS and threshold mass.
For example, in the case for the bottom quark mass we have for the PS mass
\begin{eqnarray}
  m_b^{\rm PS}(\mu_f = 2~\mbox{GeV}) &=& 
  4.163 + (0.399 - 0.191)
  + (0.199 - 0.120) 
  \nonumber\\&&
  + (0.145 - 0.114)
  + (0.1364 - 0.1368 \pm 0.0003)~\mbox{GeV}
  \nonumber\\&=&
  4.163 + 0.207 + 0.080 + 0.032 - (0.0004 \pm 0.0003)~\mbox{GeV}
  \,,
  \label{eq::mbPSMS}
\end{eqnarray}
where the second terms inside the round brackets after the first equality sign
originate from the PS-OS relation.  As expected due to the very definition of
the PS mass, one observes a significant cancellation between the coefficients
of the PS-OS and PS-$\overline{\rm MS}$ relation.  The cancellation becomes
stronger at higher loop-order. In particular, at four loops one observes a
cancellation of three significant digits, which is the reason why four digits
after the comma are provided. Note that the details of the cancellations
depend on $\mu_f$, as we will discuss in Section~\ref{sec::muf_dep}.

After the second equality sign the numbers in the round brackets are added.
One observes a nice convergence behaviour with decreasing coefficients which
has to be compared to the OS-$\overline{\rm MS}$ relation where the three- and
four-loop coefficients have the same order of magnitude,
cf. Eq.~(\ref{eq::mb}).  The four-loop coefficient in Eq.~(\ref{eq::mbPSMS})
only amounts to $-0.4$~MeV which is actually of the same order of magnitude
as the uncertainty. Note, however, that both the central value and the uncertainty
are far below the expected precision of the $\overline{\rm MS}$ bottom quark
mass within the foreseeable future.

The analogue equation to~(\ref{eq::mbPSMS}) for the top quark reads
\begin{eqnarray}
  m_t^{\rm PS}(\mu_f = 80~\mbox{GeV}) &\!=\!& 
  163.508 + (7.531 - 3.685) +  (1.607 - 0.989) 
  \nonumber\\&&
  + (0.495 - 0.403) 
  + (0.195 - 0.211 \pm 0.0004)~\mbox{GeV}
  \nonumber\\&\!=\!&
  163.508 + 3.847 + 0.618 + 0.092 - (0.016 \pm 0.0004)~\mbox{GeV}
  \,.
  \label{eq::mtPSMS}
\end{eqnarray}
Also here one observes a drastic reduction of the correction terms when
going to higher orders. In fact, the last term amounts to only 16~MeV
instead of 200~MeV in Eq.~(\ref{eq::mt}).

For the 1S mass we obtain the following perturbative
relations to the $\overline{\rm MS}$ bottom and top mass
\begin{eqnarray}
  m_b^{\rm 1S} &=& 
  4.163 + (0.399 - 0.047)
  + (0.195 - 0.072) 
  \nonumber\\&&
  + (0.139 - 0.100)
  + (0.129 - 0.137 \pm 0.0003)~\mbox{GeV}
  \nonumber\\&=&
  4.163 + 0.352 + 0.123 + 0.039 - (0.008 \pm 0.0003)~\mbox{GeV}
  \,,
  \nonumber\\
  m_t^{\rm 1S} &=& 
  163.508 + (7.531 - 0.428) +  (1.588 - 0.368) 
  \nonumber\\&&
  + (0.479 - 0.262) 
  + (0.185 - 0.174 \pm 0.0004)~\mbox{GeV}
  \nonumber\\&=&
  163.508 + 7.103 + 1.220 +  0.217 + (0.011 \pm 0.0004)~\mbox{GeV}
  \,,
  \label{eq::mbmt1SMS}
\end{eqnarray}
where the first and second number in the bracket originates from the
OS-$\overline{\rm MS}$ and OS-1S relation, respectively. Furthermore,
we order the terms according the $\varepsilon$ expansion as defined in
Refs.~\cite{Hoang:1998hm,Hoang:1998ng,Hoang:1999zc}. It is interesting
to note that at leading order (LO) (first round bracket) the contribution from the
OS-1S relation amounts only to a few per cent of the OS-$\overline{\rm
  MS}$ relation. At N$^3$LO, however, it is more than 90\% both for 
bottom and top.

Similar results to those presented in
Eqs.~(\ref{eq::mbPSMS}),~(\ref{eq::mtPSMS}) and~(\ref{eq::mbmt1SMS}) for
bottom and top are also obtained for the charm quark in case $\mu=3$~GeV is
chosen for the renormalization scale. On the other hand, in the case
when the relation
of the threshold mass to $m_c(m_c)$ is computed the four-loop term exceeds the
three-loop one.  We furthermore observe that the relations to the RS and
RS$^\prime$ masses behave very similar to the PS and 1S masses.  We refrain
from providing explicit results which are easily obtained with the help of {\tt
  RunDec}~\cite{Chetyrkin:2000yt} and {\tt CRunDec}~\cite{Schmidt:2012az}.

In practice a threshold quark mass is extracted from comparison of
experimental measurements and theory predictions.  Afterwards it is converted
to the $\overline{\rm MS}$ quark mass. In Table~\ref{tab::mqMS} we show the
results for the scale invariant $\overline{\rm MS}$ quark masses $m_q(m_q)$
($q=t,b,c$) and $m_c(3~\mbox{GeV})$ using one- to four-loop accuracy for the
conversion. The input values for the threshold masses (which are provided
at the top of each table) are chosen such that the four-loop results
agree with the input values discussed in Eq.~(\ref{eq::input}).
For the top quark a rapid convergence is observed with four-loop
contributions between 10 and 20~MeV.  The situation is similar for the bottom
quark where the four-loop term amounts to at most 8~MeV for the case of the 1S
mass.  As already mentioned above, the four-loop term for the case where
$m_c(m_c)$ is computed from the threshold masses is larger than the three-loop
contribution which is different for $m_c(3~\mbox{GeV})$ where the four-loop
term is smaller by up to a factor four.  Thus, even in this case we observe a
reasonable convergence of the perturbative series; for the PS and RS masses the
N$^3$LO corrections are even below 10~MeV.

The results in Table~\ref{tab::mqMS} show that perturbatively well-behaved
quark mass relations are obtained after introducing threshold masses. To
exploit them at third order in perturbation theory, which is
mandatory due to current precision reached for the quark masses,
it is necessary to use the four-loop relation between the on-shell and
$\overline{\rm MS}$ quark mass for the construction of the 
$\overline{\rm MS}$-threshold mass relation.


\begin{table}
{\scalefont{0.7}
\begin{tabular}{ccc}
\begin{minipage}{20em}
\begin{center}
\begin{tabular}{c|cccc}
input & $m^{\rm PS} =$ & $m^{\rm 1S} = $ & $m^{\rm RS} =$ & $m^{\rm RS^\prime}
=$ \\
\#loops  &    168.049 &    172.060 &    166.290 &    171.785\\
\hline
1       &    164.174    &    164.904    &    163.702    &    164.226 \\
2       &    163.580    &    163.727    &    163.520    &    163.591 \\
3       &    163.492    &    163.519    &    163.490    &    163.500 \\
4       &    163.508    &    163.508    &    163.508    &    163.508 \\
\hline
4 ($\times 1.002$)       &    163.507   &    163.507    &    163.507    &
163.507 \\
\end{tabular}
\end{center}
\end{minipage}
& \hspace*{4em} &
\begin{minipage}{20em}
\begin{center}
\begin{tabular}{c|cccc}
input & $m^{\rm PS} =$ & $m^{\rm 1S} = $ & $m^{\rm RS} =$ & $m^{\rm RS^\prime}
=$ \\
\#loops  &      4.481 &      4.668 &      4.364 &      4.692\\
\hline
1       &      4.266    &      4.308    &      4.210    &      4.286 \\
2       &      4.191    &      4.192    &      4.173    &      4.196 \\
3       &      4.163    &      4.155    &      4.159    &      4.165 \\
4       &      4.163    &      4.163    &      4.163    &      4.163 \\
\hline
4 ($\times 1.002$)       &      4.163   &      4.163    &      4.163    &
4.163 \\
\end{tabular}
\end{center}
\end{minipage}
\\ \\ {\scalefont{1.3} (a) $m_t(m_t)$ } && {\scalefont{1.3} (b) $m_b(m_b)$}
\\ \\
\begin{minipage}{20em}
\begin{center}
\begin{tabular}{c|cccc}
input & $m^{\rm PS} =$ & $m^{\rm 1S} = $ & $m^{\rm RS} =$ & $m^{\rm RS^\prime}
=$ \\
\#loops  &      1.130 &      1.513 &      1.035 &      1.351\\
\hline
1       &      1.255    &      1.342    &      1.249    &      1.146 \\
2       &      1.230    &      1.250    &      1.273    &      1.276 \\
3       &      1.235    &      1.214    &      1.249    &      1.250 \\
4       &      1.279    &      1.279    &      1.279    &      1.279 \\
\hline
4 ($\times 1.002$)       &      1.278   &      1.278    &      1.278
&      1.278 \\
\end{tabular}
\end{center}
\end{minipage}
&&
\begin{minipage}{20em}
\begin{center}
\begin{tabular}{c|cccc}
input & $m^{\rm PS} =$ & $m^{\rm 1S} = $ & $m^{\rm RS} =$ & $m^{\rm RS^\prime}
=$ \\
\#loops  &      1.153 &      1.545 &      1.043 &      1.357\\
\hline
1       &      1.077    &      1.261    &      1.028    &      1.074 \\
2       &      1.021    &      1.117    &      1.008    &      1.020 \\
3       &      0.993    &      1.032    &      0.992    &      0.995 \\
4       &      0.986    &      0.986    &      0.986    &      0.986 \\
\hline
4 ($\times 1.002$)       &      0.986   &      0.986    &      0.986    &
0.986 \\
\end{tabular}
\end{center}
\end{minipage}
\\ \\ {\scalefont{1.3} (c) $m_c(m_c)$ } && {\scalefont{1.3} (d) $m_c(3~\mbox{GeV})$}
\end{tabular}
}
\caption{\label{tab::mqMS}$m_q(m_q)$ ($q=t,b,c$) in GeV [see (a), (b), (c)]
  and $m_c(3~\mbox{GeV})$ (d) computed from the PS, 1S, RS
  and RS$^\prime$ quark mass using LO to N$^3$LO accuracy. The numbers in
  the last line are obtained by taking into account the uncertainty of
  the four-loop coefficient, i.e., it is increased by 0.2\%. This leads to a
  shift of at most 1~MeV.
  The factorization scales for the PS, RS and RS$^\prime$ masses are set to 
  2~GeV for bottom and charm. For the top quark we use $\mu_f=80$~GeV for
  the PS, RS and RS$^\prime$ masses.}
\end{table}


\begin{figure}[t]
  \centering
  \begin{tabular}{cc}
    \includegraphics[width=0.35\linewidth]{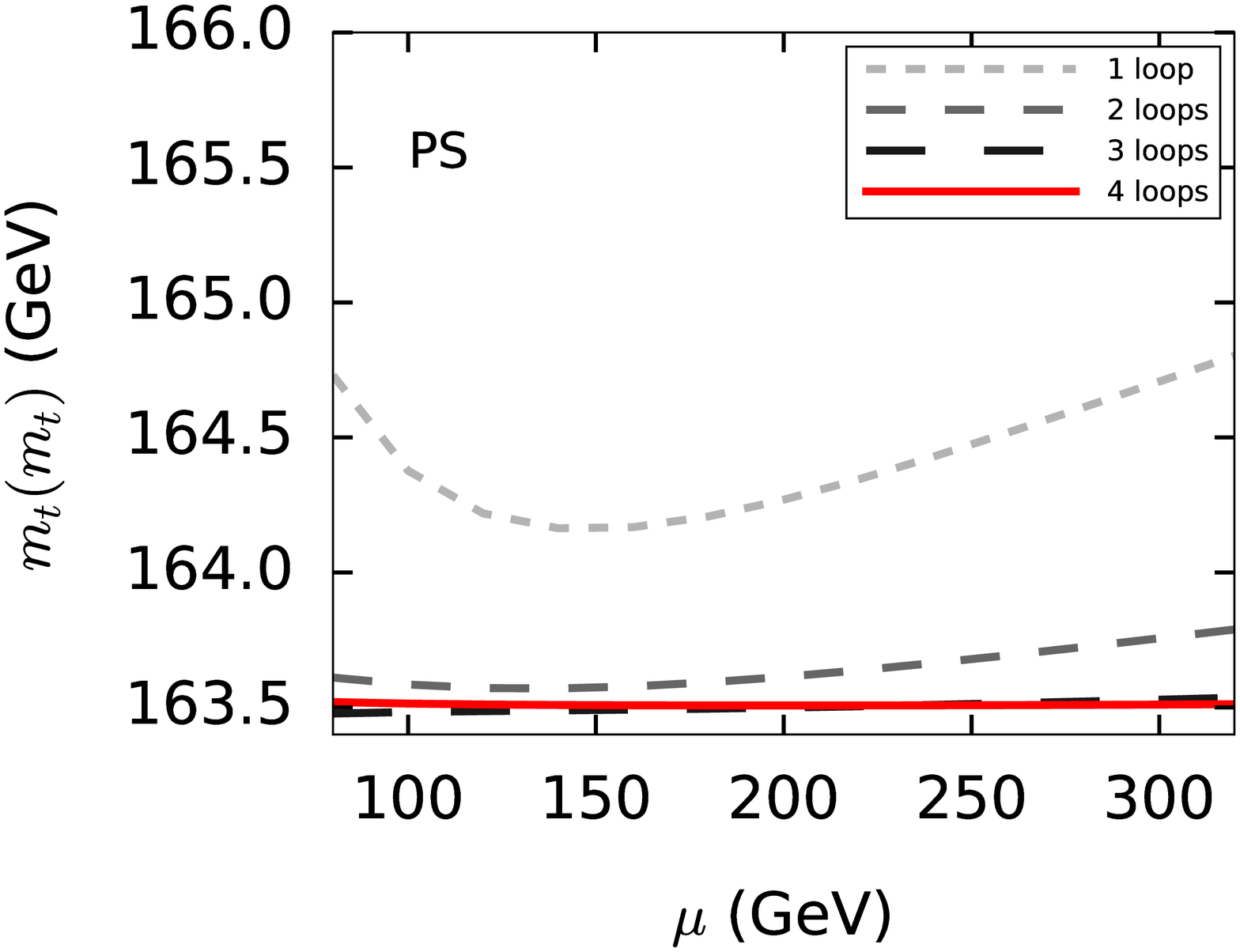}&
    \includegraphics[width=0.35\linewidth]{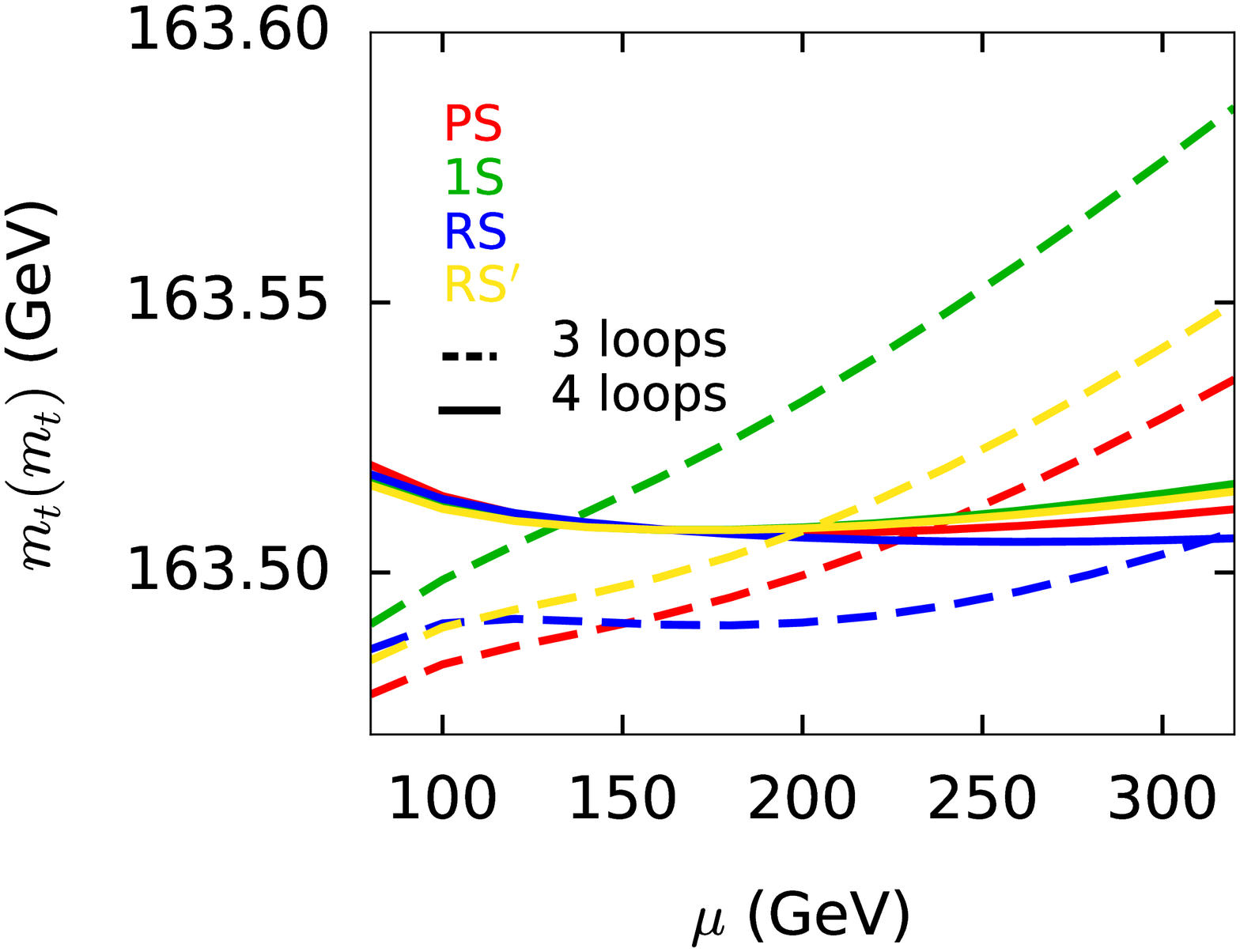}\\[-.5em]
    (a) & (b)
    \\[.5em]
    \includegraphics[width=0.35\linewidth]{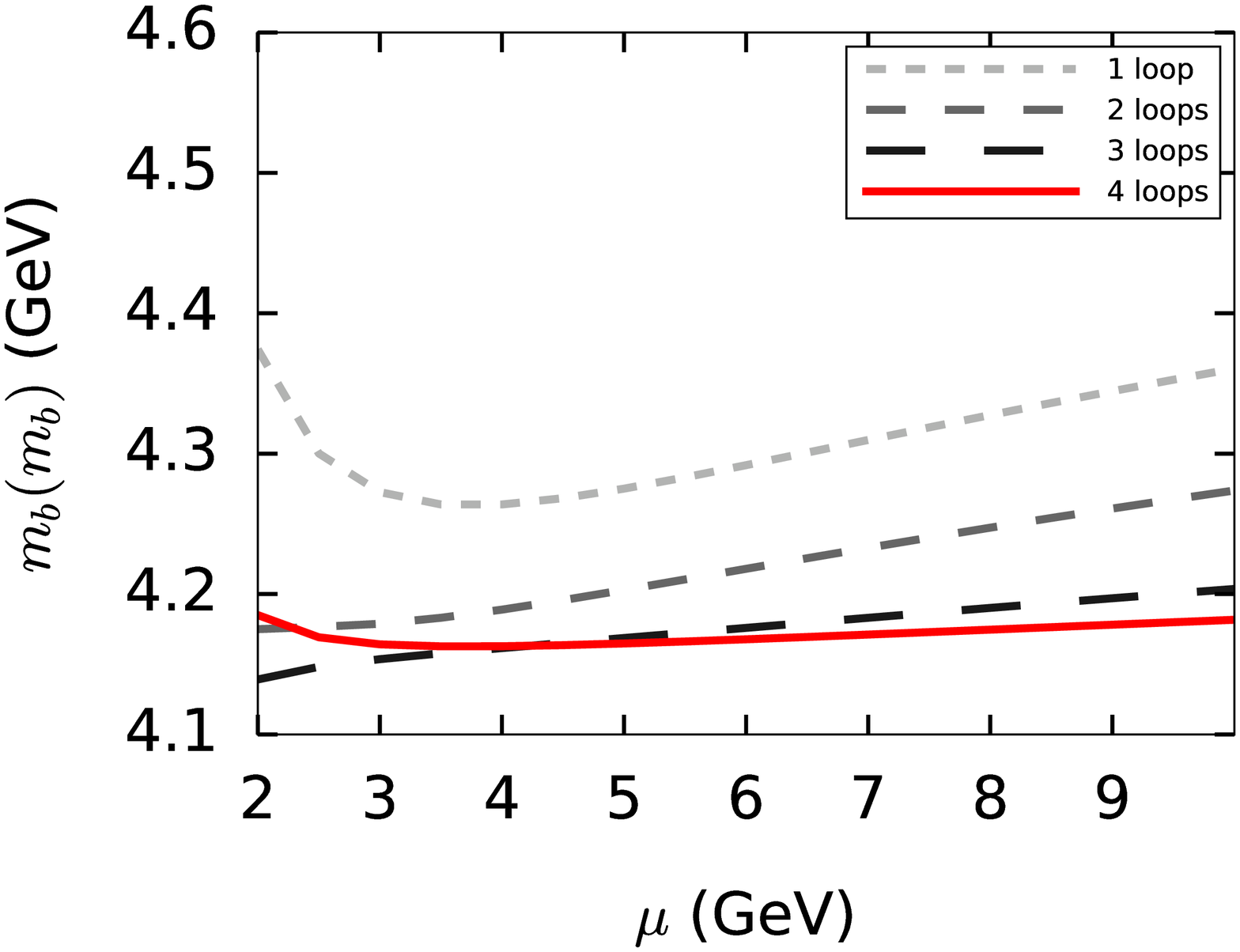}&
    \includegraphics[width=0.35\linewidth]{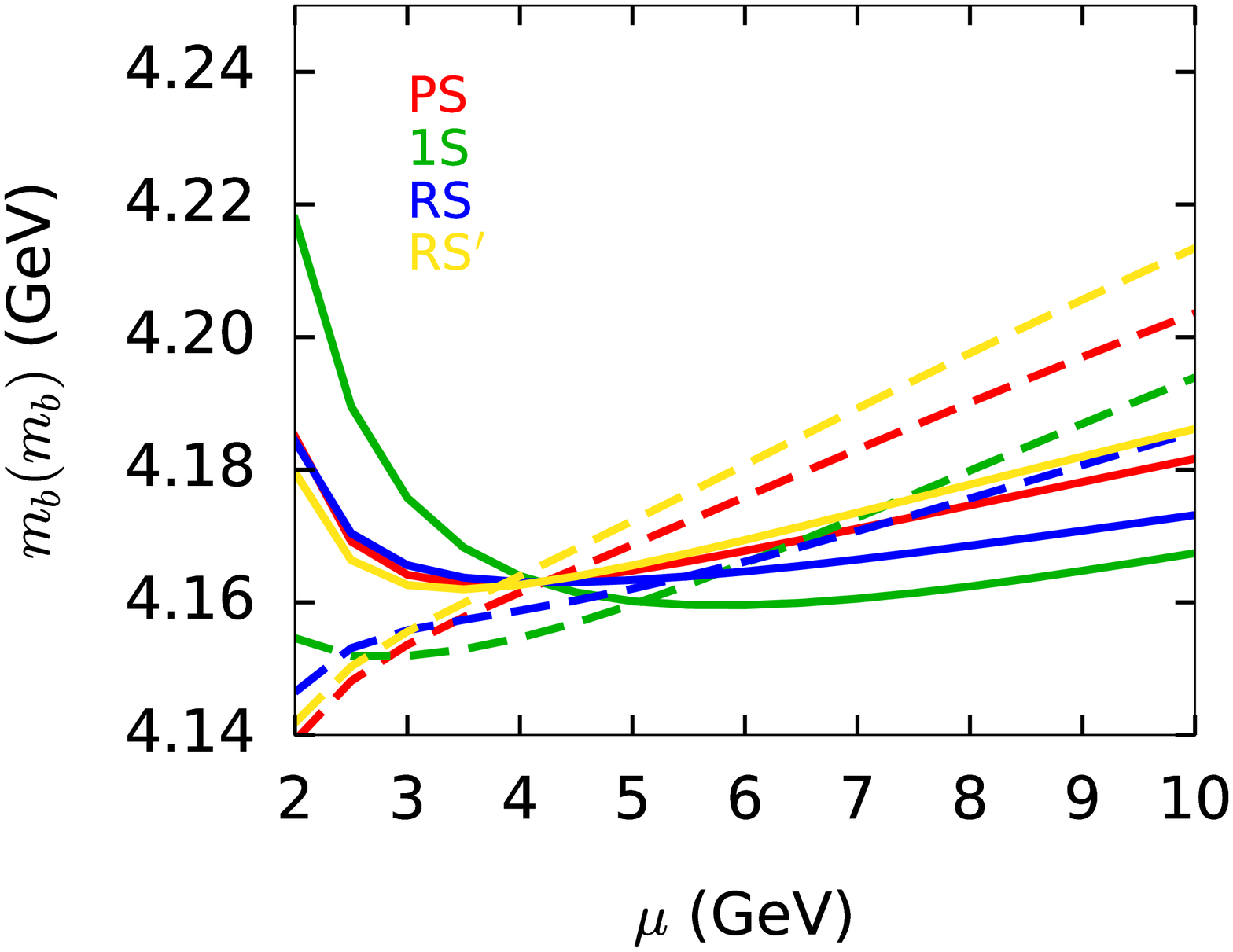}\\[-.5em]
    (c) & (d)
    \\[.5em]
    \includegraphics[width=0.35\linewidth]{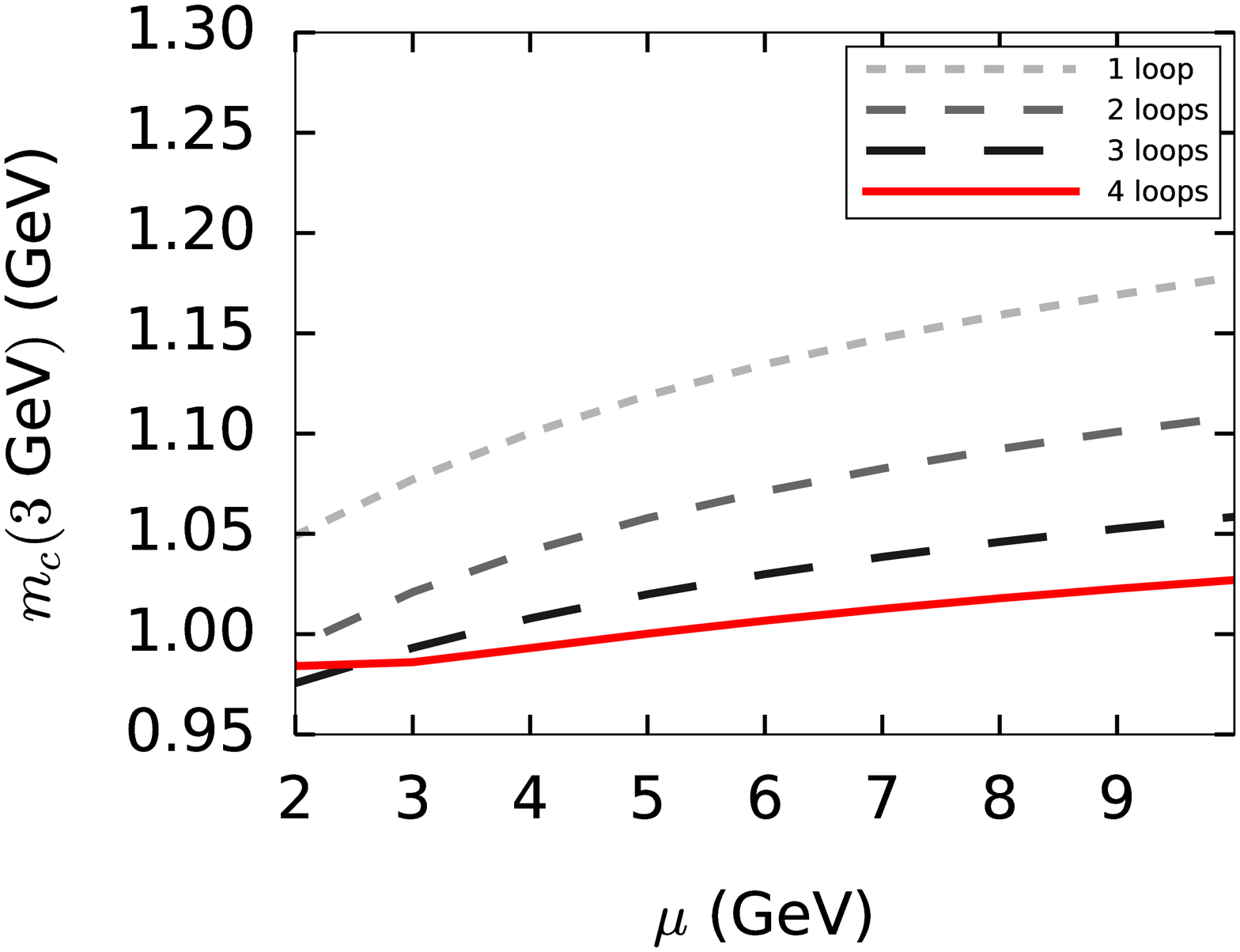}&
    \includegraphics[width=0.35\linewidth]{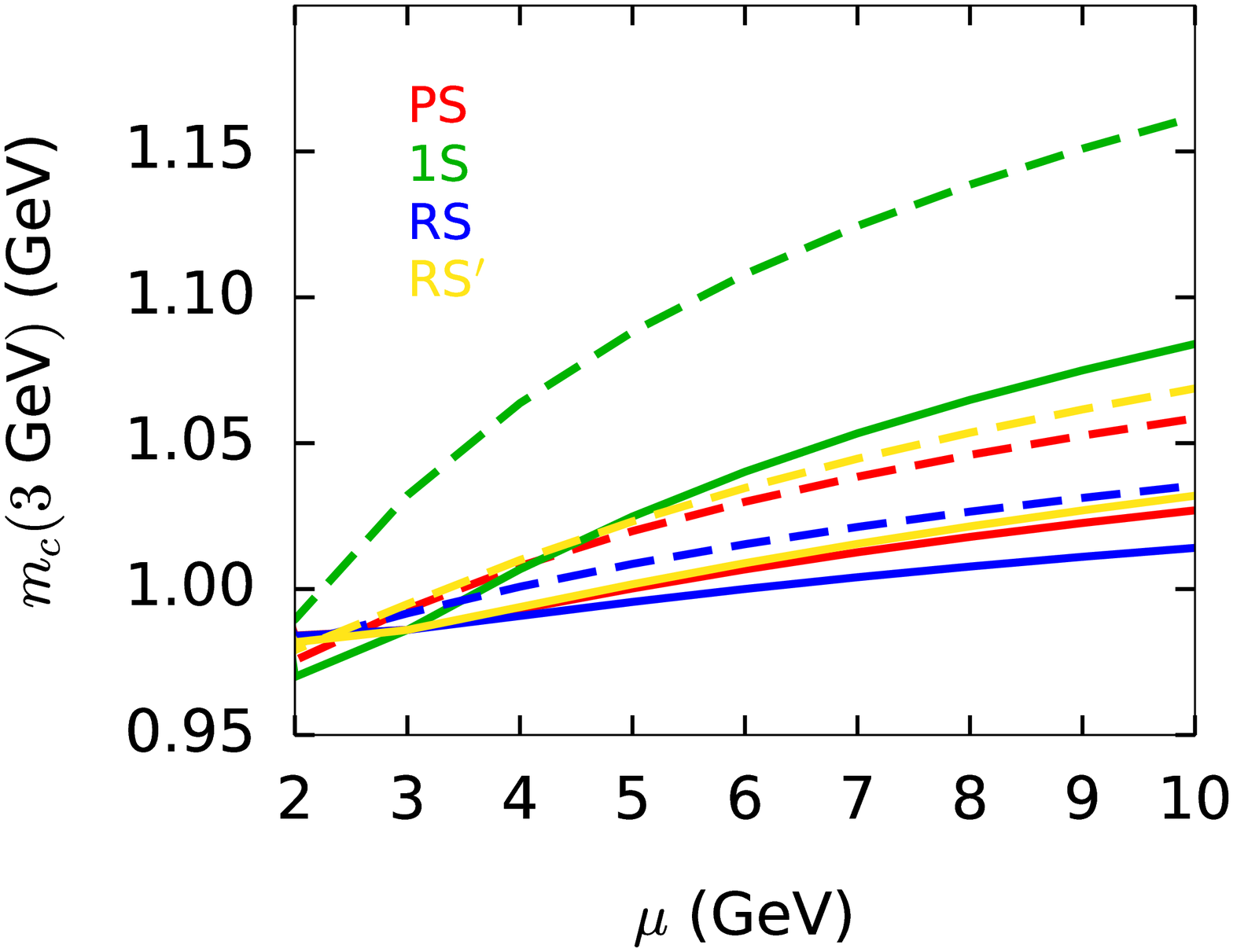}\\[-.5em]
    (e) & (f)
  \end{tabular}
  \caption{(a) $\overline{\rm MS}$ top quark mass $m_t(m_t)$ 
    computed from the PS mass with LO, NLO, NNLO and N$^3$LO accuracy
    as a function of the renormalization scale used in the 
    $\overline{\rm MS}$-threshold mass relation.
    (b) $\overline{\rm MS}$ top quark mass $m_t(m_t)$ computed from
    the PS, 1S, RS and RS$^\prime$ mass with NNLO (dashed) and N$^3$LO (solid line)
    accuracy as a function of the renormalization scale used in the
    $\overline{\rm MS}$-threshold mass relation. At the right end of the plot
    the lines from bottom to top correspond to the RS, PS, RS$^\prime$ 1S
    mass. (c)-(f) show the results for bottom and charm. For the bottom quark
    the four-loop result for the 1S mass is below (above) the others for high
    (low) values of $\mu$.}
  \label{fig::mMS-mTHR-mu_int}
\end{figure}

To obtain the results in Table~\ref{tab::mqMS} we have set the
renormalization scale in the relation between the threshold and $\overline{\rm
  MS}$ mass to the quark mass itself or to 3~GeV. As an alternative one could
also apply the conversion relation at some intermediate scale $\mu$ and then
run with four-loop accuracy in the $\overline{\rm MS}$ scheme for either
the scale invariant mass or to $\mu=3$~GeV for the charm quark. The
corresponding results are shown in Fig.~\ref{fig::mMS-mTHR-mu_int} where
$m_t(m_t)$, $m_b(m_b)$ and $m_c(3~\mbox{GeV})$ are shown as a function of the
intermediate scale $\mu$.  The panels on the left show the results for the PS
mass for the one- to four-loop analysis.  In all three cases one observes a
rapid convergence when including higher order corrections resulting in an
almost horizontal, i.e, $\mu$ independent, result at four loops.

The panels on the right compare the various threshold masses at three and four
loops.  Note that by construction the four-loop curves coincide for
$\mu=m_q(m_q)$ for top and bottom and for $\mu=3$~GeV for charm.  In all cases
one observes that the four-loop curves are significantly flatter than the
three-loop results. Particularly good results are obtained for the top quark
in panel (b) where in a large range the four-loop results lie on top of each
other. The four-loop curves in the case of the bottom quark show stronger
variations below, say, $\mu=2.5$~GeV. Here the PS, RS and RS$^\prime$ results
are quite close together whereas the 1S curve shows a quite strong rise for
$\mu\to 2$~GeV.  Note that the scale on the $y$ axis for the charm plot covers
a bigger range than for the bottom quark. Nevertheless the four-loop
curve shows a quite flat behaviour. One observes again that the 1S
curve deviates from the remaining ones.


\subsection{\label{sec::muf_dep}$\mu_f$ dependence of PS, RS and RS$^\prime$ mass}

\begin{figure}[t]
  \begin{tabular}{cc}
    \includegraphics[width=.45\textwidth]{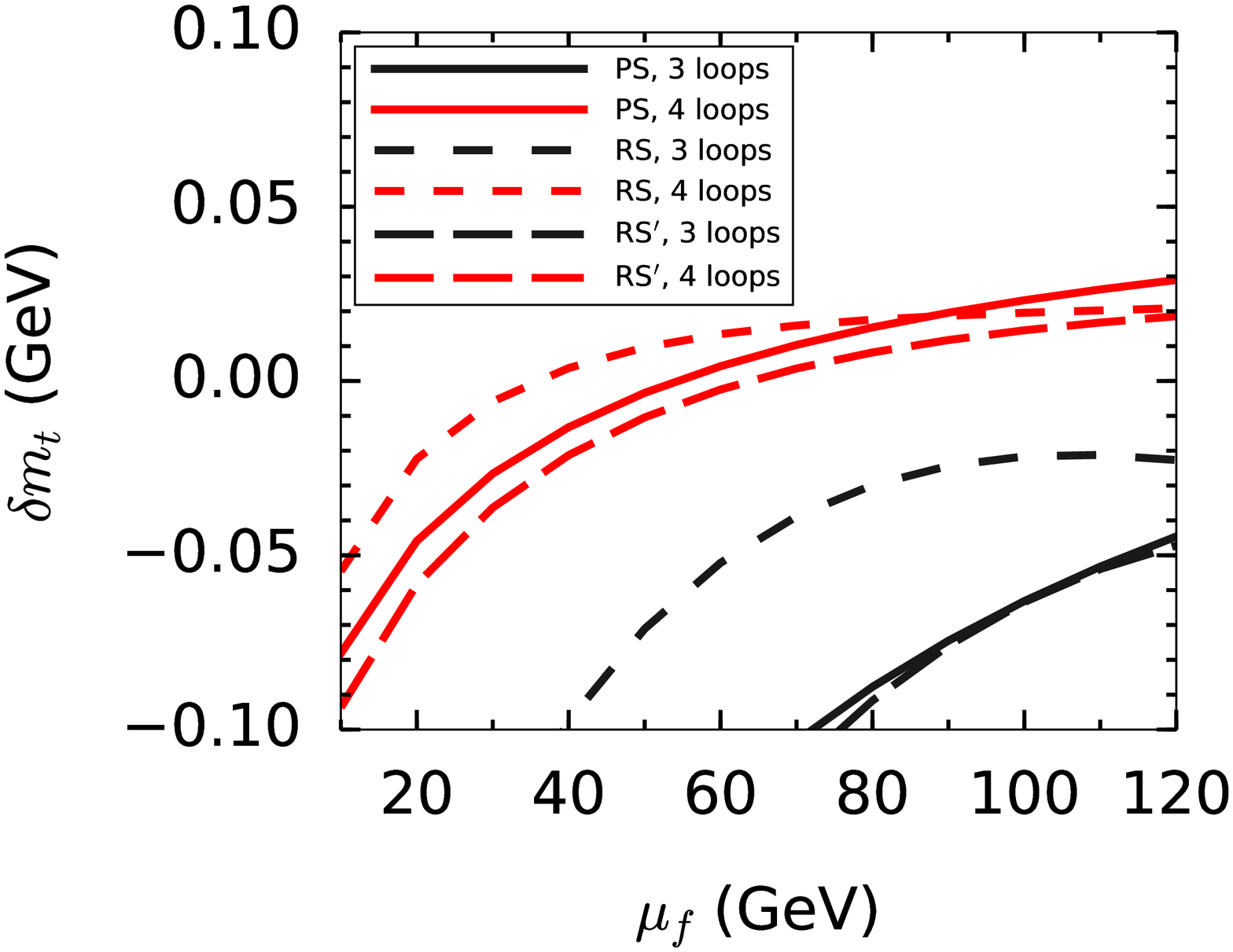} &
    \includegraphics[width=.45\textwidth]{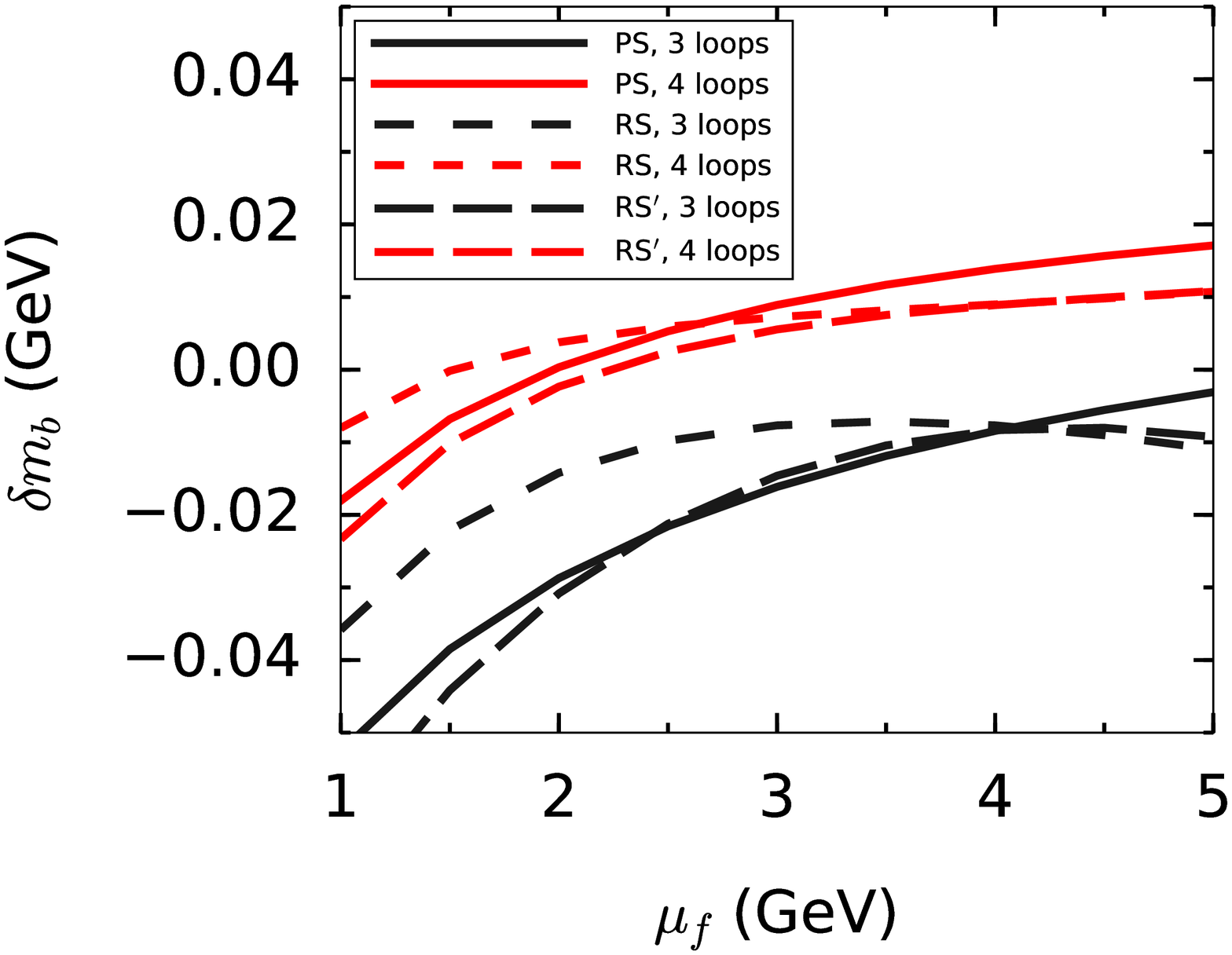}
    \\
    \includegraphics[width=.45\textwidth]{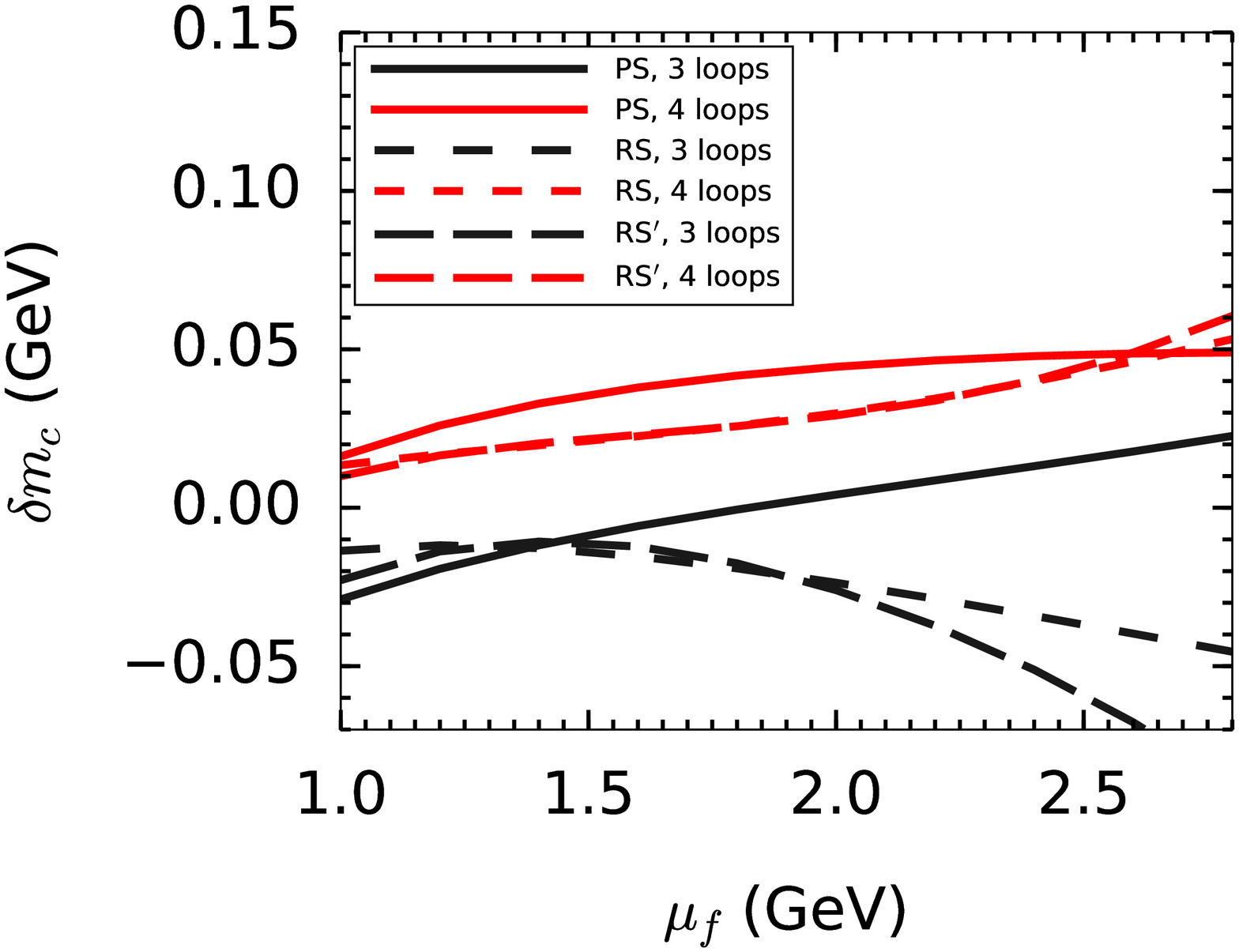} &
    \includegraphics[width=.45\textwidth]{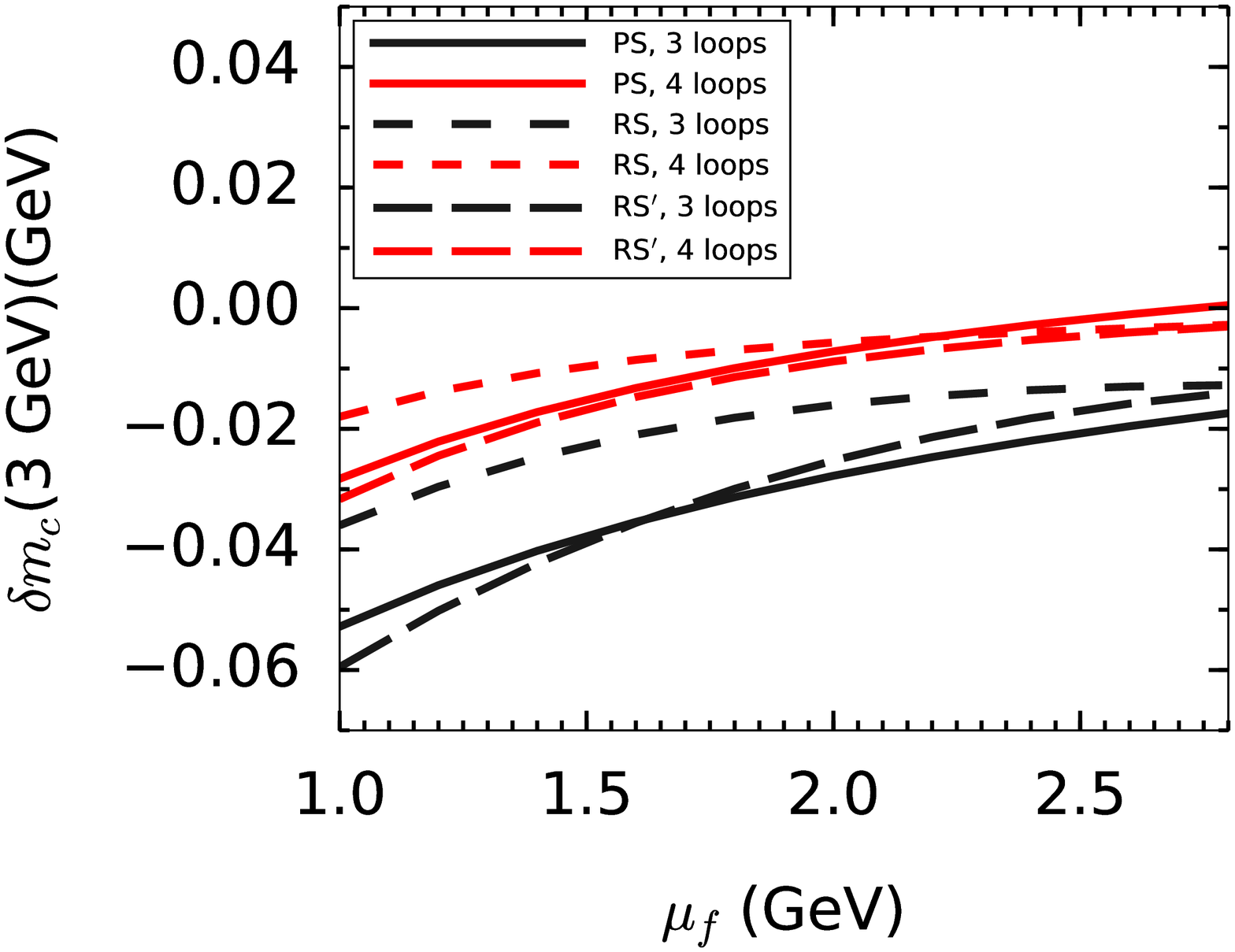}
  \end{tabular}
  \caption[]{\label{fig::mMSmuf}Three- (black, lower three curves
      in each plot) and 
    four-loop (red, upper three curves in each plot) contribution in
    transition form PS, RS and RS$^\prime$ mass to $m_q(m_q)$ and $m_c(3~\mbox{GeV})$
    as a function of the factorization scale $\mu_f$.}
\end{figure}

In this Section we study the dependence of the PS, RS and RS$^\prime$ mass on
the factorization scale $\mu_f$. To do this we use $m_t(m_t)$,
$m_b(m_b)$, $m_c(m_c)$ and $m_c(3~\mbox{GeV})$ from Eq.~(\ref{eq::input})
and compute the threshold masses for the given value of $\mu_f$ to four-loop
accuracy. This value is then used as starting point for the computation of
the $\overline{\rm MS}$ mass at one- to four-loop order as a function of
$\mu_f$.  In Fig.~\ref{fig::mMSmuf} the three- (black) and four-loop (red)
contributions to the conversion formula are plotted for the PS (solid), RS
(short dashes) and RS$^\prime$ (long dashes) masses. In the case of the top
quark the default scale for the PS mass suggested in Ref.~\cite{Beneke:1998rk}
is $\mu_f=20$~GeV. For this value the four-loop contribution amounts to about
$-50$~MeV.  One observes that the perturbative conversion formula is better
behaved for larger values of $\mu_f$. In fact, the four-loop term vanishes for
$\mu_f\approx 50$~MeV and amounts to about $+10$~MeV for $\mu_f\approx
80$~MeV, a value suggested in Ref.~\cite{Beneke:2015kwa} in the context of top
quark pair production close to threshold.  Similar conclusions also hold for
the RS and RS$^\prime$ masses.

For the bottom quark the general behaviour of the three- and four-loop
correction terms is similar to the top quark case. Here, the suggested 
default value of $\mu_f=2$~GeV~\cite{Beneke:1998rk,Pineda:2001zq} seems to be a 
good choice from the perturbative point of view.

For completeness we show in Fig.~\ref{fig::mMSmuf} the corresponding results
for the charm quark masses $m_c(m_c)$ and $m_c(3~\mbox{GeV})$.
Here, the results are less conclusive, in particular for $m_c(m_c)$. Over a
large range of $\mu_f$ the four-loop term is even larger than the three-loop
contribution which is a sign that the formalism should not be applied to
$m_c(m_c)$. The situation is better in case $m_c(3~\mbox{GeV})$ is
considered, which is probably due to the smaller values of $\alpha_s$
(which increases significantly when going from $\mu=3$~GeV to
$\mu=m_c(m_c)\approx 1.3$~GeV). For $m_c(3~\mbox{GeV})$  the four-loop
contribution is always smaller than the three-loop term, however,
it comes close to zero only for values near $\mu_f\approx 3$~GeV.


\subsection{\label{sub::zm_apinl}$c_m$ in terms of $\alpha_s^{(n_l)}$}

For certain applications (see, e.g, Ref.~\cite{Beneke:2016cbu}) it is
necessary to express the $\overline{\rm MS}$-on-shell relation in
terms of $\alpha_s^{(n_l)}$ instead of $\alpha_s^{(n_l+1)}$.
It is obtained by using the decoupling relation for 
$\alpha_s$ which is given by\footnote{The formulae of this subsection and
  the ones of the appendices (except Appendix~\ref{app::ana_res}) can be found on the website
  {\tt https://www.ttp.kit.edu/$\_${}\,media/progdata/2016/ttp16-023.tgz}.}
\begin{equation}
  \label{eq:21}
  \alpha_s^{(n_l+1)} = \zeta_{\alpha_s} \alpha_s^{(n_l)}
  \,,
\end{equation}
with
\begin{equation}
  \label{eq:22}
 \zeta_{\alpha_s} = 1 + \frac{1}{6} \frac{\alpha_s^{(n_l)}}{\pi} 
 \log\left(\frac{\mu^2}{m^2(\mu^2)}\right) + {\cal O}(\alpha_s^2)
 \,,
\end{equation}
where results up to four-loop order can be found in
Refs.~\cite{Schroder:2005hy,Chetyrkin:2005ia}. In our case we only
need three-loop corrections which have been computed for the first
time in Ref.~\cite{Chetyrkin:1997un}.  Inserting Eq.~(\ref{eq:21})
into the equation for $z_m$ leads to
\begin{equation}
  \label{eq:19}
  c_m(n_l) = c_m(n_l + 1)|_{\alpha_s^{(n_l+1)} \to
    \alpha_s^{(n_l)}} + \delta c_m(n_l) 
  \,,
\end{equation}
with
\begin{align}
%
%
 & \delta c_m^{(2)} =
\frac{l_{\mu }^2}{6}+\frac{2 l_{\mu }}{9}
\,,
\\
%
%
  &\delta c_m^{(3)} =\bigg [ 
\bigg(-\frac{\zeta _3}{18}+\frac{\pi
   ^2}{9}+\frac{117}{32}+\frac{1}{27} \pi ^2 \log
   (2)\bigg) l_{\mu }+\frac{5 l_{\mu }^3}{8}+\frac{25
   l_{\mu }^2}{9}\nonumber\\
& +\bigg\{-\frac{l_{\mu
   }^3}{36}-\frac{13 l_{\mu
   }^2}{108}+\bigg(-\frac{71}{432}-\frac{\pi
   ^2}{54}\bigg) l_{\mu }\bigg\} n_l-\frac{11}{54}
 \bigg ]\,,
\\
%
%
  &\delta c_m^{(4)} = \bigg [ 
l_{\mu } \bigg(-\frac{110 a_4}{27}-\frac{1439 \pi ^2 \zeta
   _3}{864}+\frac{107515 \zeta _3}{27648}+\frac{1975 \zeta
   _5}{432}-\frac{695 \pi ^4}{15552}+\frac{676601 \pi
   ^2}{77760}\nonumber\\
&+\frac{18532949}{373248}-\frac{55 \log
   ^4(2)}{324}-\frac{11}{81} \pi ^2 \log
   ^2(2)-\frac{271}{162} \pi ^2 \log (2)\bigg)\nonumber\\
&+n_l
   \bigg\{l_{\mu } \bigg(\frac{4 a_4}{27}-\frac{241 \zeta
   _3}{144}+\frac{61 \pi ^4}{3888}-\frac{1057 \pi
   ^2}{1296}-\frac{502145}{93312}+\frac{\log
   ^4(2)}{162}\nonumber\\
&+\frac{1}{81} \pi ^2 \log
   ^2(2)-\frac{11}{162} \pi ^2 \log
   (2)\bigg)+\bigg(-\frac{7 \zeta _3}{18}-\frac{25 \pi
   ^2}{108}-\frac{11233}{2592}-\frac{1}{54} \pi ^2 \log
   (2)\bigg) l_{\mu }^2\nonumber\\
&-\frac{83 l_{\mu
   }^4}{432}-\frac{1171 l_{\mu }^3}{864}+\frac{11 \pi
   ^2}{648}+\frac{12295}{46656}\bigg\}+\frac{83099 \zeta
   _3}{20736}+\bigg(-\frac{17 \zeta _3}{18}+\frac{19 \pi
   ^2}{18}+\frac{442177}{10368}\nonumber\\
&+\frac{19}{54} \pi ^2 \log
   (2)\bigg) l_{\mu }^2+\frac{431 l_{\mu
   }^4}{216}+\frac{8869 l_{\mu }^3}{648}+n_l^2
   \bigg\{\bigg(\frac{7 \zeta _3}{108}+\frac{13 \pi
   ^2}{648}+\frac{2353}{46656}\bigg) l_{\mu
   }+\frac{l_{\mu }^4}{216}\nonumber\\
&+\frac{13 l_{\mu
   }^3}{432}+\bigg(\frac{89}{1296}+\frac{\pi
   ^2}{108}\bigg) l_{\mu }^2\bigg\}-\frac{11 \pi
   ^2}{108}-\frac{209567}{23328}-\frac{11}{324} \pi ^2
   \log (2)
 \bigg ]
 \,,
\end{align}
with
\begin{equation}
l_\mu = \log\left(\frac{\mu^2}{{m^2(\mu)}} \right)\,, \quad
a_n = \mathrm{Li}_n\left(\frac{1}{2}\right)\,.
\label{eq::an}
\end{equation}



\section{\label{sec::concl}Conclusions}

The main result of this paper is the calculation of the four-loop coefficient
in the relation between the $\overline{\rm MS}$ and on-shell heavy quark mass.
Up to the reduction to master integrals the calculation is performed
analytically. However, most of the master integrals are only known
numerically. For QCD, we managed to obtain an uncertainty of 0.2\% for the
four-loop coefficient.

We have also computed the coefficients of the individual colour structures.
It is interesting to note that the large coefficients ($z_m^{FAAA}$ and
$z_m^{FAAL}$) are known to high precision and furthermore also have large
colour factors. Thus, they dominate the numerical result obtained after
specifying $N_c$, in particular the physical result for $N_c=3$.  Some
coefficients are known to high relative precision, others have uncertainties
of about 30\%. There is one coefficient ($z_m^{d_{FA}}$) with an uncertainty
which is larger than the central value.  Fortunately, it has only a minor
numerical contribution to $z_m$.

In this paper several applications have been discussed. Among them is the
numerical analysis of the heavy quark relation for the top, bottom and charm
quark. Furthermore, the relations between the $\overline{\rm MS}$ and several
threshold masses are investigated. We have shown that the latter have
well-behaved perturbative expansions, in particular for the top and bottom
quark. We have furthermore investigated the dependence of the PS, RS and
RS$^\prime$ masses on the factorization scale. It turns out that for bottom
and charm $\mu_f=2$~GeV is a reasonable choice. For the top quark we observe
that for $\mu_f=80$~GeV the four-loop corrections are small.  The numerical
results presented in Section~\ref{sec::appl} are easily reproduced with the
help of {\tt RunDec}~\cite{Chetyrkin:2000yt} and {\tt
  CRunDec}~\cite{Schmidt:2012az} where the latest results for the mass
relations are implemented.


\section*{Acknowledgements}

We are thankful to Tobias Huber for advice on numerical evaluation of MB
integrals and to Kirill Melnikov for
carefully reading the manuscript.
We thank the High Performance Computing Center Stuttgart (HLRS) and
the Supercomputing Center of Lomonosov Moscow State University~\cite{LMSU} for
providing computing time used for the numerical computations with {\tt
  FIESTA}.  P.M was supported in part by the EU Network HIGGSTOOLS
PITN-GA-2012-316704.  The work of V.S. was supported by the Alexander von
Humboldt Foundation (Humboldt Forschungspreis).


\newpage

\begin{appendix}


\section{\label{app::intfam}Integral families}

\begin{figure}[h]
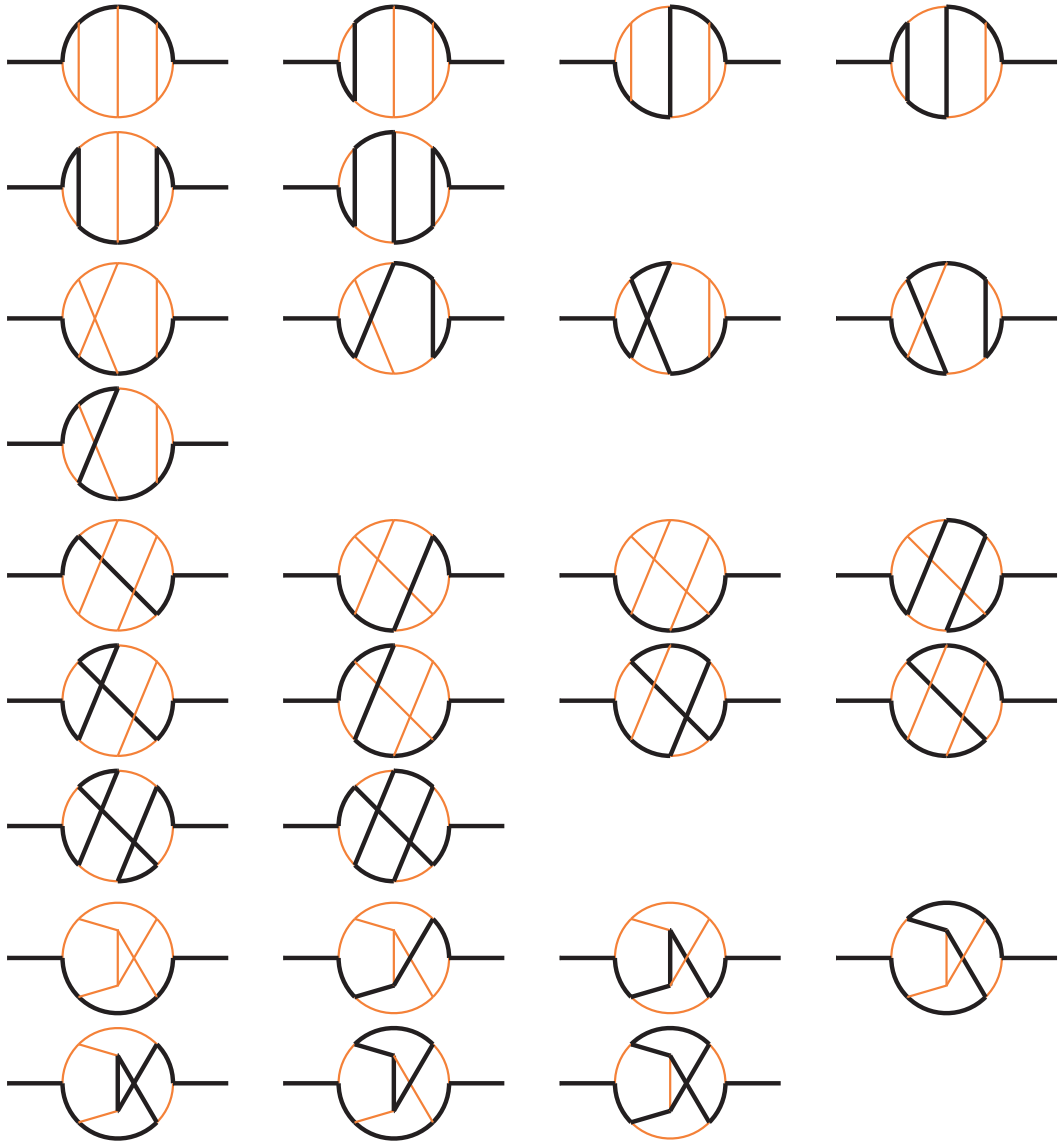

  \begin{center}
  \includegraphics[width=0.9\textwidth]{allTopos4L1.epsi}
  \includegraphics[width=0.9\textwidth]{allTopos4L2.epsi}
  \includegraphics[width=0.9\textwidth]{allTopos4L3.epsi}
  \includegraphics[width=0.9\textwidth]{allTopos4L4.epsi}
  \caption{\label{fig::intfam_ini}
    Integral families needed at four-loop order. Thick black lines
    indicate massive and thin orange lines massless particles.}
  \end{center}
\end{figure}

Graphical representation of the 102 integral families is shown in
Figs.~\ref{fig::intfam_ini} to~\ref{fig::intfam_fin}. They are obtained from
Fig.~\ref{fig::4l_prototype} by introducing a throughgoing massive line.
Note that tables are only required for 100 families  since the
colour factors of the diagrams mapped to two families are zero.

\begin{figure}[h]
  \begin{center}
  \includegraphics[width=0.9\textwidth]{allTopos4L5.epsi}
  \includegraphics[width=0.9\textwidth]{allTopos4L6.epsi}
  \includegraphics[width=0.9\textwidth]{allTopos4L7.epsi}
  \includegraphics[width=0.9\textwidth]{allTopos4L8.epsi}
  \caption{Four-loop families (continued).}
  \end{center}
\end{figure}

\begin{figure}[h]
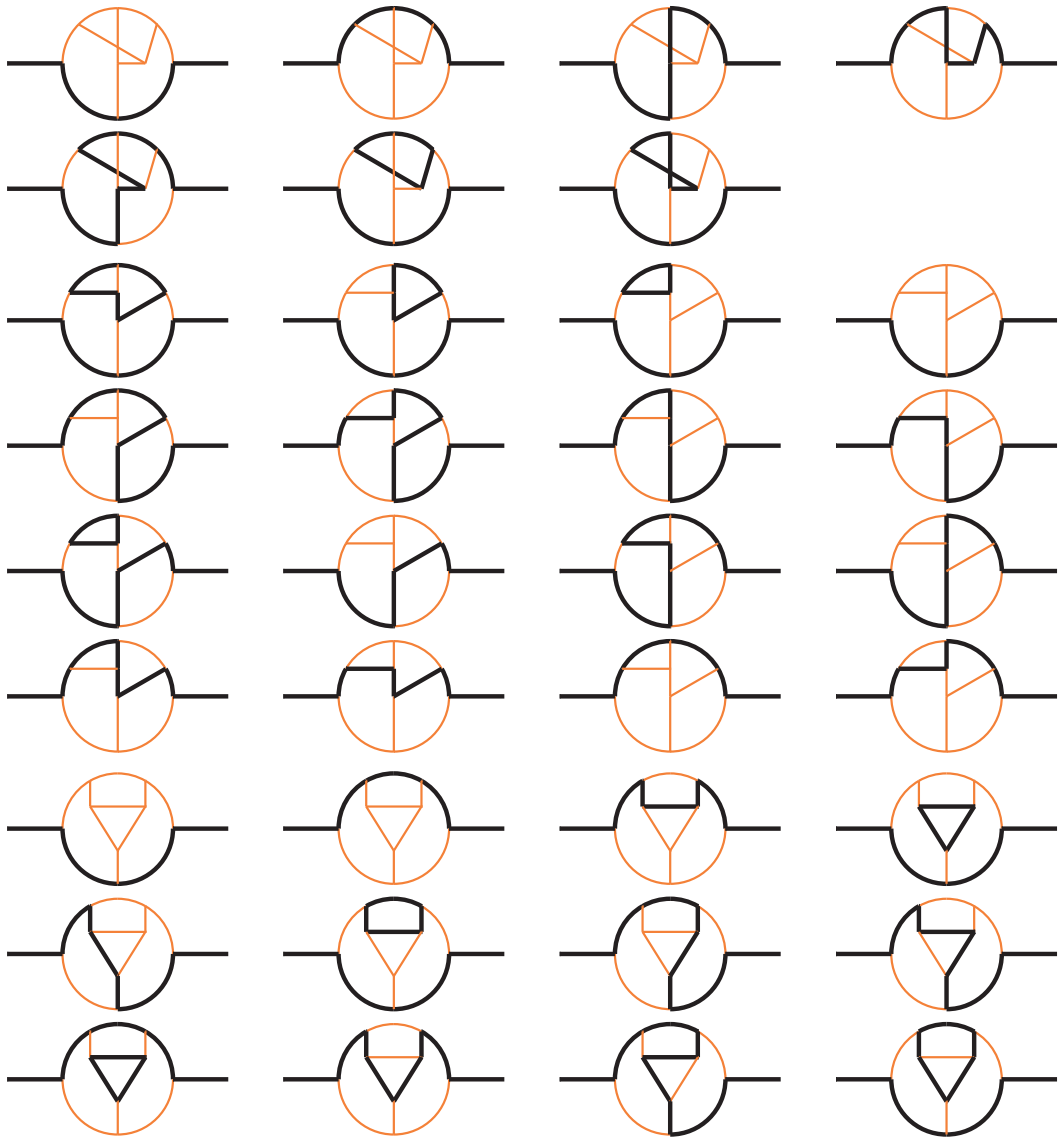

  \begin{center}
  \includegraphics[width=0.9\textwidth]{allTopos4L9.epsi}
  \includegraphics[width=0.9\textwidth]{allTopos4L10.epsi}
  \includegraphics[width=0.9\textwidth]{allTopos4L11.epsi}
  \caption{\label{fig::intfam_fin}Four-loop families (continued).}
  \end{center}
\end{figure}

\clearpage


\section{\label{app::ana_res3}Analytic results for $z_m$ up to three loops}

In this Appendix we present analytic results for $z_m$ up to three loops
including higher order terms in $\epsilon$ which might be important in case
the relation between the $\overline{\rm MS}$ and on-shell mass is used in
divergent expressions. For $\mu^2=M^2$ our results read
\begin{align}
&\nonumber z_m^{(1)} =   \left(\frac{\zeta (3)}{3}-8-\frac{\pi ^2}{6}-\frac{3 \pi
   ^4}{640}\right) \epsilon ^3 C_F+\left(\frac{\zeta
   (3)}{4}-4-\frac{\pi ^2}{12}\right) \epsilon ^2
   C_F\\
&\nonumber +\bigg(-2-\frac{\pi ^2}{16}\bigg) \epsilon  C_F-C_F \,, \\
&\nonumber z_m^{(2)} = \epsilon  \bigg(C_A C_F \bigg(6 a_4+\frac{13 \zeta
   _3}{4}-\frac{7 \pi ^4}{80}+\frac{271 \pi
   ^2}{1152}-\frac{8581}{768}+\frac{\log
   ^4(2)}{4}\\
&\nonumber +\frac{1}{2} \pi ^2 \log ^2(2)-\frac{3}{2} \pi
   ^2 \log (2)\bigg)+C_F^2 \bigg(-12 a_4-\frac{33 \zeta
   _3}{4}+\frac{7 \pi ^4}{40}\\
&\nonumber -\frac{213 \pi
   ^2}{128}-\frac{91}{256}-\frac{\log ^4(2)}{2}-\pi ^2
   \log ^2(2)+3 \pi ^2 \log (2)\bigg)\\
&\nonumber + T C_F
   \bigg(-\frac{7 \zeta _3}{2}-\frac{227 \pi
   ^2}{288}+\frac{1133}{192}+\pi ^2 \log (2)\bigg)
   n_h+\bigg(\zeta _3+\frac{97 \pi
   ^2}{288}+\frac{581}{192}\bigg) T C_F
   n_l\bigg)\\
&\nonumber +\epsilon ^2 \bigg(C_A C_F \bigg(36 a_4+36
   a_5-\frac{11 \pi ^2 \zeta _3}{16}+\frac{3929 \zeta
   _3}{288}-\frac{609 \zeta _5}{16}-\frac{1087 \pi
   ^4}{2560}\\
&\nonumber +\frac{1537 \pi
   ^2}{2304}-\frac{58543}{1536}-\frac{3 \log
   ^5(2)}{10}+\frac{3 \log ^4(2)}{2}-\pi ^2 \log ^3(2)+3
   \pi ^2 \log ^2(2)\\
&\nonumber +\frac{13}{60} \pi ^4 \log
   (2)-\frac{11}{2} \pi ^2 \log (2)\bigg)+C_F^2
   \bigg(-72 a_4-72 a_5+\frac{11 \pi ^2 \zeta
   _3}{8}-\frac{1195 \zeta _3}{32}\\
&\nonumber +\frac{609 \zeta
   _5}{8}+\frac{5119 \pi ^4}{7680}-\frac{1639 \pi
   ^2}{256}-\frac{1905}{512}+\frac{3 \log ^5(2)}{5}-3 \log
   ^4(2)+2 \pi ^2 \log ^3(2)\\
&\nonumber -6 \pi ^2 \log
   ^2(2)-\frac{13}{30} \pi ^4 \log (2)+11 \pi ^2 \log
   (2)\bigg)+T C_F n_h \bigg(-24 a_4-\frac{1273 \zeta
   _3}{72}\\
&\nonumber +\frac{93 \pi ^4}{640}-\frac{1553 \pi
   ^2}{576}+\frac{8135}{384}-\log ^4(2)-2 \pi ^2 \log
   ^2(2)+5 \pi ^2 \log (2)\bigg)\\
&\nonumber +\bigg(\frac{239 \zeta
   _3}{72}+\frac{199 \pi ^4}{1920}+\frac{643 \pi
   ^2}{576}+\frac{4079}{384}\bigg) T C_F n_l\bigg)\\
&\nonumber +C_A
   C_F \bigg(\frac{3 \zeta _3}{8}+\frac{\pi
   ^2}{12}-\frac{1111}{384}-\frac{1}{4} \pi ^2 \log
   (2)\bigg)\\
&\nonumber +C_F^2 \bigg(-\frac{3 \zeta _3}{4}-\frac{5
   \pi ^2}{16}+\frac{7}{128}+\frac{1}{2} \pi ^2 \log
   (2)\bigg)\\
&\nonumber +\bigg(\frac{143}{96}-\frac{\pi ^2}{6}\bigg)
   T C_F n_h+\bigg(\frac{71}{96}+\frac{\pi ^2}{12}\bigg)
   T C_F n_l \,,\\
&\nonumber z_m^{(3)} = 
 C_A C_F{}^2 \bigg(-\frac{4 a_4}{3}-\frac{19 \pi ^2 \zeta _3}{16}-\frac{773 \zeta _3}{96}+\frac{45 \zeta _5}{16}+\frac{65 \pi ^4}{432}+\frac{509 \pi ^2}{576}+\frac{13189}{4608} \\
& \nonumber -\frac{\log ^4(2)}{18}-\frac{31}{36} \pi ^2 \log ^2(2)-\frac{31}{72} \pi ^2 \log (2)\bigg)+C_A{}^2 C_F \bigg(\frac{11 a_4}{3}+\frac{51 \pi ^2 \zeta _3}{64}+\frac{1343 \zeta _3}{288} \\
& \nonumber -\frac{65 \zeta _5}{32}-\frac{179 \pi ^4}{3456}-\frac{1955 \pi ^2}{3456}-\frac{1322545}{124416}+\frac{11 \log ^4(2)}{72}+\frac{11}{36} \pi ^2 \log ^2(2)-\frac{115}{72} \pi ^2 \log (2)\bigg) \\
& \nonumber +T n_h \bigg(C_A C_F \bigg(-\frac{4 a_4}{3}+\frac{\pi ^2 \zeta _3}{8}-\frac{109 \zeta _3}{144}-\frac{5 \zeta _5}{8}-\frac{43 \pi ^4}{1080}-\frac{449 \pi ^2}{144}+\frac{144959}{15552}-\frac{\log ^4(2)}{18} \\
& \nonumber +\frac{1}{18} \pi ^2 \log ^2(2)+\frac{32}{9} \pi ^2 \log (2)\bigg)+C_F{}^2 \bigg(\frac{8 a_4}{3}-\frac{53 \zeta _3}{24}+\frac{91 \pi ^4}{2160}-\frac{85 \pi ^2}{108}+\frac{1067}{576}+\frac{\log ^4(2)}{9} \\
& \nonumber -\frac{1}{9} \pi ^2 \log ^2(2)+\frac{8}{9} \pi ^2 \log (2)\bigg)\bigg)+T n_l \bigg(C_A C_F \bigg(-\frac{4 a_4}{3}+\frac{89 \zeta _3}{144}+\frac{19 \pi ^4}{2160}+\frac{175 \pi ^2}{432} \\
& \nonumber +\frac{70763}{15552}-\frac{\log ^4(2)}{18}-\frac{1}{9} \pi ^2 \log ^2(2)+\frac{11}{18} \pi ^2 \log (2)\bigg)+C_F{}^2 \bigg(\frac{8 a_4}{3}+\frac{55 \zeta _3}{24}-\frac{119 \pi ^4}{2160} \\
& \nonumber +\frac{13 \pi ^2}{18}+\frac{1283}{576}+\frac{\log ^4(2)}{9}+\frac{2}{9} \pi ^2 \log ^2(2)-\frac{11}{9} \pi ^2 \log (2)\bigg)\bigg)+C_F{}^3 \bigg(-12 a_4-\frac{\pi ^2 \zeta _3}{16} \\
& \nonumber -\frac{81 \zeta _3}{16}+\frac{5 \zeta _5}{8}-\frac{\pi ^4}{48}-\frac{613 \pi ^2}{192}-\frac{2969}{768}-\frac{\log ^4(2)}{2}+\frac{1}{2} \pi ^2 \log ^2(2)+\frac{29}{4} \pi ^2 \log (2)\bigg) \\
& \nonumber +\bigg(\frac{2 \zeta _3}{9}+\frac{13 \pi ^2}{108}-\frac{5917}{3888}\bigg) T^2 C_F n_h n_l+\bigg(\frac{11 \zeta _3}{18}+\frac{4 \pi ^2}{135}-\frac{9481}{7776}\bigg) T^2 C_F n_h{}^2 \\
& \nonumber +\bigg(-\frac{7 \zeta _3}{18}-\frac{13 \pi ^2}{108}-\frac{2353}{7776}\bigg) T^2 C_F n_l{}^2 \\
 &\nonumber + \epsilon \Bigg \{
 \bigg(-\frac{\zeta _3{}^2}{2}+\frac{21}{8} \pi ^2 \log (2) \zeta _3+\frac{1621 \pi ^2 \zeta _3}{192}-\frac{1761 \zeta _3}{16}-300 a_4+\frac{103 \zeta _5}{4}-72 a_5 \\
 &\nonumber +\frac{3 \log ^5(2)}{5}+\frac{1}{8} \pi ^2 \log ^4(2)-\frac{25 \log ^4(2)}{2}-\pi ^2 \log ^3(2)-\frac{1}{8} \pi ^4 \log ^2(2)-\frac{93}{4} \pi ^2 \log ^2(2)\\
 &\nonumber -\frac{141}{160} \pi ^4 \log (2)+\frac{317}{4} \pi ^2 \log (2)-\frac{5 \pi ^6}{189}+\frac{437 \pi ^4}{1280}+3 a_4 \pi ^2-\frac{110185 \pi ^2}{6144}-\frac{15709}{512}\bigg) C_F{}^3 \\
& \nonumber +C_A \bigg(-\frac{249 \zeta _3{}^2}{16}+7 \pi ^2 \log (2) \zeta _3-\frac{3125 \pi ^2 \zeta _3}{384}-\frac{767 \zeta _3}{36}+\frac{34 a_4}{9}+\frac{5935 \zeta _5}{48}-\frac{200 a_5 }{3} \\
& \nonumber +\frac{5 \log ^5(2)}{9}+\frac{1}{3} \pi ^2 \log ^4(2)+\frac{17 \log ^4(2)}{108}+\frac{181}{54} \pi ^2 \log ^3(2)-\frac{1}{3} \pi ^4 \log ^2(2) \\
& \nonumber -\frac{4765}{216} \pi ^2 \log ^2(2)+\frac{911 \pi ^4 \log (2)}{1728}-\frac{1181}{54} \pi ^2 \log (2)-\frac{7709 \pi ^6}{60480}+\frac{10219 \pi ^4}{3240}+8 a_4 \pi ^2 \\
& \nonumber -\frac{174769 \pi ^2}{55296}+\frac{339421}{27648}\bigg) C_F{}^2+n_l{}^2 T^2 \bigg(-\frac{197 \zeta _3}{54}-\frac{23 \pi ^4}{270}-\frac{679 \pi ^2}{864}-\frac{131425}{46656}\bigg) C_F \\
& \nonumber +n_h{}^2 T^2 \bigg(\frac{16 a_4}{3}+\frac{1247 \zeta _3}{270}+\frac{2 \log ^4(2)}{9}-\frac{2}{9} \pi ^2 \log ^2(2)-\frac{8}{45} \pi ^2 \log (2)-\frac{31 \pi ^4}{1080} \\
& \nonumber +\frac{5803 \pi ^2}{7200}-\frac{2404781}{233280}\bigg) C_F+n_h n_l T^2 \bigg(\frac{16 a_4}{3}+\frac{163 \zeta _3}{27}+\frac{2 \log ^4(2)}{9}+\frac{4}{9} \pi ^2 \log ^2(2) \\
& \nonumber -\frac{16}{9} \pi ^2 \log (2)+\frac{5 \pi ^4}{72}+\frac{289 \pi ^2}{432}-\frac{314485}{23328}\bigg) C_F+C_A{}^2 \bigg(\frac{137 \zeta _3{}^2}{16}-\frac{133}{32} \pi ^2 \log (2) \zeta _3 \\
& \nonumber +\frac{43 \pi ^2 \zeta _3}{48}+\frac{7433 \zeta _3}{432}+\frac{658 a_4}{9}-\frac{2939 \zeta _5}{48}+\frac{154 a_5}{3}-\frac{77 \log ^5(2)}{180}-\frac{19}{96} \pi ^2 \log ^4(2) \\
& \nonumber +\frac{329 \log ^4(2)}{108}-\frac{77}{54} \pi ^2 \log ^3(2)+\frac{19}{96} \pi ^4 \log ^2(2)+\frac{1819}{108} \pi ^2 \log ^2(2)-\frac{187 \pi ^4 \log (2)}{4320} \\
& \nonumber -\frac{3835}{432} \pi ^2 \log (2)+\frac{9469 \pi ^6}{120960}-\frac{31319 \pi ^4}{20736}-\frac{19 a_4 \pi ^2}{4}-\frac{17873 \pi ^2}{6912}-\frac{52167985}{746496}\bigg) C_F \\
& \nonumber +n_h T \bigg(\bigg(-\frac{328 a_4}{9}-\frac{211 \pi ^2 \zeta _3}{48}-\frac{499 \zeta _3}{18}+\frac{28 \zeta _5}{3}+16 a_5 -\frac{2 \log ^5(2)}{15}-\frac{41 \log ^4(2)}{27} \\
& \nonumber +\frac{2}{9} \pi ^2 \log ^3(2)-\frac{175}{27} \pi ^2 \log ^2(2)+\frac{29}{180} \pi ^4 \log (2)+\frac{1151}{54} \pi ^2 \log (2)+\frac{5393 \pi ^4}{12960} \\
& \nonumber -\frac{136901 \pi ^2}{13824}+\frac{19129}{1152}\bigg) C_F{}^2+C_A \bigg(\frac{19 \zeta _3{}^2}{8}+\frac{21}{8} \pi ^2 \log (2) \zeta _3+\frac{41 \pi ^2 \zeta _3}{12}-\frac{44381 \zeta _3}{432} \\
& \nonumber -\frac{1228 a_4}{9}-\frac{143 \zeta _5}{12}-8 a_5 +\frac{\log ^5(2)}{15}+\frac{1}{8} \pi ^2 \log ^4(2)-\frac{307 \log ^4(2)}{54}-\frac{1}{9} \pi ^2 \log ^3(2) \\
& \nonumber -\frac{1}{8} \pi ^4 \log ^2(2)-\frac{697}{27} \pi ^2 \log ^2(2)-\frac{29}{360} \pi ^4 \log (2)+\frac{2147}{54} \pi ^2 \log (2)-\frac{41 \pi ^6}{1080} \\
& \nonumber +\frac{18151 \pi ^4}{25920}+3 a_4 \pi ^2-\frac{8677 \pi ^2}{864}+\frac{6253805}{93312}\bigg) C_F\bigg)+n_l T \bigg(\bigg(\frac{496 a_4}{9}-\frac{73 \pi ^2 \zeta _3}{48} \\
& \nonumber +\frac{491 \zeta _3}{18}-\frac{131 \zeta _5}{3}+\frac{112 a_5}{3}-\frac{14 \log ^5(2)}{45}+\frac{62 \log ^4(2)}{27}-\frac{28}{27} \pi ^2 \log ^3(2) \\
& \nonumber +\frac{124}{27} \pi ^2 \log ^2(2)-\frac{17}{540} \pi ^4 \log (2)-\frac{256}{27} \pi ^2 \log (2)-\frac{11527 \pi ^4}{25920}+\frac{89507 \pi ^2}{13824}+\frac{54083}{3456}\bigg) C_F{}^2 \\
& \nonumber +C_A \bigg(-\frac{248 a_4}{9}+\frac{13 \pi ^2 \zeta _3}{24}+\frac{2329 \zeta _3}{432}+\frac{455 \zeta _5}{24}-\frac{56 a_5 }{3}+\frac{7 \log ^5(2)}{45}-\frac{31 \log ^4(2)}{27} \\
& \nonumber +\frac{14}{27} \pi ^2 \log ^3(2)-\frac{62}{27} \pi ^2 \log ^2(2)+\frac{17 \pi ^4 \log (2)}{1080}+\frac{128}{27} \pi ^2 \log (2) \\
& \nonumber +\frac{11297 \pi ^4}{25920}+\frac{2587 \pi ^2}{864}+\frac{2963153}{93312}\bigg) C_F\bigg)
 \Bigg \}
\,,
\end{align}
with
\[
a_4 = \mathrm{Li}_4\left(\frac{1}{2}\right)\,.
\]
The logarithmic contributions can be found in Appendix~\ref{app::ren}.


\section{\label{app::ren}Renormalization scale dependence of $z_m^{(4)}$}

In this Appendix we present the dependence of $z_m(\mu)$ and 
$c_m(\mu)$ on $\log(\mu)$. The corresponding analytic expressions are easily
constructed from Eqs.~(\ref{eq::OS2MS}) and~(\ref{eq::MS2OS}) by taking the
derivative with respect to $\mu^2$ and exploiting the fact that $M$ is
$\mu$-independent. The $\mu$-dependence of $m(\mu)$ and $\alpha_s(\mu)$ is
governed by corresponding renormalization group equations which are needed
to four- and three-loop accuracy, respectively.

Our results read
\begin{align}
%
%
&  z_m^{(1),\mathrm{log}}  = -\frac{3}{4} C_F L_M\,, \\
%
%
&  z_m^{(2),\mathrm{log}}  =L_M \bigg(-\frac{185 C_A C_F}{96}+\frac{13}{24} T C_F n_h+\frac{13}{24} T C_F n_l+\frac{21
   C_F^2}{32}\bigg) \nonumber\\
& + L_M^2 \bigg\{-\frac{11 C_A C_F}{32}+\frac{1}{8} T C_F n_h+\frac{1}{8} T C_F n_l+\frac{9
   C_F^2}{32}\bigg\}\,, \\
%
%
&  z_m^{(3),\mathrm{log}}  = L_M \bigg\{ T n_h \bigg(C_A C_F \bigg(\frac{\zeta _3}{2}+\frac{\pi ^2 l_2}{6}-\frac{13 \pi
   ^2}{36}+\frac{583}{108}\bigg) \nonumber \\ 
&+C_F^2 \bigg(-\frac{\zeta _3}{4}-\frac{1}{3} \pi ^2
   l_2+\frac{\pi ^2}{3}-\frac{151}{384}\bigg)\bigg)+C_A C_F^2 \bigg(-\frac{53 \zeta
   _3}{32}+\frac{53 \pi ^2 l_2}{48}-\frac{61 \pi ^2}{96}+\frac{5813}{1536}\bigg)\nonumber \\ 
&+C_A^2 C_F
   \bigg(\frac{11 \zeta _3}{16}-\frac{11}{24} \pi ^2 l_2+\frac{11 \pi
   ^2}{72}-\frac{13243}{1728}\bigg)+T n_l \bigg(C_A C_F \bigg(\frac{\zeta _3}{2}+\frac{\pi
   ^2 l_2}{6}+\frac{7 \pi ^2}{72}+\frac{869}{216}\bigg)\nonumber \\ 
&+C_F^2 \bigg(-\frac{\zeta
   _3}{4}-\frac{1}{3} \pi ^2 l_2+\frac{7 \pi
   ^2}{48}+\frac{65}{384}\bigg)\bigg)+\bigg(\frac{\pi ^2}{18}-\frac{143}{108}\bigg) T^2
   C_F n_h n_l\nonumber \\ 
&+\bigg(\frac{\pi ^2}{9}-\frac{197}{216}\bigg) T^2 C_F n_h^2+C_F^3
   \bigg(\frac{9 \zeta _3}{16}-\frac{3}{8} \pi ^2 l_2+\frac{15 \pi
   ^2}{64}-\frac{489}{512}\bigg)+\bigg(-\frac{89}{216}-\frac{\pi ^2}{18}\bigg) T^2 C_F
   n_l^2\bigg\} \nonumber\\
&+ L_M^2 \bigg\{T n_h \bigg(\frac{373 C_A C_F}{288}-\frac{13 C_F^2}{32}\bigg)+T n_l
   \bigg(\frac{373 C_A C_F}{288}-\frac{13 C_F^2}{32}\bigg)+\frac{109}{64} C_A
   C_F^2\nonumber \\ 
&-\frac{2341 C_A^2 C_F}{1152}-\frac{13}{36} T^2 C_F n_h n_l-\frac{13}{72} T^2 C_F
   n_h^2-\frac{13}{72} T^2 C_F n_l^2-\frac{27 C_F^3}{128}\bigg\} \nonumber \\
&+ L_M^3 \bigg\{T n_h \bigg(\frac{11 C_A C_F}{72}-\frac{3 C_F^2}{32}\bigg)+T n_l
   \bigg(\frac{11 C_A C_F}{72}-\frac{3 C_F^2}{32}\bigg)+\frac{33}{128} C_A
   C_F^2\nonumber \\ 
&-\frac{121}{576} C_A^2 C_F-\frac{1}{18} T^2 C_F n_h n_l-\frac{1}{36} T^2 C_F
   n_h^2-\frac{1}{36} T^2 C_F n_l^2-\frac{9 C_F^3}{128}\bigg\}\,,
\\
%
%
&  z_m^{(4),\mathrm{log}}  = L_M \bigg\{\bigg(\frac{3 l_2{}^4}{8}-\frac{3}{8} \pi ^2 l_2{}^2-\frac{351 \pi ^2 l_2}{64}+9
   a_4+\frac{3 \pi ^2 \zeta _3}{64}+\frac{663 \zeta _3}{128}\nonumber\\
& -\frac{15 \zeta _5}{32}+\frac{\pi
   ^4}{64}+\frac{1241 \pi ^2}{512}+\frac{18505}{4096}\bigg) C_F{}^4+C_A \bigg(-\frac{4
   l_2{}^4}{3}+\frac{97}{48} \pi ^2 l_2{}^2\nonumber\\
& +\frac{7595 \pi ^2 l_2}{384}-32 a_4+\frac{23 \pi ^2
   \zeta _3}{32}-\frac{2149 \zeta _3}{256}-\frac{25 \zeta _5}{64}-\frac{49 \pi
   ^4}{288}-\frac{4677 \pi ^2}{512}-\frac{55487}{3072}\bigg) C_F{}^3\nonumber\\
& +C_A{}^2 \bigg(-\frac{77
   l_2{}^4}{288}-\frac{187}{72} \pi ^2 l_2{}^2+\frac{1123 \pi ^2 l_2}{1152}-\frac{77
   a_4}{12}-\frac{989 \pi ^2 \zeta _3}{256}-\frac{6781 \zeta _3}{256}+\frac{965 \zeta
   _5}{128}\nonumber\\
& +\frac{6257 \pi ^4}{13824}+\frac{3575 \pi ^2}{1536}+\frac{8025563}{331776}\bigg)
   C_F{}^2+\frac{2489 n_h n_l{}^2 T^3 C_F}{1296}\nonumber\\
& +n_h{}^2 n_l T^3 \bigg(-\zeta _3-\frac{3 \pi
   ^2}{20}+\frac{3677}{1296}\bigg) C_F+n_h{}^3 T^3 \bigg(-\frac{2 \zeta _3}{3}-\frac{4 \pi
   ^2}{135}+\frac{4865}{3888}\bigg) C_F\nonumber\\
& +n_l{}^3 T^3 \bigg(\frac{\zeta _3}{3}+\frac{13 \pi
   ^2}{108}+\frac{1301}{3888}\bigg) C_F+C_A{}^3 \bigg(\frac{121 l_2{}^4}{288}+\frac{121}{144}
   \pi ^2 l_2{}^2-\frac{1367 \pi ^2 l_2}{288}\nonumber\\
& +\frac{121 a_4}{12}+\frac{561 \pi ^2 \zeta
   _3}{256}+\frac{1223 \zeta _3}{96}-\frac{495 \zeta _5}{128}-\frac{1969 \pi
   ^4}{13824}-\frac{19873 \pi ^2}{13824}-\frac{4722671}{124416}\bigg) C_F\nonumber\\
& -\frac{d_{FF}
   n_h}{4 N_c}-\frac{d_{FF} n_l}{4 N_c}+\frac{15 d_{FF} n_h   \zeta _3}{8 N_c}
+\frac{15 d_{FF} n_l \zeta _3}{8 N_c}
-\frac{15 d_{FA}   \zeta _3}{16 N_c}+\frac{d_{FA}}{8 N_c}\nonumber\\
& +n_l{}^2 T^2
   \bigg(\bigg(-\frac{l_2{}^4}{9}-\frac{2}{9} \pi ^2 l_2{}^2+\frac{11 \pi ^2 l_2}{9}-\frac{8
   a_4}{3}-\frac{11 \zeta _3}{8}+\frac{11 \pi ^4}{216}-\frac{21 \pi
   ^2}{32}-\frac{50131}{20736}\bigg) C_F{}^2\nonumber\\
& +C_A \bigg(\frac{l_2{}^4}{18}+\frac{1}{9} \pi ^2
   l_2{}^2-\frac{11 \pi ^2 l_2}{18}+\frac{4 a_4}{3}-\frac{37 \zeta _3}{16}-\frac{\pi
   ^4}{216}-\frac{29 \pi ^2}{36}-\frac{15953}{2592}\bigg) C_F\bigg)
\nonumber\\
& 
+n_h n_l T^2 \bigg(\bigg(-\frac{2 l_2{}^4}{9}-\frac{1}{9} \pi ^2
   l_2{}^2+\frac{\pi ^2 l_2}{3}-\frac{16 a_4}{3}+\zeta _3+\frac{\pi ^4}{216}-\frac{\pi
   ^2}{864}-\frac{41383}{10368}\bigg) C_F{}^2\nonumber\\
& +C_A \bigg(\frac{l_2{}^4}{9}+\frac{1}{18} \pi ^2
   l_2{}^2-\frac{25 \pi ^2 l_2}{6}+\frac{8 a_4}{3}-\frac{\pi ^2 \zeta _3}{8}-\frac{\zeta
   _3}{2}+\frac{5 \zeta _5}{8}+\frac{17 \pi ^4}{432}\nonumber\\
& +\frac{1345 \pi
   ^2}{432}-\frac{26213}{1296}\bigg) C_F\bigg)+n_h{}^2 T^2
   \bigg(\bigg(-\frac{l_2{}^4}{9}+\frac{1}{9} \pi ^2 l_2{}^2-\frac{8 \pi ^2 l_2}{9}-\frac{8
   a_4}{3}+\frac{19 \zeta _3}{8}\nonumber\\
& -\frac{5 \pi ^4}{108}+\frac{1757 \pi
   ^2}{2160}-\frac{32635}{20736}\bigg) C_F{}^2+C_A \bigg(\frac{l_2{}^4}{18}-\frac{1}{18} \pi
   ^2 l_2{}^2-\frac{32 \pi ^2 l_2}{9}+\frac{4 a_4}{3}-\frac{\pi ^2 \zeta _3}{8}\nonumber\\
& +\frac{29 \zeta
   _3}{16}+\frac{5 \zeta _5}{8}+\frac{19 \pi ^4}{432}+\frac{7211 \pi
   ^2}{2160}-\frac{36473}{2592}\bigg) C_F\bigg)+n_l T \bigg(\bigg(\frac{5
   l_2{}^4}{12}-\frac{2}{3} \pi ^2 l_2{}^2\nonumber\\
& -\frac{311 \pi ^2 l_2}{48}+10 a_4+\frac{\pi ^2 \zeta
   _3}{16}+\frac{69 \zeta _3}{32}+\frac{5 \zeta _5}{4}+\frac{179 \pi ^4}{2880}+\frac{175 \pi
   ^2}{64}+\frac{883}{512}\bigg) C_F{}^3\nonumber\\
& +C_A \bigg(\frac{29 l_2{}^4}{72}+\frac{14}{9} \pi ^2
   l_2{}^2-\frac{1075 \pi ^2 l_2}{288}+\frac{29 a_4}{3}+\frac{19 \pi ^2 \zeta
   _3}{16}+\frac{391 \zeta _3}{32}-\frac{25 \zeta _5}{8}\nonumber\\
& -\frac{2567 \pi ^4}{8640}+\frac{365
   \pi ^2}{384}+\frac{12563}{10368}\bigg) C_F{}^2+C_A{}^2 \bigg(-\frac{11
   l_2{}^4}{36}-\frac{11}{18} \pi ^2 l_2{}^2+\frac{251 \pi ^2 l_2}{72}\nonumber\\
& -\frac{22
   a_4}{3}-\frac{51 \pi ^2 \zeta _3}{64}+\frac{7 \zeta _3}{32}+\frac{15 \zeta
   _5}{32}+\frac{223 \pi ^4}{3456}+\frac{1991 \pi ^2}{1152}+\frac{601319}{20736}\bigg)
   C_F\bigg)\nonumber\\
& +n_h T \bigg(\bigg(\frac{5 l_2{}^4}{12}-\frac{5}{12} \pi ^2 l_2{}^2-\frac{129 \pi
   ^2 l_2}{16}+10 a_4+\frac{\pi ^2 \zeta _3}{16}+\frac{177 \zeta _3}{32}+\frac{5 \zeta
   _5}{4}\nonumber\\
& -\frac{31 \pi ^4}{2880}+\frac{4481 \pi ^2}{1152}+\frac{991}{512}\bigg) C_F{}^3+C_A
   \bigg(\frac{29 l_2{}^4}{72}+\frac{37}{72} \pi ^2 l_2{}^2-\frac{13 \pi ^2 l_2}{96}\nonumber\\
& +\frac{29
   a_4}{3}+\frac{35 \pi ^2 \zeta _3}{32}+\frac{7 \zeta _3}{8}-\frac{85 \zeta _5}{32}+\frac{29
   \pi ^4}{4320}-\frac{527 \pi ^2}{1728}-\frac{10771}{2592}\bigg) C_F{}^2\nonumber\\
& +C_A{}^2
   \bigg(-\frac{11 l_2{}^4}{36}-\frac{11}{72} \pi ^2 l_2{}^2+\frac{139 \pi ^2
   l_2}{12}-\frac{22 a_4}{3}-\frac{29 \pi ^2 \zeta _3}{64}-\frac{57 \zeta _3}{16}-\frac{5
   \zeta _5}{4}\nonumber\\
& -\frac{239 \pi ^4}{3456}-\frac{28735 \pi ^2}{3456}+\frac{895403}{20736}\bigg)
   C_F\bigg)\bigg\} \nonumber \\
& + L_M^2 \bigg\{T^2 n_h^2 \bigg(C_A C_F \bigg(-\frac{\zeta _3}{4}-\frac{1}{12} \pi ^2
   l_2+\frac{\pi ^2}{3}-\frac{9707}{2304}\bigg)+C_F^2 \bigg(\frac{\zeta _3}{8}+\frac{\pi
   ^2 l_2}{6}\nonumber \\
& -\frac{5 \pi ^2}{24}+\frac{923}{2304}\bigg)\bigg)+T^2 n_h n_l \bigg(C_A C_F
   \bigg(-\frac{\zeta _3}{2}-\frac{1}{6} \pi ^2 l_2+\frac{5 \pi
   ^2}{24}-\frac{8123}{1152}\bigg)\nonumber \\
& +C_F^2 \bigg(\frac{\zeta _3}{4}+\frac{\pi ^2
   l_2}{3}-\frac{25 \pi ^2}{96}+\frac{383}{1152}\bigg)\bigg)+T n_h \bigg(C_A C_F^2
   \bigg(\frac{\zeta _3}{64}-\frac{103}{96} \pi ^2 l_2\nonumber \\
& +\frac{175 \pi
   ^2}{192}-\frac{10853}{2304}\bigg)+C_A^2 C_F \bigg(\frac{11 \zeta _3}{32}+\frac{11 \pi
   ^2 l_2}{24}-\frac{55 \pi ^2}{96}+\frac{6527}{512}\bigg)\nonumber \\
& +C_F^3 \bigg(\frac{3 \zeta
   _3}{32}+\frac{5 \pi ^2 l_2}{16}-\frac{31 \pi ^2}{128}+\frac{57}{256}\bigg)\bigg)+C_A
   C_F^3 \bigg(\frac{357 \zeta _3}{256}-\frac{119}{128} \pi ^2 l_2\nonumber \\
& +\frac{287 \pi
   ^2}{512}-\frac{6477}{2048}\bigg)+C_A^2 C_F^2 \bigg(-\frac{649 \zeta
   _3}{256}+\frac{649 \pi ^2 l_2}{384}-\frac{715 \pi
   ^2}{768}+\frac{373625}{36864}\bigg)\nonumber \\
& +C_A^3 C_F \bigg(\frac{121 \zeta
   _3}{128}-\frac{121}{192} \pi ^2 l_2+\frac{121 \pi ^2}{576}-\frac{337657}{27648}\bigg)+\nonumber \\
& T^2
   n_l^2 \bigg(C_A C_F \bigg(-\frac{\zeta _3}{4} -\frac{1}{12} \pi ^2 l_2-\frac{\pi
   ^2}{8}-\frac{6539}{2304}\bigg)+C_F^2 \bigg(\frac{\zeta _3}{8}+\frac{\pi ^2
   l_2}{6}-\frac{5 \pi ^2}{96}-\frac{157}{2304}\bigg)\bigg)\nonumber \\
& +T n_l \bigg(C_A C_F^2
   \bigg(\frac{\zeta _3}{64}-\frac{103}{96} \pi ^2 l_2+\frac{185 \pi
   ^2}{384}-\frac{7883}{2304}\bigg)\nonumber \\
& +C_A^2 C_F \bigg(\frac{11 \zeta _3}{32}+\frac{11 \pi ^2
   l_2}{24}+\frac{11 \pi ^2}{192}+\frac{5559}{512}\bigg)+C_F^3 \bigg(\frac{3 \zeta
   _3}{32}+\frac{5 \pi ^2 l_2}{16}-\frac{11 \pi
   ^2}{64}+\frac{3}{256}\bigg)\bigg)\nonumber \\
& +\frac{125}{144} T^3 C_F n_h
   n_l^2+\bigg(\frac{161}{144}-\frac{\pi ^2}{12}\bigg) T^3 C_F n_h^2
   n_l+\bigg(\frac{197}{432}-\frac{\pi ^2}{18}\bigg) T^3 C_F n_h^3\nonumber \\
& +C_F^4 \bigg(-\frac{27
   \zeta _3}{128}+\frac{9 \pi ^2 l_2}{64}-\frac{45 \pi
   ^2}{512}+\frac{2889}{4096}\bigg)+\bigg(\frac{89}{432}+\frac{\pi ^2}{36}\bigg) T^3 C_F
   n_l^3\bigg\} \nonumber \\
&+L_M^3 \bigg\{T^2 n_h n_l \bigg(\frac{97 C_F^2}{288}-\frac{91 C_A C_F}{72}\bigg)+T^2
   n_h^2 \bigg(\frac{97 C_F^2}{576}-\frac{91 C_A C_F}{144}\bigg)\nonumber \\
& +T n_h
   \bigg(-\frac{389}{288} C_A C_F^2+\frac{1163}{576} C_A^2 C_F+\frac{9
   C_F^3}{64}\bigg)+T^2 n_l^2 \bigg(\frac{97 C_F^2}{576}-\frac{91 C_A
   C_F}{144}\bigg)\nonumber \\
& +T n_l \bigg(-\frac{389}{288} C_A C_F^2+\frac{1163}{576} C_A^2
   C_F+\frac{9 C_F^3}{64}\bigg)-\frac{45}{64} C_A C_F^3+\frac{21361 C_A^2
   C_F^2}{9216}\nonumber \\
& -\frac{27995 C_A^3 C_F}{13824}+\frac{13}{72} T^3 C_F n_h
   n_l^2+\frac{13}{72} T^3 C_F n_h^2 n_l\nonumber \\
& +\frac{13}{216} T^3 C_F n_h^3+\frac{13}{216}
   T^3 C_F n_l^3+\frac{45 C_F^4}{1024}\bigg\} \nonumber \\
&+L_M^4 \bigg\{T^2 n_h n_l \bigg(\frac{11 C_F^2}{192}-\frac{11 C_A C_F}{96}\bigg)+T^2
   n_h^2 \bigg(\frac{11 C_F^2}{384}-\frac{11 C_A C_F}{192}\bigg)\nonumber \\
& +T n_h
   \bigg(-\frac{121}{768} C_A C_F^2+\frac{121}{768} C_A^2 C_F+\frac{9
   C_F^3}{256}\bigg)+T^2 n_l^2 \bigg(\frac{11 C_F^2}{384}-\frac{11 C_A
   C_F}{192}\bigg)\nonumber \\
& +T n_l \bigg(-\frac{121}{768} C_A C_F^2+\frac{121}{768} C_A^2
   C_F+\frac{9 C_F^3}{256}\bigg)-\frac{99 C_A C_F^3}{1024}+\frac{1331 C_A^2
   C_F^2}{6144}\nonumber \\
& -\frac{1331 C_A^3 C_F}{9216}+\frac{1}{48} T^3 C_F n_h n_l^2+\frac{1}{48}
   T^3 C_F n_h^2 n_l+\frac{1}{144} T^3 C_F n_h^3\nonumber \\
& +\frac{1}{144} T^3 C_F n_l^3+\frac{27
   C_F^4}{2048}\bigg\}
\,,
\end{align}
with 
\begin{equation}
d_{FF} = d_F^{abcd}d_F^{abcd} , \, d_{FA} = d_F^{abcd}d_A^{abcd} ,\, 
L_M = \log\left ( \frac{\mu^2}{M^2}\right ), l_2 = \log(2)\,. 
\label{eq::dFFdFA}
\end{equation}

We present the $\mu$ dependence of the four-loop term of the
inverted relation in the form
\begin{eqnarray}
  c_m^{(4),\rm log} &=& - z_m^{(4),\rm log}|_{L_M\to l_m} + \delta c_m^{(4),\rm log}
  \,,
\end{eqnarray}
where
\begin{align}
&\nonumber  \delta c_m^{(4),\rm log}
   = l_m \bigg(T n_h \bigg(C_A C_F^2 \bigg(2 a_4-\frac{3 \pi ^2 \zeta _3}{16}+\frac{89 \zeta
   _3}{48}+\frac{15 \zeta _5}{16}+\frac{l_2^4}{12}-\frac{1}{12} \pi ^2 l_2^2 \\
&\nonumber -\frac{85
   \pi ^2 l_2}{16}+\frac{43 \pi ^4}{720}+\frac{43 \pi
   ^2}{9}-\frac{95551}{20736}\bigg)+C_F^3 \bigg(-4 a_4+\frac{21 \zeta
   _3}{8}-\frac{l_2^4}{6}+\frac{1}{6} \pi ^2 l_2^2 \\
&\nonumber -\frac{11 \pi ^2 l_2}{8}-\frac{91 \pi
   ^4}{1440}+\frac{641 \pi ^2}{576}-\frac{917}{384}\bigg)\bigg)+C_A C_F^3 \bigg(2
   a_4+\frac{57 \pi ^2 \zeta _3}{32}+\frac{1635 \zeta _3}{128} \\
&\nonumber -\frac{135 \zeta
   _5}{32}+\frac{l_2^4}{12}+\frac{31}{24} \pi ^2 l_2^2+\frac{35 \pi ^2
   l_2}{192}-\frac{65 \pi ^4}{288}-\frac{827 \pi ^2}{768}+\frac{191}{3072}\bigg) \\
&\nonumber +C_A^2
   C_F^2 \bigg(-\frac{11 a_4}{2}-\frac{153 \pi ^2 \zeta _3}{128}-\frac{2779 \zeta
   _3}{384}+\frac{195 \zeta _5}{64}-\frac{11 l_2^4}{48}-\frac{11}{24} \pi ^2
   l_2^2+\frac{491 \pi ^2 l_2}{192} \\
&\nonumber +\frac{179 \pi ^4}{2304}+\frac{1831 \pi
   ^2}{2304}-\frac{16873}{165888}\bigg)+T n_l \bigg(C_A C_F^2 \bigg(2 a_4-\frac{5 \zeta
   _3}{24}+\frac{l_2^4}{12}+\frac{1}{6} \pi ^2 l_2^2 \\
&\nonumber -\frac{43 \pi ^2 l_2}{48}-\frac{19
   \pi ^4}{1440}-\frac{385 \pi ^2}{576}+\frac{62885}{20736}\bigg)+C_F^3 \bigg(-4
   a_4-\frac{33 \zeta _3}{8}-\frac{l_2^4}{6}-\frac{1}{3} \pi ^2 l_2^2 \\
&\nonumber +\frac{43 \pi ^2
   l_2}{24}+\frac{119 \pi ^4}{1440}-\frac{97 \pi
   ^2}{96}-\frac{1295}{384}\bigg)\bigg)+C_F^4 \bigg(18 a_4+\frac{3 \pi ^2 \zeta
   _3}{32}+\frac{459 \zeta _3}{64}-\frac{15 \zeta _5}{16} \\
&\nonumber +\frac{3 l_2^4}{4}-\frac{3}{4}
   \pi ^2 l_2^2-\frac{339 \pi ^2 l_2}{32}+\frac{\pi ^4}{32}+\frac{1181 \pi
   ^2}{256}+\frac{11135}{2048}\bigg) \\
&\nonumber +\bigg(-\frac{\zeta _3}{3}-\frac{25 \pi
   ^2}{144}-\frac{1861}{5184}\bigg) T^2 C_F^2 n_h n_l+\bigg(-\frac{11 \zeta
   _3}{12}-\frac{11 \pi ^2}{360}+\frac{4943}{10368}\bigg) T^2 C_F^2
   n_h^2 \\
&\nonumber +\bigg(\frac{7 \zeta _3}{12}+\frac{25 \pi ^2}{144}-\frac{8665}{10368}\bigg) T^2
   C_F^2 n_l^2\bigg) \\
&\nonumber + l_m^2 \bigg(T n_h \bigg(C_A C_F^2 \bigg(-\frac{21 \zeta _3}{32}-\frac{5}{16} \pi ^2
   l_2+\frac{65 \pi ^2}{96}-\frac{751}{144}\bigg) \\
&\nonumber +C_F^3 \bigg(\frac{3 \zeta
   _3}{16}+\frac{5 \pi ^2 l_2}{8}-\frac{25 \pi ^2}{64}-\frac{237}{512}\bigg)\bigg)+C_A
   C_F^3 \bigg(\frac{165 \zeta _3}{64}-\frac{55}{32} \pi ^2 l_2+\frac{275 \pi
   ^2}{256}-\frac{8661}{2048}\bigg) \\
&\nonumber +C_A^2 C_F^2 \bigg(-\frac{165 \zeta
   _3}{128}+\frac{55 \pi ^2 l_2}{64}-\frac{55 \pi ^2}{192}+\frac{126869}{18432}\bigg) \\
&\nonumber +T
   n_l \bigg(C_A C_F^2 \bigg(-\frac{21 \zeta _3}{32}-\frac{5}{16} \pi ^2 l_2-\frac{35 \pi
   ^2}{192}-\frac{1519}{576}\bigg) \\
&\nonumber +C_F^3 \bigg(\frac{3 \zeta _3}{16}+\frac{5 \pi ^2
   l_2}{8}-\frac{25 \pi ^2}{64}-\frac{237}{512}\bigg)\bigg)+\bigg(\frac{623}{576}-\frac{5
   \pi ^2}{48}\bigg) T^2 C_F^2 n_h n_l \\
&\nonumber +\bigg(\frac{1163}{1152}-\frac{5 \pi
   ^2}{24}\bigg) T^2 C_F^2 n_h^2+\bigg(\frac{83}{1152}+\frac{5 \pi ^2}{48}\bigg) T^2
   C_F^2 n_l^2+\frac{2169 C_F^4}{1024}\bigg) \\
&\nonumber +l_m^3 \bigg(T n_h \bigg(\frac{45 C_F^3}{128}-\frac{1061}{576} C_A C_F^2\bigg)+T n_l
   \bigg(\frac{45 C_F^3}{128}-\frac{1061}{576} C_A C_F^2\bigg)-\frac{693}{512} C_A
   C_F^3 \\
&\nonumber +\frac{14101 C_A^2 C_F^2}{4608}+\frac{67}{144} T^2 C_F^2 n_h
   n_l+\frac{67}{288} T^2 C_F^2 n_h^2+\frac{67}{288} T^2 C_F^2 n_l^2-\frac{45
   C_F^4}{256}\bigg) \\
&\nonumber + l_m^4 \bigg(-\frac{121}{384} T C_A C_F^2 n_h-\frac{121}{384} T C_A C_F^2
   n_l+\frac{1331 C_A^2 C_F^2}{3072}+\frac{11}{96} T^2 C_F^2 n_h n_l \\
& +\frac{11}{192}
   T^2 C_F^2 n_h^2+\frac{11}{192} T^2 C_F^2 n_l^2+\frac{27 C_F^4}{1024}\bigg)
   \,,
\end{align}
with
\begin{equation}
  l_m = \log\left ( \frac{\mu^2}{m^2(\mu)}\right )
\end{equation}

Note that the $\mu$ dependence at four-loop order has also been discussed in
Ref.~\cite{Kataev:2015gvt}.


\section{\label{app::CT}Counterterm contribution to $Z_m^{\rm OS}$}

In this Appendix we show the four-loop contribution to $Z_m^{\rm OS}$
introduced by the lower loop orders. To be precise we write
\begin{eqnarray}
  Z_m^{\rm OS} = \sum_{n\ge0} \left(\frac{\alpha_s(\mu)}{\pi}\right)^n
  Z_m^{{\rm OS},(n)} 
\end{eqnarray}
and split $Z_m^{{\rm OS},(4)}$ according to
\begin{eqnarray} 
  Z_m^{{\rm OS},(4)} &=& Z_m^{{\rm OS},(4)}\Big|_{\mbox{\tiny CT}} +
  Z_m^{{\rm OS},(4)}\Big|_{\mbox{\tiny genuine 4 loop}} 
  \,,
\end{eqnarray}
where $Z_m^{{\rm OS},(4)}|_{\mbox{\tiny CT}}$
contains all counterterm contributions from the renormalization of the strong
coupling constant and quark mass. For $\mu^2=M^2$ it is given by
\begin{align}
  &Z_m ^{{\rm OS},(4)}\Big|_{\mbox{\tiny CT}} = \Bigg \{ 
T^2 n_h n_l \bigg(\frac{11 C_A C_F}{48}+\frac{17
   C_F^2}{64}\bigg)+T^2 n_h^2 \bigg(\frac{11 C_A
   C_F}{96}+\frac{11 C_F^2}{64}\bigg)\nonumber\\
&+T n_h
   \bigg(-\frac{47}{64} C_A C_F^2-\frac{121}{384}
   C_A^2 C_F-\frac{189 C_F^3}{256}\bigg)+T^2 n_l^2
   \bigg(\frac{11 C_A C_F}{96}+\frac{11
   C_F^2}{96}\bigg)\nonumber\\
&+T n_l \bigg(-\frac{121}{192} C_A
   C_F^2-\frac{121}{384} C_A^2 C_F-\frac{135
   C_F^3}{256}\bigg)\nonumber\\
&+\frac{1377 C_A
   C_F^3}{1024}+\frac{1331 C_A^2
   C_F^2}{1536}
+\frac{1331 C_A^3
   C_F}{4608}-\frac{1}{24} T^3 C_F n_h
   n_l^2\nonumber\\
&-\frac{1}{24} T^3 C_F n_h^2 n_l-\frac{1}{72}
   T^3 C_F n_h^3-\frac{1}{72} T^3 C_F n_l^3 +\frac{243
   C_F^4}{256}
\Bigg \} \frac{1}{\epsilon^4} \nonumber \\
& + \Bigg \{ 
T^2 n_h n_l \bigg(\frac{235 C_A C_F}{288}+\frac{337
   C_F^2}{384}\bigg)+T^2 n_h^2 \bigg(\frac{235 C_A
   C_F}{576}+\frac{175 C_F^2}{384}\bigg)\nonumber\\
&+T n_h
   \bigg(-\frac{1399}{512} C_A C_F^2-\frac{2671 C_A^2
   C_F}{2304}-\frac{559 C_F^3}{512}\bigg)+T^2 n_l^2
   \bigg(\frac{235 C_A C_F}{576}+\frac{27
   C_F^2}{64}\bigg) \nonumber\\
&+T n_l \bigg(-\frac{325}{128} C_A
   C_F^2-\frac{2671 C_A^2 C_F}{2304}-\frac{5
   C_F^3}{4}\bigg)+\frac{1}{256} \bigg(904-3 \pi
   ^2\bigg) C_A C_F^3\nonumber\\
&+\frac{3841 C_A^2
   C_F^2}{1024}+\frac{3421 C_A^3
   C_F}{3072}-\frac{7}{48} T^3 C_F n_h
   n_l^2-\frac{7}{48} T^3 C_F n_h^2 n_l\nonumber\\
&-\frac{7}{144}
   T^3 C_F n_h^3-\frac{7}{144} T^3 C_F n_l^3-\frac{45
   C_F^4}{512}
\Bigg \} \frac{1}{\epsilon^3} \nonumber \\
& + \Bigg \{ 
T n_h \bigg(C_A C_F^2 \bigg(-\frac{5 \zeta
   _3}{16}+\frac{7 \pi ^2 l_2}{24}+\frac{157 \pi
   ^2}{256}-\frac{183953}{9216}\bigg) \nonumber \\
&+\bigg(-\frac{11}{24
   } \pi ^2 l_2+\frac{1199 \pi
   ^2}{2304}-\frac{59435}{4608}\bigg) C_A^2 C_F\nonumber \\
&+C_F^3
   \bigg(-\frac{21 \zeta _3}{16}+\frac{5 \pi ^2
   l_2}{4}-\frac{59 \pi
   ^2}{256}-\frac{35327}{3072}\bigg)\bigg)\nonumber \\
&+T^2 n_h^2
   \bigg(C_A C_F \bigg(\frac{\zeta _3}{8}+\frac{\pi ^2
   l_2}{12}-\frac{181 \pi
   ^2}{576}+\frac{7933}{1728}\bigg)+\bigg(-\frac{1}{6}
   \pi ^2 l_2-\frac{101 \pi
   ^2}{384}+\frac{10775}{2304}\bigg) C_F^2\bigg)\nonumber \\
&+T^2
   n_h n_l \bigg(C_A C_F \bigg(\frac{\zeta
   _3}{4}+\frac{\pi ^2 l_2}{6}-\frac{49 \pi
   ^2}{288}+\frac{6745}{864}\bigg)+\bigg(-\frac{1}{3} \pi
   ^2 l_2+\frac{\pi ^2}{24}+\frac{15769}{2304}\bigg)
   C_F^2\bigg)\nonumber \\
&+C_A C_F^3 \bigg(\frac{1353 \zeta
   _3}{512}-\frac{467}{256} \pi ^2 l_2+\frac{4149 \pi
   ^2}{2048}+\frac{20165}{1536}\bigg)\nonumber \\
&+C_A^2 C_F^2
   \bigg(-\frac{277 \zeta _3}{512}+\frac{143 \pi ^2
   l_2}{384}+\frac{\pi ^4}{1440}+\frac{4103 \pi
   ^2}{18432}+\frac{122785}{6144}\bigg)\nonumber \\
&+C_A^3 C_F
   \bigg(-\frac{121 \zeta _3}{128}+\frac{121 \pi ^2
   l_2}{192}-\frac{4477 \pi
   ^2}{27648}+\frac{1953781}{165888}\bigg)\nonumber \\
&+T n_l
   \bigg(C_A C_F^2 \bigg(-\frac{35 \zeta
   _3}{64}+\frac{31 \pi ^2 l_2}{96}-\frac{1963 \pi
   ^2}{2304}-\frac{10477}{768}\bigg)\nonumber \\
&+\bigg(-\frac{11}{24}
   \pi ^2 l_2-\frac{253 \pi
   ^2}{2304}-\frac{50723}{4608}\bigg) C_A^2 C_F+C_F^3
   \bigg(-\frac{33 \zeta _3}{32}+\frac{19 \pi ^2
   l_2}{16}-\frac{683 \pi
   ^2}{512}-\frac{4391}{768}\bigg)\bigg)\nonumber \\
&+T^2 n_l^2
   \bigg(C_A C_F \bigg(\frac{\zeta _3}{8}+\frac{\pi ^2
   l_2}{12}+\frac{83 \pi
   ^2}{576}+\frac{5557}{1728}\bigg)+\bigg(-\frac{1}{6}
   \pi ^2 l_2+\frac{323 \pi
   ^2}{1152}+\frac{871}{384}\bigg)
   C_F^2\bigg)\nonumber \\
&+\bigg(\frac{11 \pi
   ^2}{144}-\frac{1193}{864}\bigg) T^3 C_F n_h^2
   n_l+\bigg(-\frac{977}{864}-\frac{\pi ^2}{144}\bigg)
   T^3 C_F n_h n_l^2 \nonumber \\ 
&+\bigg(\frac{23 \pi
   ^2}{432}-\frac{1409}{2592}\bigg) T^3 C_F
   n_h^3+C_F^4 \bigg(\frac{1107 \zeta
   _3}{256}-\frac{369}{128} \pi ^2 l_2+\frac{297 \pi
   ^2}{128}+\frac{2787}{1024}\bigg)\nonumber \\
&+\bigg(-\frac{761}{259
   2}-\frac{13 \pi ^2}{432}\bigg) T^3 C_F n_l^3
\Bigg \} \frac{1}{\epsilon^2} \nonumber \\
& + \Bigg \{ 
\bigg(\frac{625 l_2^4}{128}+\frac{225}{64} \pi ^2
   l_2^2-\frac{745 \pi ^2 l_2}{16}+\frac{1875
   a_4}{16}+\frac{3 \pi ^2 \zeta _3}{64}+\frac{4695 \zeta
   _3}{64}-\frac{75 \zeta _5}{64}-\frac{2311 \pi
   ^4}{2560}\nonumber \\
&+\frac{65941 \pi
   ^2}{3072}+\frac{44617}{2048}\bigg) C_F^4+C_A
   \bigg(\frac{2557 l_2^4}{768}+\frac{2341}{384} \pi ^2
   l_2^2-\frac{1775 \pi ^2 l_2}{64}+\frac{2557
   a_4}{32}\nonumber \\
&+\frac{317 \pi ^2 \zeta _3}{64} +\frac{39451
   \zeta _3}{512}-\frac{475 \zeta _5}{32}-\frac{56531 \pi
   ^4}{46080}+\frac{169661 \pi
   ^2}{12288}+\frac{169787}{3072}\bigg) C_F^3\nonumber \\
&+C_A^2
   \bigg(-\frac{113 l_2^4}{144}+\frac{79}{288} \pi ^2
   l_2^2+\frac{10145 \pi ^2 l_2}{1152}-\frac{113
   a_4}{6}+\frac{115 \pi ^2 \zeta _3}{768}-\frac{2123
   \zeta _3}{4608}\nonumber \\
&+\frac{187 \zeta _5}{128}-\frac{5039 \pi
   ^4}{92160}+\frac{27593 \pi
   ^2}{12288}+\frac{3298805}{36864}\bigg) C_F^2\nonumber \\
&+n_l^3
   T^3 \bigg(-\frac{77 \zeta _3}{108}-\frac{611 \pi
   ^2}{2592}-\frac{22697}{15552}\bigg) C_F\nonumber \\
&+n_h n_l^2
   T^3 \bigg(-\frac{1}{3} \pi ^2 l_2+\frac{13 \zeta
   _3}{36}+\frac{25 \pi ^2}{864}-\frac{32417}{5184}\bigg)
   C_F\nonumber \\
&+n_h^3 T^3 \bigg(-\frac{1}{3} \pi ^2 l_2+\frac{193
   \zeta _3}{108}+\frac{3749 \pi
   ^2}{12960}-\frac{51857}{15552}\bigg) C_F \nonumber \\
&+n_h^2 n_l
   T^3 \bigg(-\frac{2}{3} \pi ^2 l_2+\frac{103 \zeta
   _3}{36}+\frac{2393 \pi
   ^2}{4320}-\frac{42137}{5184}\bigg) C_F\nonumber \\
&+C_A^3
   \bigg(-\frac{605 l_2^4}{576}-\frac{605}{288} \pi ^2
   l_2^2+\frac{2405 \pi ^2 l_2}{288}-\frac{605
   a_4}{24}-\frac{561 \pi ^2 \zeta _3}{256}-\frac{148433
   \zeta _3}{6912}+\frac{715 \zeta _5}{128}\nonumber \\
&+\frac{25091
   \pi ^4}{69120}+\frac{169519 \pi
   ^2}{165888}+\frac{64234201}{995328}\bigg) C_F+n_l^2
   T^2 \bigg(\bigg(\frac{5 l_2^4}{18}+\frac{5}{9} \pi ^2
   l_2^2-\frac{20 \pi ^2 l_2}{9}\nonumber \\
&+\frac{20
   a_4}{3}+\frac{2257 \zeta _3}{288}-\frac{49 \pi
   ^4}{432}+\frac{5509 \pi
   ^2}{2304}+\frac{77327}{6912}\bigg) C_F^2\nonumber \\
&+C_A
   \bigg(-\frac{5 l_2^4}{36}-\frac{5}{18} \pi ^2
   l_2^2+\frac{10 \pi ^2 l_2}{9}-\frac{10
   a_4}{3}+\frac{85 \zeta _3}{36}+\frac{41 \pi
   ^4}{1080}+\frac{4655 \pi
   ^2}{3456}+\frac{343919}{20736}\bigg)
   C_F\bigg)\nonumber \\
&+n_h^2 T^2 \bigg(\bigg(\frac{5
   l_2^4}{18}+\frac{2}{9} \pi ^2 l_2^2+\frac{43 \pi ^2
   l_2}{18}+\frac{20 a_4}{3}-\frac{369 \zeta
   _3}{32}-\frac{7 \pi ^4}{432}-\frac{79751 \pi
   ^2}{34560}+\frac{109217}{4608}\bigg) C_F^2\nonumber \\
&+C_A
   \bigg(-\frac{5 l_2^4}{36}-\frac{1}{9} \pi ^2
   l_2^2+\frac{53 \pi ^2 l_2}{9}-\frac{10
   a_4}{3}+\frac{\pi ^2 \zeta _3}{8}-\frac{721 \zeta
   _3}{72}-\frac{5 \zeta _5}{8}-\frac{23 \pi
   ^4}{2160}\nonumber \\
&-\frac{27391 \pi
   ^2}{5760}+\frac{610895}{20736}\bigg) C_F\bigg)\nonumber \\
&+n_h
   n_l T^2 \bigg(\bigg(\frac{5 l_2^4}{9}+\frac{7}{9} \pi
   ^2 l_2^2+\frac{\pi ^2 l_2}{24}+\frac{40
   a_4}{3}-\frac{79 \zeta _3}{48}-\frac{7 \pi
   ^4}{54}-\frac{143 \pi
   ^2}{216}+\frac{475145}{13824}\bigg) C_F^2\nonumber \\
&+C_A
   \bigg(-\frac{5 l_2^4}{18}-\frac{7}{18} \pi ^2
   l_2^2+7 \pi ^2 l_2-\frac{20 a_4}{3}+\frac{\pi ^2
   \zeta _3}{8}-\frac{551 \zeta _3}{72}-\frac{5 \zeta
   _5}{8}+\frac{59 \pi ^4}{2160}\nonumber \\
&-\frac{6893 \pi
   ^2}{1728}+\frac{477407}{10368}\bigg) C_F\bigg)\nonumber \\
&+n_l T
   \bigg(\bigg(-\frac{101 l_2^4}{48}-\frac{65}{24} \pi
   ^2 l_2^2+\frac{155 \pi ^2 l_2}{8}-\frac{101
   a_4}{2}-\frac{\pi ^2 \zeta _3}{16}-\frac{5045 \zeta
   _3}{128}+\frac{5 \zeta _5}{8}\nonumber \\
&+\frac{1723 \pi
   ^4}{2880}-\frac{11719 \pi
   ^2}{1024}-\frac{15091}{512}\bigg) C_F^3+C_A
   \bigg(-\frac{137 l_2^4}{288}-\frac{245}{144} \pi ^2
   l_2^2+\frac{371 \pi ^2 l_2}{144}\nonumber \\
&-\frac{137
   a_4}{12}-\frac{19 \pi ^2 \zeta _3}{16}-\frac{1831 \zeta
   _3}{72}+\frac{45 \zeta _5}{16}+\frac{6601 \pi
   ^4}{17280}-\frac{15377 \pi
   ^2}{2304}-\frac{438589}{6912}\bigg) C_F^2\nonumber \\
&+C_A^2
   \bigg(\frac{55 l_2^4}{72}+\frac{55}{36} \pi ^2
   l_2^2-\frac{883 \pi ^2 l_2}{144}+\frac{55
   a_4}{3}+\frac{51 \pi ^2 \zeta _3}{64}+\frac{1949 \zeta
   _3}{288}-\frac{65 \zeta _5}{32}\nonumber \\
&-\frac{817 \pi
   ^4}{3456}-\frac{10337 \pi
   ^2}{4608}-\frac{1610957}{27648}\bigg) C_F\bigg)+n_h T
   \bigg(\bigg(-\frac{19 l_2^4}{8}-\frac{11}{8} \pi ^2
   l_2^2\nonumber \\
&+\frac{4679 \pi ^2 l_2}{768}-57 a_4-\frac{\pi ^2
   \zeta _3}{16}-\frac{5777 \zeta _3}{1536}+\frac{5 \zeta
   _5}{8}+\frac{139 \pi ^4}{480}-\frac{13829 \pi
   ^2}{27648}\nonumber \\
&-\frac{819703}{18432}\bigg) C_F^3+C_A
   \bigg(-\frac{5 l_2^4}{18}-\frac{109}{72} \pi ^2
   l_2^2-\frac{711 \pi ^2 l_2}{32}-\frac{20
   a_4}{3}-\frac{105 \pi ^2 \zeta _3}{64}\nonumber \\
&+\frac{16697
   \zeta _3}{768}+\frac{315 \zeta _5}{64}+\frac{2261 \pi
   ^4}{8640}+\frac{83269 \pi
   ^2}{4608}-\frac{5891549}{55296}\bigg) C_F^2\nonumber \\
&+C_A^2
   \bigg(\frac{55 l_2^4}{72}+\frac{77}{72} \pi ^2
   l_2^2-\frac{67 \pi ^2 l_2}{4}+\frac{55
   a_4}{3}+\frac{29 \pi ^2 \zeta _3}{64}+\frac{6305 \zeta
   _3}{288}-\frac{5 \zeta _5}{16}\nonumber \\
&-\frac{355 \pi
   ^4}{3456}+\frac{144577 \pi
   ^2}{13824}-\frac{2188757}{27648}\bigg) C_F\bigg)
 \Bigg \} \frac{1}{\epsilon} \nonumber \\
  & + \Bigg \{ 
\bigg(-\frac{5497 l_2^5}{960}-\frac{25}{32} \pi ^2
   l_2^4+\frac{1113 l_2^4}{16}-\frac{1753}{288} \pi ^2
   l_2^3+\frac{49}{32} \pi ^4 l_2^2+\frac{3193}{24}
   \pi ^2 l_2^2\nonumber \\
&-\frac{597}{32} \pi ^2 \zeta _3
   l_2+\frac{47879 \pi ^4 l_2}{11520}-\frac{580283 \pi ^2
   l_2}{1536}+\frac{273 \zeta _3^2}{128}+\frac{3339
   a_4}{2}+\frac{5497 a_5}{8}\nonumber \\
&-\frac{30763 \pi ^2 \zeta
   _3}{768}+\frac{702863 \zeta _3}{1024}-\frac{263441
   \zeta _5}{512}+\frac{1349 \pi ^6}{7560}-\frac{50003 \pi
   ^4}{11520}-\frac{75 a_4 \pi ^2}{4}\nonumber \\
&+\frac{954883 \pi
   ^2}{9216}+\frac{1233161}{12288}\bigg) C_F^4\nonumber \\
&+C_A
   \bigg(-\frac{11719 l_2^5}{1920}-\frac{41}{32} \pi ^2
   l_2^4+\frac{7675 l_2^4}{192}-\frac{11503}{576} \pi
   ^2 l_2^3+\frac{17}{32} \pi ^4
   l_2^2+\frac{15679}{96} \pi ^2 l_2^2\nonumber \\
&-\frac{789}{32}
   \pi ^2 \zeta _3 l_2+\frac{66053 \pi ^4
   l_2}{23040}-\frac{168439 \pi ^2 l_2}{1024}+\frac{3937
   \zeta _3^2}{64}+\frac{7675 a_4}{8}+\frac{11719
   a_5}{16}\nonumber \\
&+\frac{1529 \pi ^2 \zeta _3}{384}+\frac{2917189
   \zeta _3}{6144}-\frac{895973 \zeta _5}{1024}+\frac{4763
   \pi ^6}{10080}-\frac{10254293 \pi ^4}{737280}-\frac{123
   a_4 \pi ^2}{4}\nonumber \\
&+\frac{7263359 \pi
   ^2}{73728}+\frac{518789}{2048}\bigg) C_F^3\nonumber \\
&+C_A^2
   \bigg(\frac{89 l_2^5}{144}-\frac{97}{384} \pi ^2
   l_2^4-\frac{43913 l_2^4}{3456}-\frac{595}{432} \pi
   ^2 l_2^3+\frac{169}{384} \pi ^4 l_2^2-\frac{11153
   \pi ^2 l_2^2}{1728}\nonumber \\
&-\frac{751}{128} \pi ^2 \zeta _3
   l_2-\frac{11185 \pi ^4 l_2}{6912}+\frac{169367 \pi ^2
   l_2}{1728}+\frac{4813 \zeta _3^2}{512}-\frac{43913
   a_4}{144}-\frac{445 a_5}{6}\nonumber \\
&+\frac{179791 \pi ^2 \zeta
   _3}{9216}-\frac{12445 \zeta _3}{9216}-\frac{24275 \zeta
   _5}{1024}+\frac{19679 \pi ^6}{241920}-\frac{17671879
   \pi ^4}{6635520}-\frac{97 a_4 \pi ^2}{16}\nonumber \\
&+\frac{5799935
   \pi ^2}{221184}+\frac{92980613}{221184}\bigg)
   C_F^2\nonumber \\
&+n_l^3 T^3 \bigg(-\frac{3073 \zeta
   _3}{648}-\frac{259 \pi ^4}{2160}-\frac{6035 \pi
   ^2}{5184}-\frac{620897}{93312}\bigg) C_F\nonumber \\
&+n_h n_l^2
   T^3 \bigg(\frac{5 l_2^4}{9}+\frac{10}{9} \pi ^2
   l_2^2-\frac{31 \pi ^2 l_2}{9}+\frac{40
   a_4}{3}+\frac{1319 \zeta _3}{216}-\frac{289 \pi
   ^4}{2160}\nonumber \\
&+\frac{11 \pi
   ^2}{576}-\frac{974489}{31104}\bigg) C_F+n_h^3 T^3
   \bigg(\frac{5 l_2^4}{9}+\frac{4}{9} \pi ^2
   l_2^2-\frac{83 \pi ^2 l_2}{45}+\frac{40
   a_4}{3}\nonumber \\
&+\frac{34099 \zeta _3}{3240}-\frac{167 \pi
   ^4}{2160}+\frac{220129 \pi
   ^2}{129600}-\frac{8242477}{466560}\bigg) C_F\nonumber \\
&+n_h^2
   n_l T^3 \bigg(\frac{10 l_2^4}{9}+\frac{14}{9} \pi ^2
   l_2^2-\frac{238 \pi ^2 l_2}{45}+\frac{80
   a_4}{3}+\frac{23083 \zeta _3}{1080}-\frac{197 \pi
   ^4}{2160}\nonumber \\
&+\frac{124493 \pi
   ^2}{43200}-\frac{6585109}{155520}\bigg) C_F+C_A^3
   \bigg(\frac{2783 l_2^5}{1440}+\frac{209}{384} \pi ^2
   l_2^4-\frac{5329 l_2^4}{432}\nonumber \\
&+\frac{2783}{432} \pi
   ^2 l_2^3-\frac{209}{384} \pi ^4
   l_2^2-\frac{23429}{432} \pi ^2
   l_2^2+\frac{1463}{128} \pi ^2 \zeta _3 l_2-\frac{7381
   \pi ^4 l_2}{17280}\nonumber \\
&+\frac{67979 \pi ^2
   l_2}{1728}-\frac{1507 \zeta _3^2}{64}-\frac{5329
   a_4}{18}-\frac{2783 a_5}{12}-\frac{187 \pi ^2 \zeta
   _3}{256}-\frac{3503635 \zeta _3}{41472}\nonumber \\
&+\frac{203005
   \zeta _5}{768}-\frac{104159 \pi
   ^6}{483840}+\frac{219365 \pi ^4}{41472}+\frac{209 a_4
   \pi ^2}{16}+\frac{1818263 \pi
   ^2}{331776}+\frac{1829511217}{5971968}\bigg)
   C_F\nonumber \\
&+n_h^2 T^2
   \bigg(\bigg(-\frac{l_2^5}{3}-\frac{443
   l_2^4}{108}-\frac{4}{9} \pi ^2
   l_2^3-\frac{907}{108} \pi ^2 l_2^2+\frac{11 \pi ^4
   l_2}{36}+\frac{16829 \pi ^2 l_2}{540}-\frac{886
   a_4}{9}\nonumber \\
&+40 a_5-\frac{233 \pi ^2 \zeta
   _3}{48}-\frac{1643 \zeta _3}{20}-\frac{385 \zeta
   _5}{24}+\frac{317377 \pi ^4}{414720}-\frac{730669 \pi
   ^2}{38400}+\frac{52140391}{414720}\bigg) C_F^2\nonumber \\
&+C_A
   \bigg(\frac{l_2^5}{6}+\frac{1}{8} \pi ^2
   l_2^4-\frac{233 l_2^4}{27}+\frac{2}{9} \pi ^2
   l_2^3-\frac{1}{8} \pi ^4 l_2^2-\frac{1613}{54} \pi
   ^2 l_2^2+\frac{21}{8} \pi ^2 \zeta _3 l_2\nonumber \\
&-\frac{11
   \pi ^4 l_2}{72}+\frac{27899 \pi ^2 l_2}{540}+\frac{19
   \zeta _3^2}{8}-\frac{1864 a_4}{9}-20 a_5+\frac{175
   \pi ^2 \zeta _3}{48}-\frac{665293 \zeta
   _3}{4320}\nonumber \\
&+\frac{37 \zeta _5}{48}-\frac{41 \pi
   ^6}{1080}+\frac{3851 \pi ^4}{3240}+3 a_4 \pi
   ^2-\frac{18335563 \pi
   ^2}{1036800}+\frac{31646753}{207360}\bigg)
   C_F\bigg)\nonumber \\
&+n_l^2 T^2 \bigg(\bigg(-\frac{23
   l_2^5}{45}+\frac{89 l_2^4}{27}-\frac{46}{27} \pi ^2
   l_2^3+\frac{178}{27} \pi ^2 l_2^2+\frac{61 \pi ^4
   l_2}{540}-\frac{355 \pi ^2 l_2}{27}\nonumber \\
&+\frac{712
   a_4}{9}+\frac{184 a_5}{3}-\frac{95 \pi ^2 \zeta
   _3}{48}+\frac{32747 \zeta _3}{576}-\frac{1657 \zeta
   _5}{24}-\frac{78943 \pi ^4}{414720}\nonumber \\
&+\frac{191657 \pi
   ^2}{13824}+\frac{735901}{13824}\bigg) C_F^2+C_A
   \bigg(\frac{23 l_2^5}{90}-\frac{89
   l_2^4}{54}+\frac{23}{27} \pi ^2 l_2^3-\frac{89}{27}
   \pi ^2 l_2^2\nonumber \\
&-\frac{61 \pi ^4 l_2}{1080}+\frac{355 \pi
   ^2 l_2}{54}-\frac{356 a_4}{9}-\frac{92 a_5}{3}+\frac{37
   \pi ^2 \zeta _3}{48}+\frac{14887 \zeta
   _3}{864}+\frac{1519 \zeta _5}{48}\nonumber \\
&+\frac{325 \pi
   ^4}{324}+\frac{298109 \pi
   ^2}{41472}+\frac{3159661}{41472}\bigg) C_F\bigg)\nonumber \\
&+n_h
   n_l T^2 \bigg(\bigg(-\frac{38 l_2^5}{45}-\frac{31
   l_2^4}{72}-\frac{58}{27} \pi ^2
   l_2^3-\frac{155}{36} \pi ^2 l_2^2+\frac{113 \pi ^4
   l_2}{270}+\frac{191 \pi ^2 l_2}{8}\nonumber \\
&-\frac{31
   a_4}{3}+\frac{304 a_5}{3}-\frac{41 \pi ^2 \zeta
   _3}{6}-\frac{5293 \zeta _3}{144}-\frac{1021 \zeta
   _5}{12}-\frac{20837 \pi ^4}{69120}-\frac{5063 \pi
   ^2}{864}+\frac{14881105}{82944}\bigg) C_F^2\nonumber \\
&+C_A
   \bigg(\frac{19 l_2^5}{45}+\frac{1}{8} \pi ^2
   l_2^4-\frac{185 l_2^4}{18}+\frac{29}{27} \pi ^2
   l_2^3-\frac{1}{8} \pi ^4 l_2^2-35 \pi ^2
   l_2^2+\frac{21}{8} \pi ^2 \zeta _3 l_2-\frac{113 \pi
   ^4 l_2}{540}\nonumber \\
&+\frac{2255 \pi ^2 l_2}{36}+\frac{19 \zeta
   _3^2}{8}-\frac{740 a_4}{3}-\frac{152 a_5}{3}+\frac{53
   \pi ^2 \zeta _3}{12}-\frac{65105 \zeta
   _3}{432}+\frac{389 \zeta _5}{12}-\frac{41 \pi
   ^6}{1080}\nonumber \\
&+\frac{27 \pi ^4}{16}+3 a_4 \pi
   ^2-\frac{254659 \pi
   ^2}{20736}+\frac{4764781}{20736}\bigg) C_F\bigg)\nonumber \\
&+n_h
   T \bigg(\bigg(\frac{57 l_2^5}{20}+\frac{1}{8} \pi ^2
   l_2^4-\frac{2641 l_2^4}{256}+\frac{11}{4} \pi ^2
   l_2^3-\frac{1}{8} \pi ^4 l_2^2-\frac{187}{128} \pi
   ^2 l_2^2+\frac{21}{8} \pi ^2 \zeta _3 l_2\nonumber \\
&-\frac{521
   \pi ^4 l_2}{160}+\frac{823 \pi ^2 l_2}{256}-\frac{\zeta
   _3^2}{2}-\frac{7923 a_4}{32}-342 a_5+\frac{5743 \pi
   ^2 \zeta _3}{192}-\frac{107213 \zeta
   _3}{4608}\nonumber \\
&+\frac{6503 \zeta _5}{32}-\frac{5 \pi
   ^6}{189}+\frac{63667 \pi ^4}{184320}+3 a_4 \pi
   ^2+\frac{2023643 \pi
   ^2}{165888}-\frac{26730911}{110592}\bigg) C_F^3\nonumber \\
&+C_A
   \bigg(\frac{37 l_2^5}{45}-\frac{19}{96} \pi ^2
   l_2^4+\frac{12683 l_2^4}{384}+\frac{503}{108} \pi
   ^2 l_2^3+\frac{19}{96} \pi ^4 l_2^2+\frac{127015
   \pi ^2 l_2^2}{1152}-\frac{133}{32} \pi ^2 \zeta _3
   l_2\nonumber \\
&+\frac{5041 \pi ^4 l_2}{8640}-\frac{18911 \pi ^2
   l_2}{72}-\frac{3219 \zeta _3^2}{128}+\frac{12683
   a_4}{16}-\frac{296 a_5}{3}-\frac{2437 \pi ^2 \zeta
   _3}{256}+\frac{284675 \zeta _3}{512}\nonumber \\
&+\frac{134183 \zeta
   _5}{768}+\frac{7 \pi ^6}{216}-\frac{103817 \pi
   ^4}{55296}-\frac{19 a_4 \pi ^2}{4}+\frac{5872135 \pi
   ^2}{82944}-\frac{183814805}{331776}\bigg)
   C_F^2\nonumber \\
&+C_A^2 \bigg(-\frac{209
   l_2^5}{180}-\frac{13}{24} \pi ^2 l_2^4+\frac{433
   l_2^4}{18}-\frac{319}{108} \pi ^2
   l_2^3+\frac{13}{24} \pi ^4 l_2^2+\frac{1183}{12}
   \pi ^2 l_2^2-\frac{91}{8} \pi ^2 \zeta _3
   l_2\nonumber \\
&+\frac{1243 \pi ^4 l_2}{2160}-\frac{10241 \pi ^2
   l_2}{72}+\frac{65 \zeta _3^2}{32}+\frac{1732
   a_4}{3}+\frac{418 a_5}{3}-\frac{937 \pi ^2 \zeta
   _3}{96}+\frac{430415 \zeta _3}{1152}\nonumber \\
&-\frac{9431 \zeta
   _5}{96}+\frac{22097 \pi ^6}{120960}-\frac{53237 \pi
   ^4}{11520}-13 a_4 \pi ^2+\frac{347143 \pi
   ^2}{10368}-\frac{197507311}{497664}\bigg)
   C_F\bigg)\nonumber \\
&+n_l T \bigg(\bigg(\frac{383
   l_2^5}{120}+\frac{1}{8} \pi ^2 l_2^4-\frac{685
   l_2^4}{24}+\frac{275}{36} \pi ^2 l_2^3-\frac{1}{8}
   \pi ^4 l_2^2-\frac{166}{3} \pi ^2
   l_2^2+\frac{21}{8} \pi ^2 \zeta _3 l_2\nonumber \\
&-\frac{257 \pi
   ^4 l_2}{180}+\frac{579 \pi ^2 l_2}{4}-\frac{\zeta
   _3^2}{2}-685 a_4-383 a_5+\frac{3169 \pi ^2 \zeta
   _3}{192}-\frac{245419 \zeta _3}{768}+\frac{23615 \zeta
   _5}{64}\nonumber \\
&-\frac{5 \pi ^6}{189}+\frac{484063 \pi
   ^4}{184320}+3 a_4 \pi ^2-\frac{1185613 \pi
   ^2}{18432}-\frac{152405}{1024}\bigg) C_F^3\nonumber \\
&+C_A
   \bigg(\frac{175 l_2^5}{144}+\frac{1}{3} \pi ^2
   l_2^4-\frac{1661 l_2^4}{432}+\frac{1199}{216} \pi
   ^2 l_2^3-\frac{1}{3} \pi ^4 l_2^2-\frac{3247}{108}
   \pi ^2 l_2^2+7 \pi ^2 \zeta _3 l_2\nonumber \\
&+\frac{5 \pi ^4
   l_2}{54}-\frac{385 \pi ^2 l_2}{54}-\frac{249 \zeta
   _3^2}{16}-\frac{1661 a_4}{18}-\frac{875
   a_5}{6}-\frac{1763 \pi ^2 \zeta _3}{384}-\frac{87917
   \zeta _3}{576}+\frac{85783 \zeta _5}{384}\nonumber \\
&-\frac{7709
   \pi ^6}{60480}+\frac{2344277 \pi ^4}{829440}+8 a_4 \pi
   ^2-\frac{643145 \pi
   ^2}{13824}-\frac{151453}{512}\bigg) C_F^2\nonumber \\
&+C_A^2
   \bigg(-\frac{253 l_2^5}{180}-\frac{19}{96} \pi ^2
   l_2^4+\frac{3913 l_2^4}{432}-\frac{253}{54} \pi ^2
   l_2^3+\frac{19}{96} \pi ^4 l_2^2+\frac{6235}{216}
   \pi ^2 l_2^2-\frac{133}{32} \pi ^2 \zeta _3
   l_2\nonumber \\
&+\frac{671 \pi ^4 l_2}{2160}-\frac{14093 \pi ^2
   l_2}{432}+\frac{137 \zeta _3^2}{16}+\frac{3913
   a_4}{18}+\frac{506 a_5}{3}-\frac{89 \pi ^2 \zeta
   _3}{48}+\frac{23219 \zeta _3}{1152}-\frac{8791 \zeta
   _5}{48}\nonumber \\
&+\frac{9469 \pi ^6}{120960}-\frac{392849 \pi
   ^4}{103680}-\frac{19 a_4 \pi ^2}{4}-\frac{264949 \pi
   ^2}{20736}-\frac{134979991}{497664}\bigg) C_F\bigg)
 \Bigg \}  \nonumber 
\\
&\nonumber  + \xi \Bigg [
\bigg \{ T n_h \bigg(\frac{1}{256} C_A C_F{}^2-\frac{9
   C_F{}^3}{128}\bigg)+\frac{477 C_A
   C_F{}^3}{2048}-\frac{9}{128} T C_F{}^3 n_l+\frac{297
   C_F{}^4}{512}  \bigg \} \frac{1}{\epsilon^4} \\
&\nonumber  +\bigg \{ T n_h \bigg(-\frac{5}{768} C_A C_F{}^2-\frac{75
   C_F{}^3}{256}\bigg)+\bigg(\frac{2271}{2048}-\frac{3
   \pi ^2}{512}\bigg) C_A C_F{}^3 \\ 
&\nonumber  -\frac{75}{256} T
   C_F{}^3 n_l+\frac{999 C_F{}^4}{1024}  \bigg \} \frac{1}{\epsilon^3} \\
& \nonumber 
+\bigg \{
\bigg(\frac{9 \zeta _3}{1024}-\frac{\pi
   ^4}{5760}+\frac{1}{1024}\bigg) C_A{}^2 C_F{}^2+T n_h
   \bigg(\bigg(\frac{31}{768}+\frac{\pi ^2}{768}\bigg)
   C_A C_F{}^2 \\
&\nonumber +\bigg(\frac{9 \pi
   ^2}{128}-\frac{913}{512}\bigg) C_F{}^3\bigg)+C_A
   C_F{}^3 \bigg(-\frac{9 \zeta _3}{32}+\frac{9 \pi ^2
   l_2}{64}+\frac{37 \pi
   ^2}{2048}+\frac{1387}{256}\bigg)\\
&\nonumber+C_F{}^4
   \bigg(\frac{27 \zeta _3}{64}-\frac{9}{32} \pi ^2
   l_2+\frac{225 \pi
   ^2}{512}+\frac{12069}{2048}\bigg)+\bigg(-\frac{697}{51
   2}-\frac{9 \pi ^2}{128}\bigg) T C_F{}^3 n_l
\bigg \} \frac{1}{\epsilon^2} \\
&\nonumber  +\bigg \{
C_A C_F{}^3 \bigg(-\frac{27 a_4}{8}-\frac{1245 \zeta
   _3}{512}-\frac{9 l_2{}^4}{64}-\frac{9}{32} \pi ^2
   l_2{}^2+\frac{15 \pi ^2 l_2}{16}+\frac{21 \pi
   ^4}{512}+\frac{377 \pi
   ^2}{6144} \\
&\nonumber+\frac{80197}{4096}\bigg)+C_F{}^4
   \bigg(\frac{27 a_4}{4}+\frac{639 \zeta _3}{128}+\frac{9
   l_2{}^4}{32}+\frac{9}{16} \pi ^2 l_2{}^2-\frac{15 \pi
   ^2 l_2}{8}-\frac{63 \pi ^4}{640} \\
&\nonumber+\frac{1659 \pi
   ^2}{1024}+\frac{49519}{4096}\bigg)+\bigg(-\frac{1}{192
   } \pi ^2 \zeta _3+\frac{51 \zeta _3}{1024}-\frac{7
   \zeta _5}{512}-\frac{383 \pi ^4}{552960}+\frac{3 \pi
   ^2}{1024}+\frac{43}{3072}\bigg) C_A{}^2 C_F{}^2 \\
&\nonumber+T n_h
   \bigg(\bigg(\frac{5 \zeta _3}{192}-\frac{5 \pi
   ^2}{2304}-\frac{97}{768}\bigg) C_A C_F{}^2+C_F{}^3
   \bigg(\frac{33 \zeta _3}{16}-\frac{9}{16} \pi ^2
   l_2+\frac{111 \pi
   ^2}{256}-\frac{6899}{1024}\bigg)\bigg) \\
&\nonumber+\bigg(-\frac{1
   5 \zeta _3}{32}-\frac{75 \pi
   ^2}{256}-\frac{4955}{1024}\bigg) T C_F{}^3 n_l
\bigg \} \frac{1}{\epsilon}\\
&\nonumber + T n_h \bigg(C_F{}^3 \bigg(\frac{27 a_4}{2}+\frac{739 \zeta
   _3}{64}+\frac{9 l_2{}^4}{16}+\frac{9}{8} \pi ^2
   l_2{}^2-\frac{51 \pi ^2 l_2}{16}-\frac{9 \pi
   ^4}{128}+\frac{997 \pi
   ^2}{512}-\frac{54873}{2048}\bigg)\\
&\nonumber +\bigg(-\frac{25
   \zeta _3}{576}-\frac{\pi ^4}{768}+\frac{31 \pi
   ^2}{2304}+\frac{1309}{2304}\bigg) C_A
   C_F{}^2\bigg)+C_A C_F{}^3 \bigg(-\frac{45
   a_4}{2}-\frac{81 a_5}{4}+\frac{37 \pi ^2 \zeta
   _3}{128}\\
&\nonumber-\frac{6083 \zeta _3}{512}+\frac{5157 \zeta
   _5}{256}+\frac{27 l_2{}^5}{160}-\frac{15
   l_2{}^4}{16}+\frac{9}{16} \pi ^2 l_2{}^3-\frac{15}{8}
   \pi ^2 l_2{}^2-\frac{63 \pi ^4 l_2}{640}+\frac{143 \pi
   ^2 l_2}{32}\\
&\nonumber+\frac{533 \pi ^4}{2048}+\frac{2297 \pi
   ^2}{3072}+\frac{609965}{8192}\bigg)+C_F{}^4 \bigg(45
   a_4+\frac{81 a_5}{2}-\frac{45 \pi ^2 \zeta
   _3}{64}+\frac{7077 \zeta _3}{256}-\frac{5481 \zeta
   _5}{128}\\
&\nonumber-\frac{27 l_2{}^5}{80}+\frac{15
   l_2{}^4}{8}-\frac{9}{8} \pi ^2 l_2{}^3+\frac{15}{4} \pi
   ^2 l_2{}^2+\frac{63 \pi ^4 l_2}{320}-\frac{143 \pi ^2
   l_2}{16}-\frac{699 \pi ^4}{2560}+\frac{17449 \pi
   ^2}{2048}+\frac{433933}{8192}\bigg)\\
&\nonumber+\bigg(-\frac{25
   \zeta _3{}^2}{512}-\frac{115 \pi ^2 \zeta
   _3}{9216}+\frac{265 \zeta _3}{1024}+\frac{391 \zeta
   _5}{3072}-\frac{5 \pi ^6}{6912}-\frac{349 \pi
   ^4}{110592}+\frac{91 \pi
   ^2}{3072}+\frac{125}{1024}\bigg) C_A{}^2
   C_F{}^2\\
&\nonumber+\bigg(-\frac{125 \zeta _3}{64}-\frac{9 \pi
   ^4}{128}-\frac{697 \pi
   ^2}{512}-\frac{35505}{2048}\bigg) T C_F{}^3 n_l
\Bigg ]\\
&\nonumber + \xi^2 \Bigg [
\bigg \{
\frac{27 C_F{}^4}{1024}-\frac{27 C_A C_F{}^3}{2048}
\bigg \} \frac{1}{\epsilon^4}\\
&\nonumber +\bigg \{
\frac{27 C_F{}^4}{512}-\frac{135 C_A C_F{}^3}{4096}
\bigg \} \frac{1}{\epsilon^3}\\
&\nonumber +\bigg \{
\left(-\frac{1071}{4096}-\frac{27 \pi ^2}{2048}\right) C_A
   C_F{}^3+\left(\frac{117}{256}+\frac{27 \pi
   ^2}{1024}\right) C_F{}^4
\bigg \} \frac{1}{\epsilon^2}\\
&\nonumber  +\bigg \{
\left(-\frac{45 \zeta _3}{512}-\frac{135 \pi
   ^2}{4096}-\frac{2117}{4096}\right) C_A
   C_F{}^3+\left(\frac{45 \zeta _3}{256}+\frac{27 \pi
   ^2}{512}+\frac{169}{256}\right) C_F{}^4
\bigg \} \frac{1}{\epsilon}\\
&\nonumber
+ \left(-\frac{225 \zeta _3}{1024}-\frac{27 \pi
   ^4}{2048}-\frac{1071 \pi
   ^2}{4096}-\frac{14895}{4096}\right) C_A
   C_F{}^3\\
& +\left(\frac{45 \zeta _3}{128}+\frac{27 \pi
   ^4}{1024}+\frac{117 \pi
   ^2}{256}+\frac{1531}{256}\right) C_F{}^4
\Bigg ]
\,,
\end{align}
where $a_n$ and $l_2$ are given in Eqs.~(\ref{eq::an})
and~(\ref{eq::dFFdFA}), respectively.
The QCD gauge parameter $\xi$ is defined via the gluon propagator
\begin{eqnarray}
  D_g^{\mu\nu}(q) &=& -i\,
  \frac{g^{\mu\nu}- \xi \frac{q^\mu q^\nu}{q^2}}
  {q^2+ i\varepsilon}
  \,.
\end{eqnarray}


\section{\label{app::ana_res}Analytic results for $z_m$}

In this Appendix we repeat for convenience the coefficients of
the colour structures presented in Subsection~\ref{sub::SUN}
which are known analytically~\cite{Lee:2013sx}. They are given by
\begin{align}
  &\nonumber  Z_m^{FLLL} =  
   \frac{317 \zeta_3}{432}
  + \frac{71 \pi^4}{4320}
  + \frac{89 \pi^2}{648}
  + \frac{42979}{186624} \,, \\
  &\nonumber  Z_m^{FLLH} =  \frac{5\zeta_3}{144}-\frac{19\pi^4}{480}+\frac{\pi ^2}{6}+\frac{128515}{62208} \,, \\
  &\nonumber  Z_m^{FFLL} = \frac{2
    l_2^5}{45}-\frac{11
    l_2^4}{27}+\frac{4}{27}\pi^2
  l_2^3-\frac{22}{27}\pi
  ^2l_2^2+\frac{31}{540}\pi^4
  l_2+\frac{103}{54}\pi^2
  l_2 \\ 
&\nonumber  \quad -\frac{88a_4}{9}-\frac{16
    a_5}{3}
  +\frac{305\zeta_5}{48}+\frac{3\pi^2\zeta_3}{8}-\frac{2839\zeta_3}{576}+\frac{3683\pi^4}{51840}-\frac{5309\pi
    ^2}{3456}-\frac{2396921}{497664} \,,
\\
  &\nonumber  Z_m^{FALL} =-\frac{1}{45}
  l_2^5+\frac{11
    l_2^4}{54}-\frac{2}{27}\pi^2
  l_2^3+\frac{11}{27}\pi
  ^2l_2^2-\frac{31\pi^4
    l_2}{1080}-\frac{103}{108}\pi
  ^2l_2+\frac{44a_4}{9}\\
  & \quad  +\frac{8
    a_5}{3}
  -\frac{41\zeta_5}{24}-\frac{13\pi^2\zeta_3}{48}-\frac{3245\zeta_3}{576}-\frac{
4723\pi
    ^4}{51840}-\frac{527
    \pi^2}{384}-\frac{2708353}{497664} \,,
\end{align}
where $a_n$ and $l_2$ are given in Eqs.~(\ref{eq::an})
and~(\ref{eq::dFFdFA}), respectively.


\section{\label{app::ZmMS}$Z_m^{\overline{\rm MS}}$ for general SU$(N_c)$
  gauge group}

In this Appendix we present $Z_m^{\overline{\rm MS}}$ up to four-loop
order~\cite{Chetyrkin:1997dh,Vermaseren:1997fq} expressed in terms of
SU$(N_c)$ colour factors. It has been obtained from the quark mass anomalous
dimension given in~\cite{Vermaseren:1997fq}.
\begin{align}
&Z_m^{\overline{\mathrm{MS}}} = 
1 -\frac{3 C_F }{4 \epsilon }\frac{\alpha _s}{\pi }\nonumber\\
&+\bigg\{\bigg[\frac{9 C_F^2}{32}+\frac{11 C_A C_F}{32}-\frac{1}{8} n_f T C_F\bigg]\frac{1}{\epsilon ^2}+\bigg[-\frac{3 C_F^2}{64}-\frac{97 C_A C_F}{192}+\frac{5}{48} n_f T C_F\bigg]\frac{1}{\epsilon }\bigg\} \left( \frac{\alpha _s}{\pi }\right)^2
\nonumber\\
&+\bigg\{\bigg[-\frac{9 C_F^3}{128}-\frac{33}{128} C_A C_F^2-\frac{121}{576} C_A^2 C_F-\frac{1}{36} n_f^2 T^2 C_F+\bigg(\frac{3 C_F^2}{32}+\frac{11 C_A C_F}{72}\bigg) n_f T\bigg]\frac{1}{\epsilon ^3}\nonumber\\
&\quad+\bigg[\frac{9 C_F^3}{256}+\frac{313}{768} C_A C_F^2+\frac{1679 C_A^2 C_F}{3456}+\frac{5}{216} n_f^2 T^2 C_F+\bigg(-\frac{29 C_F^2}{192}-\frac{121 C_A C_F}{432}\bigg) n_f T\bigg]\frac{1}{\epsilon ^2}\nonumber\\
&\quad+\bigg[-\frac{43 C_F^3}{128}+\frac{43}{256} C_A C_F^2-\frac{11413 C_A^2 C_F}{20736}+\frac{35 n_f^2 T^2 C_F}{1296} \nonumber\\
&\quad\quad+n_f T \bigg(\bigg(\frac{23}{96}-\frac{\zeta _3}{4}\bigg) C_F^2+C_A \bigg(\frac{\zeta _3}{4}+\frac{139}{1296}\bigg) C_F\bigg)\bigg]\frac{1}{\epsilon }\bigg\} \left(\frac{\alpha _s}{\pi }\right)^3\nonumber\\
&+\bigg\{\bigg[\frac{27 C_F^4}{2048}+\frac{99 C_A C_F^3}{1024}+\frac{1331
  C_A^2 C_F^2}{6144}+\frac{1331 C_A^3 C_F}{9216}-\frac{1}{144} n_f^3 T^3 C_F
\nonumber\\&\quad\quad
+\bigg(\frac{11 C_F^2}{384}+\frac{11 C_A C_F}{192}\bigg) n_f^2T^2
+\bigg(-\frac{9 C_F^3}{256}-\frac{121}{768} C_A C_F^2-\frac{121}{768} C_A^2 C_F\bigg) n_f T\bigg]\frac{1}{\epsilon ^4}\nonumber\\
&\quad+\bigg[-\frac{27 C_F^4}{2048}-\frac{23}{128} C_A C_F^3-\frac{1285 C_A^2 C_F^2}{2304}-\frac{25201 C_A^3 C_F}{55296}+\frac{5}{864} n_f^3 T^3 C_F\nonumber\\
&\quad\quad+\bigg(-\frac{37 C_F^2}{576}-\frac{43 C_A C_F}{384}\bigg) n_f^2 T^2+\bigg(\frac{23 C_F^3}{256}+\frac{943 C_A C_F^2}{2304}+\frac{1981 C_A^2 C_F}{4608}\bigg) n_f T\bigg]\frac{1}{\epsilon ^3}\nonumber\\
&\quad+\bigg[\frac{2073 C_F^4}{8192}+\frac{527 C_A C_F^3}{4096}+\frac{97661 C_A^2 C_F^2}{221184}+\frac{236333 C_A^3 C_F}{331776}+\frac{35 n_f^3 T^3 C_F}{5184}\nonumber\\
&\quad\quad+n_f^2 T^2 \bigg(\bigg(\frac{1001}{13824}-\frac{\zeta _3}{16}\bigg) C_F^2+C_A \bigg(\frac{\zeta _3}{16}+\frac{217}{4608}\bigg) C_F\bigg)+n_f T \bigg(\bigg(\frac{3 \zeta _3}{16}-\frac{275}{1024}\bigg) C_F^3\nonumber\\
&\quad\quad+C_A \bigg(-\frac{\zeta _3}{64}-\frac{10933}{27648}\bigg) C_F^2+C_A^2 \bigg(-\frac{11 \zeta _3}{64}-\frac{7009}{13824}\bigg) C_F\bigg)\bigg]\frac{1}{\epsilon ^2}\nonumber\\
&\quad+\bigg[\bigg(\frac{21 \zeta _3}{64}+\frac{1261}{8192}\bigg) C_F^4+C_A \bigg(-\frac{79 \zeta _3}{256}-\frac{15349}{12288}\bigg) C_F^3+C_A^2 \bigg(\frac{19 \zeta _3}{128}-\frac{55 \zeta _5}{128}+\frac{34045}{36864}\bigg) C_F^2\nonumber\\
&\quad\quad +n_f^3 T^3 \bigg(\frac{83}{10368}-\frac{\zeta _3}{72}\bigg) C_F+C_A^3 \bigg(-\frac{709 \zeta _3}{4608}+\frac{55 \zeta _5}{128}-\frac{70055}{73728}\bigg) C_F\nonumber\\
&\quad\quad +n_f^2 T^2 \bigg(\bigg(\frac{5 \zeta _3}{32}-\frac{\pi
  ^4}{960}-\frac{19}{1728}\bigg) C_F^2 +C_A \bigg(-\frac{5 \zeta
  _3}{32}+\frac{\pi ^4}{960}-\frac{671}{41472}\bigg) C_F\bigg) 
\nonumber\\&\quad\quad 
+\frac{d_{FA}}{N_c} \bigg(\frac{1}{32}-\frac{15 \zeta  _3}{64}\bigg)
+n_f \bigg(\frac{d_{FF}}{N_c} \bigg(\frac{15 \zeta _3}{32}-\frac{1}{16}\bigg)+T \bigg(\bigg(-\frac{69 \zeta _3}{128}+\frac{15 \zeta _5}{32}+\frac{35}{384}\bigg) C_F^3\nonumber\\
&\quad\quad+C_A \bigg(-\frac{23 \zeta _3}{64}-\frac{5 \zeta _5}{64}+\frac{11
  \pi ^4}{3840}+\frac{8819}{27648}\bigg) C_F^2\nonumber\\ 
&\quad\quad+C_A^2 \bigg(\frac{671 \zeta _3}{768}-\frac{25 \zeta _5}{64}-\frac{11 \pi ^4}{3840}+\frac{65459}{165888}\bigg) C_F\bigg)\bigg)\bigg]\frac{1}{\epsilon }\bigg\} \left(\frac{\alpha _s}{\pi }\right)^4 
\,,
\end{align}
where $d_{FF}$ and $d_{FA}$ are defined in Eq.~(\ref{eq::dFFdFA}).


\end{appendix}

\end{document}